\documentstyle[12pt]{article}
\begin{document}

\def\pslash{\not{{\hskip-.08cm} p}}
\def\kslash{\not{{\hskip-.08cm} k}}
\def\dqq{{dq^+ d^2q_\perp \over 16\pi^3 q^+}}
\def\dkk{{dk^+ d^2k_\perp \over 16\pi^3 k^+}}
\def\eg{{\it e.g.}}
\def\ie{{\it i.e.}}
\def\order{{\cal O}}
\def\calN{{\cal N}}
\def\abinit{{\it ab initio}}
\def\lt{{\mathaccent "7E \lambda}}
\def\lzero{{\Lambda_0}}
\def\lone{{\Lambda_1}}
\def\Lzero{{\Lambda_0^2 \over P^+}}
\def\Lone{{\Lambda_1^2 \over P^+}}
\def\Lam{{\Lambda^2 \over P^+}}
\def\dqt{d\tilde{q}}
\def\dkt{d\tilde{k}}
\def\Dslash{\not{{\hskip-.12cm} D}}
\def\psit{{\tilde{\psi}}}
\def\At{{\tilde{A}}}
\def\logpp{\ln\Bigl({p_1^+ p_2^+ \over \epsilon^2}\Bigr)}
\def\logq{\ln\Bigl({q^+ \over \epsilon}\Bigr)}
\def\loglmu{\ln\Bigl({\Lambda \over \mu}\Bigr)}
\def\epslash{\not{{\hskip-.08cm} \epsilon}}
\def\h0{{\cal H}^\Lambda_0}
\def\v{{\it v}^\Lambda}
\def\H{{\cal H}^\Lambda}
\def\V{{\cal V}^\Lambda}


\pagenumbering{roman}
\setcounter{page}{0}

\title{Hamiltonian Light-Front Field Theory and Quantum
Chromodynamics\footnote{Corrected version of invited lectures presented
at {\it Hadrons 94}, Gramado, Brasil, April, 1994.}}
\author{Robert J. Perry
\\Department of Physics\\The Ohio State University\\Columbus, OH 43210}
\date{June 25, 1997}
\maketitle

\begin{abstract}
\hspace{.10in}

Light-front coordinates offer a scenario in which a constituent
picture of hadron structure can emerge from QCD.  From the point of
view of field theory, there are severe difficulties with any
Hamiltonian formulation.  From the point of view of QCD it is
difficult to see how a constituent picture can emerge at any energy.
The field theoretic difficulties lead us to introduce cutoffs that
violate Lorentz covariance and gauge invariance, and a new
renormalization group formalism based on a similarity transformation
is used to identify cutoff dependence.  Counterterms restore
symmetries broken by the cutoffs, and renormalization group coupling
coherence is used to fix all counterterms required here. The
counterterms contain functions of longitudinal momentum fractions that
severely complicate renormalization, but they also offer possible
resolutions of apparent contradictions between the constituent picture
and QCD.

The similarity transformation and coupling coherence are applied to
QED; and it is shown that the resultant Hamiltonian leads to standard
lowest order bound state results, with the Coulomb interaction
emerging naturally. The same techniques are applied to QCD and with
physically motivated assumptions it is shown that a simple confinement
mechanism appears.  Bare `masses' for quarks and gluons diverge even
when the cutoff is finite,  leading to the question of how hadrons
with finite mass emerge.  Divergences in the spin-independent two-body
interaction between all partons exactly cancel these mass divergences
in color singlet states.  The issue of confinement in the lowest order
analysis is then reduced to a determination of the residual
interactions, which produce logarithmic potentials in both
longitudinal and transverse directions.

\end{abstract}

\newpage

\tableofcontents

\newpage
\pagenumbering{arabic}


\section{Credits and Outline}

I must begin with an apology for the inadequate references to the many
papers upon which this work is based, and to the alternative
light-front approaches that are not discussed.  Avaroth Harindranath
has compiled an excellent list\cite{hari1} of early references on
light-front field theory\cite{dirac}, the infinite momentum
frame\cite{weinberg}, light-front current algebra\cite{current}, and
the light-cone gauge\cite{bassetto}.

There are many relationships between the program I discuss and DLCQ
(discretized light-cone quantization)\cite{schladming}, but I discuss
DLCQ little. Initial work on solving problems in QCD using light-front
field theory was done by Brodsky and Lepage, with Brodsky and Pauli
later developing DLCQ.  There were certainly many developments
preceding this work, but very few explicit calculations existed
before the DLCQ program was launched.  This program has been carried
forward by many people.

In these lectures I outline a program designed to solve QCD at low
energies, {\bf if} it is possible to develop a reasonable few-body {\it
approximation} for hadrons.  We  know that it is possible to develop a
constituent quark {\it model} (CQM)\cite{cqm}, and that Feynman's
parton model\cite{parton} is consistent with QCD at high energies; but
it is not clear that we can actually derive a constituent picture as
an approximation for the QCD bound state problem.

Theorists have appreciated the significance of the infinite momentum
frame\cite{weinberg} for current algebra\cite{current} and the parton
model\cite{parton} for many years, and light-front QCD in light cone
gauge has been studied as a promising ground for non-perturbative
calculations\cite{schladming,qcd2}.  However, little progress has
been made on Hamiltonian light-front QCD in 3+1 dimensions outside of
perturbation theory.  We blame this impasse on an understandable
assumption that non-perturbative phenomena such as confinement and
chiral symmetry breaking should arise naturally from the canonical QCD
Hamiltonian.  While this may be true in a formal sense if we can use a
regulator that maintains gauge invariance, a constituent picture will
not arise in this case; and there is little indication that
non-perturbative light-front Hamiltonian QCD is tractable if we use
such regulators.

There are several people who have participated in the development of
new ideas that may circumvent this impasse, guided by Ken Wilson.
Important insights have been provided by explicit calculations in
perturbative QED\cite{lfqed} and QCD\cite{Pe 93a,Zh 93a}. Wei-Min
Zhang and Avaroth Harindranath have provided a good introduction to
canonical light-front QCD with many calculations, and I refer the
reader to their work\cite{Zh 93a} for important details not found
here. Daniel Mustaki has studied chiral symmetry in light-front
coordinates, and surveyed significant differences between light-front
and equal-time chiral symmetry that provide insight into how a
constituent picture with spontaneously broken chiral symmetry can arise
in light-front coordinates\cite{mustaki}. But the significant step
towards tractable non-perturbative QCD calculations has just been
taken\cite{nonpert}.

A major new technical development required by this program is the {\it
similarity transformation} developed by Stan G{\l}azek and Ken
Wilson\cite{glazek1,glazek2}.  This transformation leads to a new
renormalization group for Hamiltonians that is not plagued by problems
associated with small energy denominators which other transformations
produce\cite{bloch,perryrg}.  A second technical development that
plays a central role in the program I outline is coupling
coherence\cite{coupcoh}.  Coupling coherence fixes all new
counterterms in the QCD Hamiltonian perturbatively, and can be
exploited for cutoffs in the regime of asymptotic freedom.

The major ingredients missing from the program are mechanisms for
confinement and chiral symmetry breaking that can be self-consistently
reproduced by a non-perturbative renormalization
group\cite{wilsonrg}.  A means of implementing absolute confinement
has been proposed for study\cite{nonpert}, but it has serious
problems. In this paper I propose a new confinement mechanism that
emerges naturally from a perturbative renormalization group.

In section 2 I outline the program and its underlying assumptions.
Section 3 introduces light-front coordinates and the new form
renormalization takes.  As a simple example I sketch a solution to the
bound state problem in the Schwinger model in section 4.  I turn to
3+1 dimensions in section 5 and use perturbation theory to clarify
some of the renormalization problems we face.

The entire approach requires Wilson's renormalization group as an
alternative to the canonical formulation of Hamiltonians, so I outline
Wilson's ideas in section 6.  The renormalization group indicates how
the Hamiltonian runs when cutoffs are lowered, and coupling coherence
assumes that some Hamiltonians remain invariant in form, with only a
few canonical variables running independently with the cutoffs.  This
idea and some simple examples are developed in section 7.  The
similarity transformation provides the engine that runs a Hamiltonian
renormalization group that is at least capable of producing
well-behaved perturbative expansions of the running Hamiltonian, and
it is sketched in section 8.  The renormalization group itself takes a
new form in light-front coordinates, as sketched in section 9; with
qualitatively different scaling behavior in transverse and
longitudinal dimensions leading to functions of longitudinal momenta
in relevant and marginal operators.  Section 10 contains details
required for the study of gauge theories.

In section 11 I study QED and show how the Bohr binding energy for
positronium can be derived in this program.  Many details required for
the study of QCD appear in this section, where I also begin to discuss
how QED and QCD differ.  In section 12 asymptotic freedom is
discussed, and the precise cancellation of light-front infrared
divergences required by perturbation theory is encountered.

Section 13 contains the most important new result in this paper.  It
is shown that the QCD Hamiltonian produced by a perturbative
renormalization group and coupling coherence contains confining
interactions that may survive at low energies.  Finally, a discussion
of what remains for the future is given in section 14.


\section{Motivation, Assumptions, and Strategy}

We start with the {\it hypothesis} that it is possible to {\it derive} a
constituent picture for hadrons from QCD, with hadronic states and
masses emerging from an equation

\begin{equation}
H \mid \Psi\rangle = E \mid \Psi \rangle,
\end{equation}

\noindent where,

\begin{equation}
\mid \Psi \rangle = \phi_{q\bar{q}} \mid q\bar{q} \rangle +
\phi_{q\bar{q}g} \mid q\bar{q}g \rangle + \cdot\cdot\cdot.
\end{equation}

\noindent $E$ is an eigenvalue that contains the invariant mass of the
state, and I use shorthand notation for the Fock space components of
the state (with $q$ being a quark, $\bar{q}$ an antiquark, and $g$ a
gluon) to avoid writing complete expressions that merely obscure the
point.

If a constituent picture emerges from QCD it will be possible to
obtain reasonable approximations, under conditions clarified below,
using only a few Fock space components in a trial state.  This
approach will then immediately yield masses, and given the states and
properly renormalized observables ({\it e.g.,} the electromagnetic
current operator), we can compute any physical quantity.

A major virtue of this approach is that {\it it stays close to
physical intuition}.  The high price is that it is not even vaguely as
mathematically elegant as covariant perturbation theory.  We sacrifice
mathematical elegance and explicit symmetry for a drastic reduction in
the number of degrees of freedom that must be treated
non-perturbatively.  If a finite dimensional Hamiltonian can be found,
a wealth of powerful numerical methods can be unleashed to diagonalize
it.  Moreover, we can use variational methods developed to analyze
nonrelativistic many-body Hamiltonians.

If this Hamiltonian is actually derived from QCD, it must lead to
asymptotic freedom and the perturbative high-energy behavior that
asymptotic freedom implies.  It is hard to imagine how these results
arise naturally from a few degrees of freedom. They do not, but the
constituent picture must arise without ignoring the many-body
mechanisms required to achieve asymptotic freedom.

The treatment of a field theoretic Hamiltonian must be tailored to the
problem at hand.  There are two important steps for the low-energy
bound state problem. First, a cutoff must be introduced, and a
renormalization group used to find a cutoff Hamiltonian, $H^\Lambda$,
capable of giving cutoff-independent results.  We must adjust the
cutoff to a value most convenient for the calculation, as is done in
perturbative QCD, where one chooses a renormalization scale set by
external momenta in the problem.

Second, we can always write

\begin{equation}
H^{\Lambda}=\H_0+\V,
\end{equation}

\noindent and choose $\H_0$ however we please. The key is
to choose $\H_0$ reasonably\cite{nonpert}.  If
$\H_0$ is chosen incorrectly, an expansion in powers of
$\V$ will not converge, even asymptotically. $\H_0$
must produce reasonable zeroth order results and cannot be a free
Hamiltonian if we hope to solve bound state problems.

In QCD it is difficult to find a zeroth order approximation, other than
the full Hamiltonian, that leads to reasonable bound state results
without violating explicit gauge invariance.  Quantum mechanical
logic, our inability to solve non-perturbative problems in an infinite
dimensional Fock space, and our desire to derive a constituent picture
dictate that explicit gauge invariance be violated; so {\it we allow
gauge invariance to become a hidden symmetry}\cite{nonpert}.

When considering the development of a constituent Hamiltonian
approximation for low energy QCD, we immediately face enormous
problems.  The first class of problems emerge from the viewpoint of
field theory.

First, Fock space is too big.  A complete field theory must be capable
of solving problems that require a Fock space which is an infinite sum
of cross products of infinite dimensional Hilbert spaces.  We will
never build a digital computer to hold such a space, and
approximations should require the explicit non-perturbative use of a
small part of Fock space.

Consider two simple ways to reduce Fock space.  First, we can simply
truncate the space by throwing away states.  In field theories this
leads to divergent sensitivity to the precise form of the truncation,
and removing this sensitivity can be as hard as solving the full field
theory.  Second, we can remove or systematically reduce couplings
between states that differ drastically in `energy.' The similarity
transformation\cite{glazek1,glazek2} allows us to exploit this second
possibility and avoids many severe problems that arise in any quantum
mechanical many-body problem when states are removed directly ({\it
e.g.}, intruder states).  In either case, the absurd size of Fock
space is dealt with by introducing a cutoff.

After a cutoff is introduced we must address the fact that when
$H^\Lambda_0$ is chosen to be a free Hamiltonian, low energy states
receive important contributions from states of arbitrarily large free
energy ({\it e.g.}, perturbation theory diverges).  Without
renormalization the introduction of any cutoff produces divergent
errors.

As a simple indication of the importance of high energy states,
consider second order perturbation theory for the energy of state
$\mid \phi_m \rangle$.  The second order shift is

\begin{equation}
\delta E_m = \sum_{n \ne m} {\mid \langle \phi_{0m} \mid V \mid \phi_{0n}
\rangle \mid^2 \over E_{0m}-E_{0n}}.
\end{equation}

\noindent For local field theories in 3+1 dimensions this sum diverges
as $n \rightarrow \infty$.  The matrix elements between low and high
energy states do not fall off rapidly enough when the couplings are
local, and the free energies do not increase rapidly enough with free
relativistic dispersion relations.  As suggested above, we can
introduce a cutoff that stops this sum at a large value of $n$; but
this leads to severe problems.  An alternative exploited by the
similarity transformation is to cut off the matrix elements, so that
states with arbitrarily large energy difference do not couple.  Both
of these cutoffs render the sum finite; but to obtain cutoff
independent results, we must move the dominant effects of direct
couplings removed by the cutoff to the Hamiltonian itself. This is the
objective of renormalization theory, and we will need a new form of
renormalization appropriate to our problem.

A further problem arises when we consider the ground state of any
field theory.  If the ground state ({\it i.e.}, the vacuum) has a
complicated Fock space representation, there is no chance that
low-lying excitations built on the physical ground state will have
simple Fock space wave functions. Every wave function will include the
vacuum.  This is the primary motivation for choosing coordinates in
which vacuum and valence states naturally separate, light-front
coordinates.  This does not solve the standard QCD vacuum problem, but
it enables us to reformulate the vacuum problem so that it can be
attacked with renormalization theory\cite{nonpert}.

Finally, all regulators at our disposal violate Lorentz covariance and
gauge invariance. There was a time when it was thought that these
symmetries must be maintained explicitly even if we wish only to
obtain unique answers in perturbation theory\cite{schwinger}.  It is
well understood now that this is not the case, and I give examples
below that show how we can hide these symmetries and recover them in
physical results.  An interesting example of how one can restore gauge
invariance in Lagrangian QCD after a Pauli-Villars regulator breaks it
is provided by 't Hooft's original demonstration that Yang-Mills
theories are renormalizable\cite{thooftone}.

There is an additional list of problems from the point of view of QCD.
In QCD hadrons are bound states of infinitely many quarks and gluons.
Confinement and spontaneous chiral symmetry breaking apparently occur
because hadrons are excitations on a complicated vacuum.  Mechanisms
such as flux tubes developing between a quark and antiquark as they
separate require arbitrarily many gluons.  But in the CQM hadrons are
bound states of a few quarks.  The parton model offers some clues to
the resolution of this difficulty, but it gets around much of this
problem by considering only high energy phenomena and simply
parameterizing the low energy structure of hadrons.  The only way that
we can resolve this apparent contradiction is to move the effects of
almost all quarks and gluons out of the low energy state vectors and
into the Hamiltonian and other observables.  This is the standard view
of how the CQM should be related to QCD.

In QCD gauge invariance implies that gluons are massless to all orders
in perturbation theory, and chiral symmetry implies that the up and
down quarks have masses that are much less than the typical hadronic
mass scale, $\Lambda_{QCD}$.  But in the CQM all constituents have
masses of hadronic scale, and hadron masses are primarily due to the
constituent masses.  To derive a constituent picture we must find a
non-perturbative mechanism that generates $\Lambda_{QCD}$ as a natural
constituent scale. We violate important symmetries and will see that
quarks and gluons are thereby forced to have large bare masses even in
perturbation theory.  While these bare masses lead to zero (or small)
physical masses in perturbation theory, we cannot trust perturbation
theory as the cutoff approaches $\Lambda_{QCD}$; and any bare mass
left at this scale may play the same role as a constituent mass in the
CQM\cite{nonpert}.

Even if partons do not gain a constituent mass through
renormalization, the confinement mechanism itself produces a hadronic
mass scale.  This is illustrated in the bag model\cite{bag}, where
massless partons confined to a spherical cavity of radius $R$ have a
kinetic energy that is ${\cal O}(1/R)$.  This kinetic energy serves
the same purpose as a constituent mass, producing the bulk of a
hadron's mass.

In QCD the quark-gluon and gluon-gluon coupling approaches one as the
cutoff approaches $\Lambda_{QCD}$, and since gluons are massless and
up and down quarks are light, it is hard to understand how
interactions at low and intermediate energies will not lead to copious
production of partons.  In the CQM, Zweig's rule\cite{zweig} states
that such interactions are dominated by quark rearrangement; and
though this rule does not always hold, it is much better than one
would expect from perturbative QCD intuitions.  Zweig's rule does
emerge from QCD in an expansion in inverse powers of the number of
colors, and the answer in QCD is that non-perturbative effects lead to
behavior that is radically different from what is indicated by
perturbation theory.  We must find a natural mechanism for this
non-perturbative behavior even though we consider only a few
constituents at the lowest energy scale.

Finally, there are conferences devoted to the QCD vacuum
problem; whereas there is no vacuum problem in the CQM.  In the CQM
we are given no direct clue to where the complicated vacuum structure
has gone, but the parton model provides clues to how we want to
approach the vacuum problem.  Using light-front coordinates we first
isolate the vacuum, and we then try to isolate its effects in
reasonable counterterms.  In other words, {\it we work in the broken
symmetry phase of the theory}.

The answer that we invoke for almost all problems is {\bf
renormalization}.  In our approach renormalization differs
significantly from manifestly covariant perturbative renormalization,
such as old-fashioned Feynman perturbation theory using dimensional
regularization. We introduce cutoffs that violate symmetries which
normally prevent a constituent picture from arising, and we use
renormalization to remove cutoff dependence from physical results.  As
part of the renormalization procedure we remove the degrees of freedom
that can mix with the bare vacuum to form a complicated physical
vacuum, and we must add `counterterms' to restore vacuum effects to
physical results.  The isolation of these degrees of freedom from the
constituents is possible in light-front coordinates, and our
discussion of how a constituent picture can arise from QCD is tied to
these coordinates.


\section{Light-Front Coordinates and Renormalization}

Nonrelativistic physical intuition is naturally expressed in
equal-time coordinates, where initial conditions are specified at a
fixed time as measured in the center-of-mass rest frame and uses a
Hamiltonian to determine the time-evolution of states.  In these
coordinates microcausal interactions force the bare vacuum to mix with
states that have zero total momentum.  Individual particle velocities
must be less than the velocity of light,  but few-body states in which
partons move with small velocity in opposite directions mix with the
bare vacuum, which means that valence-like states mix with the bare
vacuum.  In perturbative calculations we can readily isolate and
remove disconnected effects, but if the vacuum has non-perturbative
structure it is not typically possible to separate its contribution
from a valence contribution in equal-time coordinates.  This leads to
potentially insurmountable problems when trying to derive a
constituent picture.

Light-front coordinates are intrinsically Minkowskian.  The
light-front {\it time} coordinate is

\begin{equation}
x^+=x^0+x^3 \;,
\end{equation}

\noindent and the light-front {\it longitudinal space} coordinate is

\begin{equation}
x^-=x^0-x^3 \;.
\end{equation}

\noindent The scalar product is

\begin{equation}
a \cdot b = {1 \over 2} a^+ b^- + {1 \over 2} a^- b^+ - {\bf a}_\perp
\cdot {\bf b}_\perp,
\end{equation}

\noindent where ${\bf a}_\perp$ and ${\bf b}_\perp$ are the transverse
components of the four-vectors.  This means that the light-front
energy (i.e., the momentum conjugate to light-front time) is $p^-$ and
the longitudinal momentum is $p^+$.

Physical particle trajectories satisfy the kinematic relativistic
constraint

\begin{equation}
p^+ \ge 0 \;,
\end{equation}

\noindent because all velocities are equal to or less than the
velocity of light.  Since longitudinal momentum is conserved, the only
states that can mix with the zero momentum bare vacuum are those in
which every bare parton has identically zero longitudinal momentum.
This makes it trivial to isolate the states which mix with the vacuum.

For a free particle of mass $m$, the light-front energy is

\begin{equation}
p^-={ {\bf p}_\perp^2+m^2 \over p^+} \;.
\end{equation}

\noindent Except when ${\bf p}_\perp=0$ and $m=0$, this energy is
infinite for $p^+=0$.  In massless theories there is a set of measure
zero that has a finite energy even when $p^+=0$, which may be
important in QCD.  This leads to a very interesting situation in which
all states that mix with the bare vacuum (except one set of states in
massless theories) have infinite free energy.  If we remove these states we
have some hope that we can replace their effects using renormalization
theory.

This is our primary motivation for using light-front coordinates. {\it
Vacuum degrees of freedom can be isolated, and we may be able to
replace them with `effective' interactions using renormalization
theory.}  In this case we can have complicated vacuum effects ({\it
i.e.}, complicated effective interactions) and still obtain a
constituent picture.  We assume that this is possible, and {\it we
always work in the `broken symmetry phase' of the theory}, in which
the vacuum is trivial and `vacuum-induced' interactions are
responsible for symmetry breaking.  This strategy is motivated by the
renormalization group\cite{wilsonrg}, which indicates that in some
cases a small number of operators dominate the evolution of the
Hamiltonian ({\it e.g.}, marginal and relevant operators) over many
orders of magnitude.  In this situation we can simply add these
operators to the Hamiltonian and adjust their strengths to the
critical values at which the symmetry is actually hidden rather than
broken explicitly.  This should be a highly nontrivial task, because
non-perturbative effects dominate the small-x physics near the QCD
vacuum.

Transverse and longitudinal coordinates have different scaling
properties and must be assigned different dimensions, as first
observed by Wilson\cite{nonpert}.  The easiest way to appreciate this
formally is to note that longitudinal scaling is a boost, and this
symmetry cannot be broken; while transverse scale invariance is broken
by masses and by coupling renormalization. Furthermore, longitudinal
locality is violated even in the free Hamiltonian and we cannot assume
longitudinal locality when constructing the Hamiltonian with the aid
of the renormalization group.  On the other hand, it may be possible
to maintain transverse locality, at least to all orders in
perturbation theory; but `vacuum counterterms' in particular may
violate transverse locality.

This offers a way in which $\Lambda_{QCD}$ can emerge as the natural
bound state scale.  $\Lambda_{QCD}$ can set the transverse scale at
which nonlocal counterterms from the vacuum or from renormalization of
infrared longitudinal divergences become important, because these
counterterms are allowed to explicitly contain $\Lambda_{QCD}$.  In
this case, if we want to learn how $\Lambda_{QCD}$ managed to appear
in these counterterms, either the vacuum problem must be solved or
infrared longitudinal divergences must be renormalized
non-perturbatively.

In this paper I assume that transverse locality is maintained to the
extent permitted by the renormalization group transformation.
Transverse locality cannot be perfectly manifest, because cutoffs
remove momentum scales required to resolve transverse position.  This
process produces direct interactions that are typically short-ranged
in the transverse direction.  In gauge theories  infrared (small
longitudinal momentum) divergences allow these interactions to become
long-ranged, and confining.

I will develop a light-front renormalization group, and show how it
can exploit the existence of separate natural transverse and
longitudinal momentum scales.  The natural transverse scale relevant
to low-energy bound states is $\Lambda_{QCD}$, while the only natural
longitudinal scale for a single hadron is the hadron longitudinal
momentum itself. In the many-hadron problem there may not be a single
scale, and it can be difficult to resolve the valence partons in a
small-x hadron without resolving the wee partons in a large-x hadron.

This leaves us with three renormalization problems that are signalled
by the divergence of the free energy:

\begin{itemize}
\item{(i)} ${\bf p}_\perp^2 \rightarrow \infty$;

\item{(ii)} $p^+ \rightarrow 0$;

\item{(iii)} $p^+=0$.
\end{itemize}

\noindent The divergences from (i) are `ultraviolet,' producing
counterterms that are local in the transverse direction.  While the
counterterm structure is richer than what is required in manifestly
covariant perturbation theory, in the perturbative regime we
discover that masses are relevant, the chiral symmetry breaking term
in the canonical QCD Hamiltonian is a relevant quark-gluon coupling,
and other gauge interactions are marginal\cite{nonpert}.

The divergences from (ii) are `infrared,' producing counterterms that
are nonlocal at least in the longitudinal direction and which may also
be nonlocal in the transverse direction.  These infrared divergences
do not appear in Feynman perturbation theory in covariant gauges. They
seem to cancel perturbatively if a free energy cutoff suppresses
infrared behavior before the residual singularities are either
directly canceled by counterterms or allowed to cancel against one
another. Coupling coherence fixes the infrared divergent part of the
Hamiltonian as a power series in the gauge coupling, allowing us to
next consider re-summations that preserve cancellation of all infrared
divergences.

Finally, (iii) is the vacuum problem discussed above.  We cannot
readily study the effects of removing zero modes as part of a limiting
procedure, but we consider (iii) to be intricately tied to (ii).  We
focus on (ii) and note that the limit of arbitrarily small
longitudinal momenta will exist only if we include appropriate
counterterms `from' (iii).  The simplest example is spontaneous
symmetry breaking, where tachyons appear if we do not add symmetry
breaking interactions to the symmetric Hamiltonian.  This leads to the
possibility that these counterterms will introduce new free
parameters.  If there are only a few operators that need to be
considered, this may not be a severe penalty.

The simplest approach to these renormalization problems is to assume
transverse locality and use power counting, which is valid in the
perturbative regime, to identify all relevant and marginal ({\it
i.e.}, renormalizable) operators.  These operators are not restricted
by rotational symmetry, which is broken by the cutoff. They are not
restricted by gauge invariance, which is also violated by the cutoff.
And they are not restricted by chiral symmetry because it is broken by
hidden vacuum structure.  In a Euclidean renormalization group
analysis power counting typically removes all but a finite number of
interactions from consideration; but in a light-front renormalization
group transverse and longitudinal directions scale separately, and
transverse scaling runs the cutoff.  The longitudinal scale is not
important for the perturbative classification of operators, and as a
result each allowed operator includes a function of the dimensionless
ratios of all available longitudinal momenta, including any
longitudinal momentum scale introduced by the cutoff.  In other words,
{\it entire functions appear in the relevant and marginal operators},
and we must compute these functions.  The simplest way to adjust the
new operators is to fix covariance and gauge invariance in the
physical observables.

A more ambitious approach is to develop a non-perturbative light-front
renormalization group that can be used to identify the important
operators directly, even when the Hamiltonian is not
approximately free.  We might then use coupling coherence to fix the
Hamiltonian if the Hamiltonian trajectory can be computed back to the
neighborhood of the Gaussian fixed point.  Coupling coherence is
described at length below, but the basic idea is simple.  As the
cutoffs change the Hamiltonian must change in a precise manner to
maintain the values of all physical observables.  One interesting
possibility is that the Hamiltonian will not change at all, and this
is called a fixed point.  The fixed point is typically central in
perturbative analyses.  A richer possibility is that a finite number
of constants that appear in the Hamiltonian, couplings and masses,
change with the cutoff while the Hamiltonians functional dependence on
these constants does not change.  The Hamiltonian reproduces itself
exactly in form, but the relative strengths of operators change
because a few couplings and masses explicitly depend on the cutoff,
and this happens because transverse scale invariance is a broken
symmetry.

This approach is simply too ambitious right now, so I employ a
compromise analysis.  I start with a perturbative renormalization
group analysis, near a critical Gaussian fixed point.  This may give a
reasonable approximation to the Hamiltonian for small coupling, but
will not reveal how chiral symmetry breaking appears.  Coupling
coherence then fixes the Hamiltonian as a power series in the coupling,

\begin{equation}
H^\Lambda = h_0 + g_\Lambda h_1 + g_\Lambda^2 h_2 +
\cdot\cdot\cdot \;,
\end{equation}

\noindent where the operators $h_i$ do not depend on the cutoff unless
there is a physical mass. I truncate this series at a fixed order, and
study the resultant Hamiltonian non-perturbatively, even though it has
been derived perturbatively.  Here I use bound state perturbation
theory, which requires that the Hamiltonian be divided, $H^\Lambda =
\H_0+\V$.  In principle we can choose any $\h0$ we
please by adding and subtracting operators, but I will choose $\h0$
directly from $H^\Lambda$ in this paper, and simply move some operators to
$\V$ so that they can be treated perturbatively.

The assumption with which we started is that a constituent picture
arises from QCD.  At high energies the constituent picture requires
extremely large numbers of constituents, partons, and $\H_0$ is the
free Hamiltonian.  As the cutoff is lowered, the phase space available
for virtual partons decreases, and eventually approaches a mass
gap that signals the complete breakdown of naive perturbation theory.
I assume that the constituent picture survives this transition, and
that the number of partons required to approximate hadronic structure
actually decreases.  This is possible if interactions that do not
change particle number dominate the low-energy dynamics.  I assume
that this happens, so we include strong interactions that are diagonal
in parton number in $\H_0$, and all operators that change parton
number are put in $\V$.  In particular, a confining interaction
appears directly in $H$ when coupling coherence is used, and it relies
on infrared singularities that appear only when parton number is
conserved.

As a third step in this practical approach we should test the
division of the Hamiltonian by making sure that perturbative
corrections are small.  This will reveal flaws in the Hamiltonian
itself, which is estimated perturbatively, and will hopefully reveal
promising non-perturbative corrections to the renormalization group
evolution of the Hamiltonian that will allow us to iteratively improve
the renormalization group part of the calculation.


\section{The Schwinger Model}

The original Schwinger model is just QED in 1+1 dimensions with a
massless charged fermion.  It can be solved 
analytically\cite{schwing,lowenstein}.  Charged particles are confined
because the Coulomb
interaction is linear and there is only one physical particle, a
massive neutral scalar particle with no self-interactions.  The Fock
space content of the physical states depends crucially on the
coordinate system and gauge, and it is only in light-front coordinates
that a simple constituent picture emerges\cite{bergknoff,mccartor}.
Similar results are found for QCD in 1+1 
dimensions\cite{thoofttwo,1+1}.

The massive Schwinger model is a conceptually simple extension in
which the fermion is given a bare mass\cite{coleman1,coleman2}. It is
not analytically solvable, but it has a much richer phenomenological
structure than the massless model.  For small masses (which is
equivalent to strong coupling since the coupling has the dimension of
mass and it is the ratio $e/m$ that is important) there is still a
single stable massive boson; however, it experiences self-interactions
that produce bound states in addition to non-trivial scattering
states.  As the mass increases ({\it i.e.}, the coupling decreases)
additional stable bosons and their bound and scattering states appear,
eventually leading to a spectrum that has an extremely complicated
bosonic description.

The Schwinger model was first studied in Hamiltonian light-front field
theory by Bergknoff\cite{bergknoff}.  My description of the model
follows his closely, and I recommend his paper to the reader. Later
work using DLCQ\cite{eller1,yung}, the lattice\cite{crewther},
light-front integral equations\cite{ma}, and light-front
Tamm-Dancoff\cite{mo} improve these calculations in some cases; but
they offer no new physical insight that is not illustrated well in
Bergknoff's calculations.

Before I proceed, I must warn the reader that I will completely ignore
the vacuum of the model and assume that all excitations can be
understood without any need to directly study the vacuum.  Many
theorists disagree with this assumption, but to my knowledge no one
has shown any direct effect of the vacuum on states with finite
momentum.  If such effects can be found, the Schwinger model might
offer a testing ground for the idea that effects attributed to the
vacuum in Lagrangian field theory can be treated as interactions in
light-front Hamiltonian field theory with zero modes removed.  I
should also warn the reader that 1+1 dimensional theories have many
special features that may make them practically useless as a testing
ground for ideas intended for QCD in 3+1 dimensions.  For my purposes
it is enough that the massive Schwinger model illustrates many basic
elements of light-front field theory.

Bergknoff showed that the physical boson in the light-front massless
Schwinger model in light-cone gauge is a pure electron-positron
state.  This is an amazing result in a strong-coupling theory of
massless bare particles, and it illustrates how a constituent picture
may arise in QCD.  The electron-positron pair is confined by the
linear Coulomb potential. The kinetic energy vanishes in the massless
limit, and the potential energy is minimized by a wave function that
is flat in momentum space, as one might expect since a linear
potential produces a state that is as localized as possible (given
kinematic constraints due to the finite velocity of light) in position
space.

In order to solve this theory I must first set up a large number of
details.  I recommend that for a first reading these details be
skimmed, because the general idea is more important than the detailed
manipulations.  The Lagrangian for the theory is

\begin{equation}
{\cal L} = \overline{\psi} \bigl( i \not{{\hskip-.08cm}\partial} -m
\bigr) \psi - e \overline{\psi} \gamma_\mu \psi A^\mu - {1 \over 4}
F_{\mu \nu} F^{\mu \nu} \;,
\end{equation}

\noindent where $F_{\mu \nu}$ is the electromagnetic field strength
tensor.  I choose light-cone gauge,

\begin{equation}
A^+=0 \;.
\end{equation}

\noindent In this gauge we avoid ghosts, so that the Fock space has a
positive norm.  This is absolutely essential if we want to apply
intuitive techniques from many-body quantum mechanics.

Many calculations are simplified by the use of a chiral representation
of the Dirac gamma matrices, so in this section I will use:

\begin{equation}
\gamma^0=\left(\begin{array}{cc}
                0 & -i \\ i & 0
                \end{array}\right)~~,~~~
                                 \gamma^1=\left(\begin{array}{cc}
                                                0 & i \\ i & 0
                                           \end{array}\right)\;,
\end{equation}

\noindent which leads to the light-front coordinate gamma matrices,

\begin{equation}
\gamma^+=\left(\begin{array}{cc}
                0 & 0 \\ 2 i & 0
                \end{array}\right)~~,~~~
                                 \gamma^-=\left(\begin{array}{cc}
                                                0 & -2 i \\ 0 & 0
                                           \end{array}\right)\;.
\end{equation}

In light-front coordinates the fermion field $\psi$
contains only one dynamical degree of freedom, rather than two.
To see this, first define the projection operators,

\begin{equation}
\Lambda_+={1 \over 2} \gamma^0 \gamma^+=\left(\begin{array}{cc}
                                               1 & 0 \\ 0 & 0
                                               \end{array}\right)~~~~~
 \Lambda_-={1 \over 2} \gamma^0 \gamma^-=\left(\begin{array}{cc}
                                                0 & 0 \\ 0 & 1
                                           \end{array}\right)\;.
\end{equation}

\noindent Using these operators split the fermion field into two
components,

\begin{equation}
\psi=\psi_+ + \psi_-=\Lambda_+ \psi + \Lambda_- \psi \;.
\end{equation}

\noindent The two-component Dirac equation in this gauge is

\begin{equation}
\biggl( {i \over 2} \gamma^+ \partial^- + {i \over 2} \gamma^-
\partial^+ -m - {e \over 2} \gamma^+ A^- \biggr) \psi = 0 \;;
\end{equation}

\noindent which can be split into two one-component equations,

\begin{equation}
i \partial^- \psit_+ = -i m \psit_- + e A^- \psit_+ \;,
\end{equation}

\begin{equation}
i \partial^+ \psit_- = i m \psit_+ \;.
\end{equation}

\noindent Here $\psit_\pm$ refers to the non-zero component of
$\psi_\pm$.

The equation for $\psi_+$, involves the light-front time derivative,
$\partial^-$; so $\psi_+$ is a dynamical degree of freedom that must
be quantized.  On the other hand, the equation for $\psi_-$ involves
only spatial derivatives, so $\psi_-$ is a constrained degree of
freedom that should be eliminated in favor of $\psi_+$.  Formally,

\begin{equation}
\psit_-={m \over \partial^+} \psit_+ \;.
\end{equation}

\noindent This equation is not well-defined until boundary conditions
are specified so that $\partial^+$ can be inverted.  I will eventually
define this operator in momentum space using a cutoff, but I want to
delay the introduction of a cutoff until a calculation requires it.

I have chosen the gauge so that $A^+=0$, and the equation for $A^-$
is

\begin{equation}
-{1 \over 4} \bigl(\partial^+\bigr)^2 A^- = e \psi_+^\dagger \psi_+ \;.
\end{equation}

\noindent $A^-$ is also a constrained degree of freedom, and we can
formally eliminate it,

\begin{equation}
A^-=-{4 e \over \bigl(\partial^+\bigr)^2} \psi_+^\dagger \psi_+ \;.
\end{equation}

We are now left with a single dynamical degree of freedom, $\psi_+$,
which we can quantize at $x^+=0$,

\begin{equation}
\bigl\{\psi_+(x^-),\psi_+^\dagger(y^-)\bigr\} = \Lambda_+
\delta(x^--y^-) \;.
\end{equation}

\noindent We can introduce free particle creation and annihilation
operators and expand the field operator at $x^+=0$,

\begin{equation}
\psit_+(x^-) = \int_{k^+ > 0} {dk^+ \over 4\pi} \Biggl[ b_k e^{-i k
\cdot x} + d_k^\dagger e^{i k \cdot x} \Biggr] \;,
\end{equation}

\noindent with,

\begin{equation}
\bigl\{b_k,b_p^\dagger\} = 4 \pi \delta(k^+-p^+) \;.
\end{equation}

\noindent In order to simplify notation, I will often write $k$ to
mean $k^+$.  If I need $k^-=m^2/k^+$, I will specify the superscript.

The next step is to formally specify the Hamiltonian.  I start with
the canonical Hamiltonian,

\begin{equation}
H = \int dx^- \Bigl( H_0 + V \Bigr) \;,
\end{equation}

\begin{equation}
H_0 = \psi_+^\dagger \Biggl({m^2 \over i\partial^+}\Biggr)
\psi_+ \;,
\end{equation}

\begin{equation}
V= -2 e^2 \psi_+^\dagger \psi_+ \Biggl({1 \over
\partial^+}\Biggr)^2 \psi_+^\dagger \psi_+ \;.
\end{equation}

\noindent To actually calculate:

\begin{itemize}
\item{replace $\psi_+$ with its expansion in terms of $b_k$ and $d_k$},
\item{normal-order},
\item{throw away constants},
\item{drop all operators that require $b_0$ and $d_0$}.
\end{itemize}

\noindent The free part of the Hamiltonian becomes

\begin{equation}
H_0=\int_{k>0} {dk \over 4\pi} \Biggl({m^2 \over k}\Biggr)
\bigl(b_k^\dagger b_k+d_k^\dagger d_k\bigr) \;.
\end{equation}

When ${\cal V}$ is normal-ordered, we encounter new one-body
operators,

\begin{equation}
H'_0={e^2 \over 2\pi} \int_{k>0} {dk \over 4\pi}
\Biggl[\int_{p>0} dp \biggl( {1 \over (k-p)^2} - {1 \over (k+p)^2}
\biggr)\Biggr] \bigl(b_k^\dagger b_k+d_k^\dagger d_k\bigr) \;.
\end{equation}

\noindent This operator contains a divergent momentum integral.  From
a mathematical point of view we have been sloppy and need to carefully
add boundary conditions and define how $\partial^+$ is inverted.
However, I want to apply physical intuition and even though no
physical photon has been exchanged to produce the initial interaction,
I will act as if a photon has been exchanged and everywhere an
`instantaneous photon exchange' occurs I will cut off the momentum.
In the above integral I insist,

\begin{equation}
|k-p|>\epsilon \;.
\end{equation}

\noindent Using this cutoff we find that

\begin{equation}
H'_0={e^2 \over \pi} \int{dk \over 4\pi} \biggl({1 \over
\epsilon} - {1 \over k} + {\cal O}(\epsilon) \biggr) \bigl(b_k^\dagger
b_k + d_k^\dagger d_k \bigr) \;.
\end{equation}

Comparing this result with the original free Hamiltonian, we see that
a divergent mass-like term appears; but it does not have the same
dispersion relation as the bare mass. Instead of depending on the
inverse momentum of the fermion, it depends on the inverse momentum
cutoff, which cannot appear in any physical result.  There is also a
finite shift in the bare mass, with the standard dispersion relation.

The normal-ordered interactions are

\begin{eqnarray}
V'= 2 e^2 \int {dk_1 \over 4\pi}\cdot\cdot\cdot{dk_4 \over 4\pi}
4\pi\delta(k_1+k_2-k_3-k_4) \nonumber \\
~~~~~~~~~~~ \Biggl\{ -{2 \over (k_1-k_3)^2} b_1^\dagger d_2^\dagger d_4
b_3 +
{2 \over (k_1+k_2)^2} b_1^\dagger d_2^\dagger d_3 b_4 \nonumber \\
~~~~~~~~~~~~~~-{1 \over (k_1-k_3)^2}
\bigl(b_1^\dagger b_2^\dagger b_3 b_4 +
d_1^\dagger d_2^\dagger d_3 d_4\bigr) +\cdot\cdot\cdot \Biggr\} \;.
\end{eqnarray}

\noindent I do not display the interactions that involve the creation
or annihilation of electron/positron pairs, which are important for
the study of multiple boson eigenstates.

The first term in $V'$ is the electron-positron interaction.
The longitudinal momentum cutoff I introduced above requires
$|k_1-k_3| > \epsilon$, so in position space we encounter a potential
which I will naively define with a Fourier transform that ignores the
fact that the momentum transfer cannot exceed the momentum of the
state,

\begin{eqnarray}
v(x^-) &=& 4 q_1 q_2 \int_{-\infty}^\infty {dk \over 4\pi}\; {1 \over
k^2}\;
\theta(|k|-\epsilon) \; exp\bigl(-{i \over 2} k x^-\bigr) \nonumber \\
&=& {q_1 q_2 \over \pi} \; \Biggl[ {2 \over \epsilon} - {\pi \over 2}
|x^-| + {\cal O}(\epsilon) \Biggr] \;.
\end{eqnarray}

\noindent This potential contains a linear Coulomb potential that we
expect in two dimensions, but it also contains a divergent constant
that is negative for unlike charges and positive for like charges.

In charge neutral states the infinite constant in $V'$ is {\it
exactly} canceled by the divergent `mass' term in $H'_0$. This
Hamiltonian assigns an infinite energy to states with net charge, and
a finite energy as $\epsilon \rightarrow 0$ to charge zero states.
This does not imply that charged particles are confined, but the
linear potential prevents charged particles from moving to arbitrarily
large separation except as charge neutral states.  The confinement
mechanism I propose for QCD in 3+1 dimensions shares many features
with this interaction.

I would also like to mention that even though the interaction between
charges is long-ranged, there are no van der Waals forces in 1+1
dimensions.  It is a simple geometrical calculation to show that all
long range forces between two neutral states cancel exactly.  This
does not happen in higher dimensions, and if we use long-range
two-body operators to implement confinement we must also find
many-body operators that cancel the strong long-range van der Waals
interactions.

Given the complete Hamiltonian in normal-ordered form we can study
bound states.  A powerful tool for the initial study of bound states
is the variational wave function.  In this case, we can begin with a
state that contains a single electron-positron pair,

\begin{equation}
|\Psi(P)\rangle = \int_0^P {dp \over 4\pi} \phi(p) b_p^\dagger
d_{P-p}^\dagger |0\rangle \;.
\end{equation}

\noindent The norm of this state is

\begin{equation}
\langle \Psi(P')|\Psi(P)\rangle = 4\pi P \delta (P'-P) \Biggl\{{1
\over P} \int_0^P {dp \over 4\pi} |\phi(p)|^2 \Biggr\}\;,
\end{equation}

\noindent where the factors outside the brackets provide a covariant
plane wave normalization for the center-of-mass motion of the bound
state, and the bracketed term should be set to one.

The expectation value of the one-body operators in the Hamiltonian is

\begin{equation}
\langle\Psi|H_0+H'_0|\Psi\rangle = {1 \over P} \int {dk
\over 4\pi} \Biggl[{m^2-e^2/\pi \over k}+ {m^2-e^2/\pi \over P-k}+
{2 e^2 \over \pi \epsilon} \Biggr] |\phi(k)|^2 \;,
\end{equation}

\noindent and the expectation value of the normal-ordered interactions
is

\begin{equation}
\langle\Psi|V'|\Psi\rangle = -{4 e^2 \over P} \int' {dk_1 \over
4\pi} {dk_2 \over 4\pi} \Bigl[{1 \over (k_1-k_2)^2}-{1 \over
P^2}\Biggr] \phi^*(k_1) \phi(k_2) \;,
\end{equation}

\noindent where I have dropped the overall plane wave norm. The prime
on the last integral indicates that the range of integration in which
$|k_1-k_2|<\epsilon$ must be removed.  By expanding the integrand
about $k_1=k_2$, one can easily confirm that the $1/\epsilon$
divergences cancel.

The easiest case to study is the massless Schwinger model.  With $m=0$
the energy is minimized when

\begin{equation}
\phi(k)=\sqrt{4\pi} \;,
\end{equation}

\noindent and the invariant-mass is

\begin{equation}
M^2={e^2 \over \pi} \;.
\end{equation}

This type of simple analysis can be used to show that {\it this
electron-positron state is actually the exact ground state of the
theory} with momentum $P$, and that bound states do not interact with
one another.  The wave function for the relative motion of the
constituents becomes more complicated when the fermion is massive,
because it must vanish when either fermion momentum vanishes or the
kinetic energy diverges.  Extremely accurate results can be obtained
in this case using polynomial basis functions\cite{mo}; however, I do
not want to pursue these calculations further here.

The primary purpose of introducing the Schwinger model is to
illustrate that bound state center-of-mass motion is easily separated
from relative motion in light-front coordinates, and that standard
quantum mechanical techniques can be used to analyze the relative
motion of charged particles once the Hamiltonian is found.  It is
intriguing that even when the fermion is massless, the states are
constituent states in light-cone gauge and in light-front
coordinates.  This is not true in other gauges and coordinate systems.
This happens because the charges screen one another perfectly, and
this is exactly how we hope to see a constituent picture emerge from
QCD in 3+1 dimensions.

The success of light-front field theory in 1+1 dimensions can
certainly be downplayed, but it should be emphasized that no other
method on the market is as powerful for bound state problems in 1+1
dimensions.

The most significant barriers to using light-front field theory to
solve low energy QCD are not encountered in 1+1 dimensions.  The
Schwinger model is super-renormalizable, so we completely avoid
serious ultraviolet divergences.  There are no transverse directions,
and we are not forced to introduce a cutoff that violates rotational
invariance, because there are no rotations.  Confinement results from
the Coulomb interaction, and chiral symmetry is not spontaneously
broken.

\section{Scalar Field Theory in 3+1 Dimensions}

The primary objective of this section is to show that cutoffs
inevitably violate covariance, and to initiate the discussion of how
to find counterterms. I also list details required for scalar field
theory calculations and example one-loop calculations.  I will spend
more time talking about problems than solutions, but anyone who
actually wants to do calculations in light-front field theory should
get their hands dirty with scalar field theory.

Canonical light-front scalar field theory is discussed by Chang, Root
and Yan\cite{chaone,chatwo}, who derive many of the results below. The
Lagrangian for scalar field theory is independent of the coordinates,

\begin{equation}
{\cal L} = {1 \over 2} \partial^\mu \phi \partial_\mu \phi - {1 \over
2} \mu^2 \phi^2 - {\lambda \over 4!} \phi^4 \;.
\end{equation}

\noindent The commutation relation for the boson field is

\begin{equation}
[\phi(x^{+}, x^{-},\vec x_\perp), \partial^{+} \phi(x^{+}, y^{-},
\vec y_\perp) ] =
i \delta^3(x-y) \;.
\end{equation}

In order to derive the Hamiltonian and other Poincar{\'e} generators,
one typically begins with the energy-momentum
tensor\cite{chaone,chatwo}. While it is certainly possible to derive a
formal, ill-defined expression for the complete tensor, I merely list
the canonical Hamiltonian,

\begin{equation}
H = \int dx^-d^2x_\perp \biggl[{1 \over 2}
\;\;\phi(x) \biggl(-\partial_\perp^2+\mu^2\biggr) \phi(x) +
{\lambda \over 4!} \phi^4(x)
\biggr] \;.
\end{equation}

This provides a formal definition of the Hamiltonian, but it is not
possible to complete calculations until this operator is regulated.
The boson field can be expanded in terms of a free particle basis,

\begin{equation}
\phi(x) = \int \dqq \;\theta(q^+)\;\biggl[ a(q) e^{-iq \cdot x} +
a^\dagger(q) e^{iq \cdot x}\biggr] \;,
\end{equation}

\noindent with,

\begin{equation}
[a(q),a^\dagger(q')] = 16 \pi^3 q^+ \delta^3(q-q') \;.
\end{equation}

\noindent If we use

\begin{equation}
\phi(x)=\int{d^4k \over
(2\pi)^3}\;\delta(k^2-\mu^2)\;\phi(k)\;e^{-ik\cdot x} \;,
\end{equation}

\noindent we find that

\begin{equation}
\phi(q)=a(q)\;\;\;[q^+>0]\;\;\;\;,\;\;\;\;
\phi(q)=a^\dagger(-q)\;\;[q^+<0] \;.
\end{equation}

The free part of the Hamiltonian is

\begin{eqnarray}
H_0 &=\int dx^-d^2x_\perp {1 \over 2} \;\phi(x) \biggl(-\partial_\perp
^2+\mu^2\biggr) \phi(x) \nonumber \\
&= \int {dq^+ d^2q_\perp \over 16\pi^3 q^+}\; \Bigl({q_\perp
^2+\mu^2 \over q^+} \Bigr) \;a^\dagger(q) a(q) \;.
\end{eqnarray}

\noindent The interaction term is more complicated and I do not expand
it in terms of creation and annihilation operators.  Fock space
eigenstates of the free Hamiltonian are

\begin{equation}
|q_1,q_2,...\rangle = a^\dagger(q_1) a^\dagger(q_2) \cdot\cdot\cdot
|0\rangle \;,
\end{equation}

\noindent with the normalization being

\begin{equation}
\langle k|q\rangle = 16 \pi^3 q^+ \delta(k^+-q^+)\delta^2(k_\perp -
q_\perp) \;.
\end{equation}

\noindent Completeness implies that

\begin{equation}
1= |0\rangle \langle 0| \;+\;\int \dqq |q\rangle \langle q| \;+\;
{1 \over 2!} \int \dqq \int \dkk |q,k\rangle \langle q,k| \;+\;
\cdot\cdot\cdot \;,
\end{equation}

\noindent so we can write the free Hamiltonian as

\begin{eqnarray}
&H_0 = \int \dqq \Bigl({q_\perp^2+\mu^2 \over q^+} \Bigr) |q\rangle
\langle q| \nonumber \\ &+\; {1 \over 2!}
\int \dqq \int \dkk \Bigl({q_\perp^2+\mu^2 \over q^+}
+ {k_\perp^2+\mu^2 \over k^+} \Bigr) |q,k\rangle \langle q,k|
+\; \cdot\cdot\cdot \;.
\end{eqnarray}

In order to provide further orientation, let me consider the problem
of computing the Green's functions for this theory.  This problem in
equal-time field theory is discussed in many textbooks, and the
light-front formalism is nearly identical\cite{qcd2}. To find the
Green's functions of a theory, at $x^+=0$ choose the initial state
$\mid i(0) \rangle$, allow it to evolve in light-front time to
$x^+=\tau$, and compute its overlap with a second state in the $x^+=0$
basis,  $\mid f(0) \rangle$.  We split the Hamiltonian into a free
part $H_0$ and an interaction $V$, and find that

\begin{eqnarray}
\langle f(0) \mid i(\tau) \rangle &= (16 \pi^3) \delta^3(P_f-P_i)\;
G(f,i;\tau) = \langle f \mid e^{-i H \tau/2} \mid i \rangle
\nonumber \\
&= i \int {d\epsilon \over 2 \pi} e^{-i \epsilon \tau /2}
(16 \pi^3) \delta^3(P_f-P_i)\; G(f,i;\epsilon) \;.
\end{eqnarray}

\noindent This definition differs slightly from that given in some
other places. It is then straightforward to demonstrate that

\begin{eqnarray}
(16 \pi^3) \delta^3(P_f-P_i)\;
G(f,i;\epsilon) ~~~=~~~ \langle f \mid {1
\over \epsilon-H+i0_+} \mid i \rangle ~~~~~~~~~~~~~~~~~~~~~~
\nonumber \\
~~~~~~~~~~~~~~~~= \langle f \mid {1 \over \epsilon-H_0+i0_+} +
{1 \over \epsilon-H_0+i0_+}
V {1 \over \epsilon-H_0+i0_+} \nonumber \\
~~~~~~~~~~~~~~~~~~~~+ {1 \over \epsilon-H_0+i0_+} V {1 \over
\epsilon-H_0+i0_+} V {1 \over \epsilon-H_0+i0_+} + \cdot \cdot \cdot
\mid i \rangle \;.
\end{eqnarray}

\noindent Operator products are evaluated by inserting a complete set
of eigenstates of $H_0$ between interactions, using Eq. (51), and
using

\begin{equation}
H_0 \mid q_1,q_2,...\rangle = \Bigl[\sum_i q_i^-\Bigr] \mid
q_1,q_2,...\rangle \;,
\end{equation}

\begin{equation}
q_i^- = {q_{\perp i}^2+\mu^2 \over q_i^+} \;,
\end{equation}

\noindent to replace operators occurring in the denominators with
c-numbers.

Divergences arise from high energy states that are created and
annihilated by adjacent $V$'s, for example, because there are an
infinite number of such states. All divergences in perturbation theory
come from intermediate states (internal lines) that have large free
energy, so divergences occur in diagrams containing internal lines
that carry large transverse momentum (`ultraviolet' divergences)
and/or small longitudinal momentum (`infrared' divergences).

Given a light-front Hamiltonian, $H=H_0+V$, we can determine the
rules for constructing time-ordered perturbation theory diagrams. The
diagrammatic rules allow us to evaluate all terms that occur in the
expansion for the Green's functions.  For $\phi^4$ canonical field
theory the diagrammatic rules for time-ordered connected Green's
functions are:

\begin{itemize}
\item Draw all allowed time-ordered diagrams with the quantum
numbers of the specified initial and final states on the external
legs.  Assign a separate momentum $k^\mu$ to each internal and
external line, setting $k^-=(k_\perp^2+\mu^2)/k^+$ for each line. The
momenta are directed in the direction time flows.

\item For each intermediate state there is a factor
$\bigl(\epsilon-\sum_i k_i^-+i0_+\bigr)^{-1}$, where the sum is over
all particles in the intermediate state.

\item Integrate $\int {dk^+ d^2k_\perp \over 16 \pi^3 k^+}\;
\theta(k^+)$ for each internal momentum.

\item For each vertex associate a factor of $16 \pi^3 \lambda \;
\delta(K^+_{in}-K^+_{out})\;\delta^2(K^\perp_{in}-K^\perp_{out})$,
where $K_{in}$ is the sum of momenta entering a vertex and $K_{out}$
is the sum leaving.

\item Multiply the contribution of each time-ordered diagram by a
symmetry factor $1/S$, where $S$ is the order of the permutation group
of the internal lines and vertices leaving the diagram unchanged with
the external lines fixed.
\end{itemize}

\noindent To obtain the Green's function, propagators for the incoming
and outgoing states must be added, and one must remove an overall
factor of $16 \pi^3~\delta^3(P_f-P_i)$.

Given the Hamiltonian and commutation relations we can study bound and
scattering states perturbatively.  As a first simple example, consider
the physical boson itself.  The zeroth order approximation for the
state is the bare boson, and there is no first order shift in the
energy because the $\phi^4$ interaction has no one-body matrix
elements after zero modes are removed ({\it i.e.}, no tadpoles).  The
second-order shift is

\begin{equation}
\delta E_m = \sum_{n\ne m} {|\langle m|V|n \rangle|^2 \over E_{0 m} -
E_{0 n}} \;,
\end{equation}

\noindent where $E_0$ is the free energy.  Diagrammatically this
corresponds to the `setting sun' diagram.  When the bare mass is zero
the shift is

\begin{equation}
\delta E(P) = {\lambda^2 \over 3! P^+} \int dk_1 dk_2 dk_3~ 16\pi^3
\delta^3(P-k_1-k_2-k_3) \Biggl[{P_\perp^2 \over P^+}-{k_{1 \perp}^2
\over k_1^+}-{k_{2 \perp}^2 \over k_2^+}-{k_{3 \perp}^2 \over k_3^+}
\Biggr]^{-1} \;,
\end{equation}

\noindent where,

\begin{equation}
dk_i={dk_i^+ d^2k_{i\perp} \over 16\pi^3 k_i^+} \;.
\end{equation}

This expression simplifies if we introduce the natural Jacobi
variables for the relative motion of the three bosons with which the
bare boson mixes.  These are $x_i$ and ${\bf r}_i$, which are defined
by

\begin{equation}
k_i^+=x_i P^+ \;,\;\;{\bf k}_{i \perp}=x_i {\bf P}_\perp+{\bf r}_i \;.
\end{equation}

\noindent In these variables,

\begin{eqnarray}
\delta E(P)&=&-{\lambda^2 \over 3! P^+} \int {dx_1 d^2r_1
\over 16\pi^3 x_1}
\cdot\cdot\cdot {dx_3 d^2r_3 \over 16\pi^3 x_3} \Biggl[{r_1^2 \over
x_1}+{r_2^2 \over x_2}+{r_3^2 \over x_3} \Biggr]^{-1} \nonumber \\ &&
~~~~~~~~~~16\pi^3 \delta(1-x_1-x_2-x_3) \delta^2({\bf r}_1+{\bf
r}_2+{\bf r}_3) \;.
\end{eqnarray}

This integral diverges.  To make further progress we need to modify
the calculation so that we can work with finite expressions.  Anyone
with training in field theory  immediately considers using dimensional
regularization, which typically violates fewer symmetries than other
regulators.  However, we need to be careful when choosing regulators,
because our ultimate goal is to solve non-perturbative problems.
Dimensional regularization can only be applied when all integrals that
diverge are explicitly displayed when the non-perturbative problem is
set up.  This does not happen, for example, if we study the
physical boson by diagonalizing the Hamiltonian in a Fock space
representation.  It is the process of diagonalization that generates
integrals, and it is not clear how dimensional regularization can be
applied before diagonalization.

Another regulator we might consider is Pauli-Villars, which violates
gauge invariance in non-abelian gauge theories (unless the method of
higher derivatives\cite{faddeev} is used), but preserves symmetries in
other cases.  What does this look like if we are diagonalizing a
Hamiltonian?  In Pauli-Villars we must add particles whose masses are
taken to infinity at the end of a calculation and whose couplings
allow them to cancel divergences caused by the mixing of physical
degrees of freedom.  This means that the size of Fock space is
drastically increased, which is a disaster for our non-perturbative
calculations.  It also means that we cannot use a positive norm space
and hermitian Hamiltonian, because we cannot arrange divergences to
cancel in this case.

While someone may come up with a clever idea, at this point {\it the only
way to regulate the above divergence} in a manner that can later be
applied directly in non-perturbative Hamiltonian calculations {\it is
to use a cutoff}.  Moreover, {\it every cutoff violates manifest
Lorentz covariance and gauge invariance}.

There are a large number of cutoffs that we might consider.  The first
class of cutoffs are functions of individual particle momenta.  For
example, we could introduce a cutoff on every transverse momentum, so
that $|{\bf p}_\perp|<\Lambda$.  This cutoff leads to counterterms
that are local in the transverse direction, but it violates rotational
and transverse boost symmetries, so that the counterterms it produces
also violate these symmetries.  Furthermore, in gauge theories this
cutoff leaves severe divergences coming from small longitudinal
momenta.  Any cutoff that depends only on single particle momenta will
violate both boost and rotational symmetries.

If possible, we would like to choose a cutoff that maintains a maximal
number of kinematic symmetries; because we can then use these
symmetries to limit the number of counterterms even in a
non-perturbative calculation.  In light-front coordinates, this means
that we want to form cutoffs using Jacobi coordinates similar to those
above.  These coordinates do not have to be functions of the total
momentum of the state, but they must be functions of the momenta of
several particles, such as those directly participating in an
interaction.  This still leaves a wide range of cutoffs to consider,
but the cutoffs I ultimately employ fall into this class.

There is an additional criterion that becomes more apparent when we
use a perturbative renormalization group.  Ultimately we are going to
introduce a cutoff and then try to adjust the Hamiltonian so that
results do not depend on the explicit value of the cutoff.  This is
most easily done in perturbation theory, and I complete perturbative
renormalization group calculations to illustrate this. Perturbative
calculations lead to reasonable results only when {\it the cutoff is a
function of the free energy variables}.  

This constraint has a simple
origin.  If the cutoff is not a function of the energies that appear
in perturbative energy denominators, such as those above, the
denominator can change sign inside the range of integration and
either produce new divergences or extreme sensitivity to the region of
phase space in which the denominator changes sign. This is a very old
problem in quantum mechanics, where we know that even small
perturbations can lead to large effects in degenerate or nearly
degenerate subspaces.  Since I want to use a perturbative
renormalization group to approximate the cutoff-dependent Hamiltonian,
I will use cutoffs that depend on the free energies of particles.
There are still several cutoffs we might consider.  One simple choice
is a cutoff on individual particle energies.  This type of cutoff
violates boost invariance, and it does not avoid the problem of small
energy denominators.  A second choice is a cutoff on the sum of single
particle energies in a given state.  This is simply related to the
invariant-mass cutoff used in many DLCQ calculations and in many of my
previous calculations.  For example, we can use

\begin{equation}
\sum_i {{\bf p}_{i\perp}^2+m_i^2 \over p_i^+} < {{\bf
P}_{\perp}^2+\Lambda^2 \over P^+} \;,
\end{equation}

\noindent where ${\bf P}_\perp$ is the total transverse momentum of
the state and $P^+$ is the total longitudinal momentum.

For the second-order energy shift we are studying, this last cutoff
avoids the small energy denominator problem while apparently
maintaining all kinematic symmetries in light-front coordinates.
However, if there are spectators or we study higher order
corrections, energy denominators can become small.  The reason is
simple.  When a cutoff on the state energies is used, it is possible
for the incoming and intermediate state energies to simultaneously
approach the cutoff.  The only way to avoid this problem is to {\it
use a cutoff that employs the energy differences which actually appear
in the energy denominators in perturbation theory.}

The energy denominators in perturbation theory, both standard
time-ordered perturbation theory and the perturbative renormalization
group, involve energy differences that result from interaction
vertices.  In order to avoid small energy denominators we need to
cut off the change in energy that occurs at individual vertices.  In
terms of matrix elements, if $|E_{in}-E_{out}|>\Lambda$ we want to
force $\langle out|V|in\rangle$ to vanish.  For the above second-order
energy shift this cutoff requires

\begin{equation}
{\Lambda^2 \over P^+}>\mid {P_\perp^2 \over P^+}-{k_{1 \perp}^2
\over k_1^+}-{k_{2 \perp}^2 \over k_2^+}-{k_{3 \perp}^2 \over k_3^+}
\mid \;.
\end{equation}

\noindent In order to use such an energy cutoff we must choose a
transverse momentum scale, $\Lambda$, and a longitudinal momentum
scale.  I have chosen the longitudinal momentum scale to be the total
longitudinal momentum, $P^+$, but the presence of a longitudinal
momentum scale in the cutoff leads to interesting questions that I
will not address here.

This cutoff leads in Jacobi coordinates to the integral,

\begin{eqnarray}
\delta E(P)&=&-{\lambda^2 \over 3! P^+} \int {dx_1 d^2r_1
\over 16\pi^3 x_1}
\cdot\cdot\cdot {dx_3 d^2r_3 \over 16\pi^3 x_3}\;
\Biggl[{r_1^2 \over x_1}+{r_2^2 \over x_2}+{r_3^2 \over x_3}
\Biggr]^{-1} \nonumber \\ & &
16\pi^3 \delta(1-x_1-x_2-x_3)
\delta^2({\bf r}_1+{\bf r}_2+{\bf r}_3)
\theta\Biggl(\Lambda^2-{r_1^2 \over x_1}-{r_2^2 \over x_2}-{r_3^2
\over x_3} \Biggr) \;.
\end{eqnarray}

\noindent A simple change of variables reveals that this result is
proportional to $\Lambda^2$ with no additional $\Lambda$ dependence.
The only dependence on the incoming momentum is $1/P^+$, and if we had
chosen a different longitudinal momentum scale in the cutoff, we would
find this new scale in $\delta E(P)$ rather than $P^+$.  So we have

\begin{equation}
\delta E(P)=-{c \lambda^2 \Lambda^2 \over P^+} \;,
\end{equation}

\noindent where $c$ is a finite constant.  We assumed that the bare
mass is zero to get this result, which represents a negative
mass-squared shift.  Since negative mass-squared is unphysical, we need
to add a `mass' counterterm to the Hamiltonian to cancel this `mass'
shift.  If we want to do non-perturbative calculations we have to add a
bare boson mass before computing the mass shift, and adjust the bare
mass to obtain the desired physical mass non-perturbatively.  These
ideas become clearer in the renormalization group.

\section{Wilson's Renormalization Group}

This section follows a recent article on the light-front
renormalization group\cite{perryrg}, and is intended to impart the
motivation for using a renormalization group and the basic elements of
renormalization group calculations.

In classical mechanics the state of a system is completely specified
by a fixed number of coordinates and momenta. The objective of
classical mechanics is to compute the state as a function of time,
given initial conditions.  The state is not regarded as fundamental;
rather a Hamiltonian that governs the time evolution of the state is
regarded as fundamental.  In nonrelativistic quantum mechanics, we
must generalize the definition of the state, so that it is specified
by a ket in a state space, and we must drastically alter the theory
of measurement; but it remains possible to specify a Hamiltonian that
governs the time evolution of the state.  In both cases the time
evolution of the state is a trajectory in a Hilbert space, and the
trajectory is determined by a Hamiltonian that must be discovered by
fitting data. In principle, we would like to further generalize this
procedure for relativistic field theory; however, any straightforward
generalization that maintains locality leads to divergences that
produce mathematical ambiguities. To make mathematically meaningful
statements we must introduce an {\it ad hoc} regulator, which I assume
is a cutoff for the reasons discussed in the last section. With a
cutoff in place, physical results can be derived as limits of
sequences of finite quantities.  The renormalization group provides
methods for constructing such limits that are more powerful than
old-fashioned Feynman perturbation theory.

If all interactions are weak and of nearly constant strength over the
entire range of scales that affect an observable, we can use standard
perturbation theory to compute the observable; however, if either of
these conditions is not met, we cannot directly compute observables
with realistic Hamiltonians. This problem is easily appreciated by
considering a simple spin system in which 1024 spins are each allowed
to take two values.  The Hamiltonian for this system is a
$2^{1024}\times2^{1024}$ matrix, and this matrix cannot generally be
diagonalized directly.  This matrix is infinitely smaller than the
matrices we must consider when solving a relativistic field theory.
If the interactions remain weak over all scales of interest, but
change in strength significantly, we can use the perturbative
renormalization group. If the interactions become strong over a large
number of scales of interest, a non-perturbative renormalization group
must be developed.  A final possibility is that the interactions are
weak over almost all scales, becoming strong only over a few scales of
interest.  In this case, the perturbative renormalization group can be
used to eliminate the perturbative scales; after which we can use
some other method to solve the remaining Hamiltonian.

The introduction of an {\it ad hoc} cutoff in field theory complicates
the basic algorithm for computing the time evolution of a state,
because we must somehow remove any dependence on the cutoff from
physical matrix elements.  This complication is so severe that it has
caused field theorists to essentially abandon many of the most
powerful tools employed in nonrelativistic quantum mechanics (\eg, the
Schr{\"o}dinger picture).  How can we make reasonable estimates in
relativistic quantum mechanics? How can we guarantee that results are
independent of the cutoff?  How can we find a sequence of Hamiltonians
that depends on the cutoff in a manner that leads to correct results
as the cutoff approaches its limit?  These are the type of questions
that led Wilson to completely reformulate the original Gell-Mann--Low
renormalization group formalism.

Wilson adopted the same general strategy familiar from the study of
the time evolution of states, adding a layer of abstraction to the
original classical mechanics problem to compute `Hamiltonian
trajectories'. The strategy is universal in physics, but the layer of
abstraction leads to a great deal of confusion. In analogy to a
formalism that yields the evolution of a state as time changes, he
developed a formalism that yields the evolution of a Hamiltonian as
the cutoff changes.  In quantum mechanics a state is represented by an
infinite number of coordinates in a Hilbert space, and the Hamiltonian
is a linear operator that generates the time evolution of these
coordinates.   In the renormalization group formalism, the existence
of a space in which the Hamiltonian can be represented by an infinite
number of coordinates is assumed, and the cutoff evolution of these
coordinates is given by the {\it renormalization group
transformation}. The Hamiltonian is less fundamental than the
renormalization group transformation, which can be used to construct
trajectories of Hamiltonians. Transformations that lower the cutoff
by different amounts are members of a renormalization `group,' which
is typically a semi-group since such transformations cannot be inverted.

Each term in a Hamiltonian can often be written as a spatial integral
whose integrand is a product of derivatives and field operators.  The
definition of the Hamiltonian space might be a set of rules that
delimit all allowed operators. These operators should be thought of as
unit vectors, and the coefficients in front of these operators as
coordinates.  This type of operator is not usually bounded, and this
is a source of divergences in field theory.  To regulate these
divergences the cutoff is included directly in the definition of the
space of Hamiltonians.  The cutoff one chooses has drastic effects on
the renormalization group.  One familiar example of a cutoff is the
lattice, which replaces spatial integrals by sums over discrete
points.  The facts that the Hamiltonian can be represented by
coordinates that correspond to the strengths of specific operators,
and that these operators are all regulated by a cutoff that is part of
the definition of the space, is all that is important at this point.

Given a space of cutoff Hamiltonians, the next step is to construct a
suitable transformation.  This is slightly subtle and is usually the
most difficult conceptual step in a renormalization group analysis,
because we must find a transformation that manages to alter the
cutoff without changing the space in which the Hamiltonian lies.
These two requirements seem mutually contradictory at first.

To see how these difficulties are usually averted, let me again use
the lattice as an example.  A typical lattice transformation consists
of two steps.  In the first step one reduces the number of lattice
points, typically by grouping them into blocks and thereby doubling
the lattice spacing; and one computes a new effective Hamiltonian on
the new lattice.  At this point the lattice has changed, so the space
in which the Hamiltonian lies has changed.  The second step in the
transformation is to rescale distances using a change of variables, so
that the lattice spacing is returned to its initial value, while one
or more distance units are changed.  After both steps are completed
the lattice itself remains unchanged, if it has an infinite volume,
but the Hamiltonian changes.  This shows one way to alter a cutoff
without leaving the initial space of Hamiltonians. Numerically the
cutoff does not change, but the units in which the cutoff is measured
change.

When a similarity transformation\cite{glazek1,glazek2} is used, it is
obvious that the space of Hamiltonians does not change because the
space upon which the Hamiltonian acts does not change.  If we consider
the Hamiltonian to be an infinite dimensional matrix, a cutoff on the
interaction matrix elements restricts how far non-zero matrix elements
can lie off the diagonal.  A similarity transformation that lowers the
cutoff brings the matrix towards the diagonal.  It is not necessary to
introduce a rescaling transformation, but there are definite reasons
to do so in a perturbative analysis, as discussed later.

Returning to the lattice example, to obtain the continuum limit ({\it
i.e.}, remove the cutoff) we need to let the lattice spacing approach
zero; but the transformation increases the lattice spacing as measured
in the original distance units. While no inverse transformation that
directly decreases the lattice spacing exists, we can obtain the limit
in which the lattice spacing goes to zero by considering a sequence of
increasingly long trajectories. Instead of fixing the initial lattice
spacing, we fix the lattice spacing at the end of the trajectory and
we construct a sequence of trajectories in which the initial lattice
spacing is steadily decreased. We then directly seek a limit for the
last part of the trajectory as it becomes infinitely long, by studying
the sequence of increasingly long trajectories.  This is an algorithm
employed by Wilson for non-perturbative renormalization group
calculations\cite{wilsonrg}.

It is possible to study the cutoff limit more directly when a
reasonable perturbative approximation exists. In this case, the
renormalization group transformation can be approximated by an
infinite number of coupled equations for the evolution of a subset of
coordinates that are asymptotically complete, and these equations can
be inverted to allow direct study of the Hamiltonian trajectory as the
cutoff increases or decreases\cite{wegner}. If it can be shown that
all but a finite number of coordinates remain smaller than some chosen
magnitude, it may be possible to approximate the trajectory by simply
ignoring the smallest coordinates, retaining an increasing number of
coordinates only as we increase the accuracy of the approximation.
In this case the task of approximating a trajectory of renormalized
Hamiltonians is reduced to the task of solving a finite number of
coupled nonlinear difference equations.

Given a transformation $T$ that maps a subspace of Hamiltonians into
the space of Hamiltonians, with the possibility that some Hamiltonians
are mapped to Hamiltonians outside the original space, we study
$T[H]$.  We can apply the transformation repeatedly, and construct a
trajectory of Hamiltonians, with the $l$-th point on the trajectory
being

\begin{equation}
H_l = T^l[H_0] \;.
\end{equation}

\noindent Any infinitely long trajectory that remains inside the space
is called a trajectory of renormalized Hamiltonians.  If the
trajectory is infinitely long, a Hamiltonian with a finite cutoff
gives the same results as a Hamiltonian with an infinite cutoff; {\it
i.e.,} it accurately describes physics at all scales even though
infinitely many scales do not explicitly appear.  It is assumed that
the trajectory is completely determined by the initial Hamiltonian,
$H_0$, and $T$; however, the dependence on $H_0$ is usually not
explicitly indicated. Moreover, we will see later that boundary
conditions may be specified in a much more general fashion.

Perturbative renormalization group analyses typically begin with the
identification of at least one fixed point, $H^*$.  A {\it fixed
point} is defined to be any Hamiltonian that satisfies the condition

\begin{equation}
H^*=T[H^*] \;.
\end{equation}

\noindent For perturbative renormalization groups the search for such
fixed points is relatively easy; however, in non-perturbative studies
such a search typically involves a difficult numerical trial and error
calculation\cite{wilsonrg}. If $H^*$ contains no interactions (\ie, no
terms with a product of more than two field operators), it is called
{\it Gaussian}. If $H^*$ has a massless eigenstate, it is called {\it
critical}. If a Gaussian fixed point has no mass term, it is a {\it
critical Gaussian} fixed point. If it has a mass term, this mass must
typically be infinite, in which case it is a {\it trivial Gaussian}
fixed point.  In lattice QCD the trajectory of renormalized
Hamiltonians stays near a critical Gaussian fixed point until the
lattice spacing becomes sufficiently large that a transition to
strong-coupling behavior occurs.  If $H^*$ contains only weak
interactions, it is called {\it near-Gaussian}, and we may be able to
use perturbation theory both to identify $H^*$ and to accurately
approximate trajectories of Hamiltonians near $H^*$. Of course, once
the trajectory leaves the region of $H^*$, it is generally necessary to
switch to a non-perturbative calculation of subsequent evolution.  If
$H^*$ contains a strong interaction, we must use non-perturbative
techniques to find $H^*$, but it may still be possible to produce
trajectories near the fixed point using perturbation theory.  The
perturbative analysis in this case includes the interactions in $H^*$
to all orders, treating only the deviations from these interactions in
perturbation theory.

Consider the immediate neighborhood of the fixed point, and assume
that the trajectory remains in this neighborhood.  This assumption
must be justified {\it a posteriori}, but if it is true we should write

\begin{equation}
H_l=H^*+\delta H_l \;,
\end{equation}

\noindent and consider the trajectory of small deviations $\delta H_l$.

As long as $\delta H_l$ is `sufficiently small,' we can use a
perturbative expansion in powers of $\delta H_l$, which leads us to
consider

\begin{equation}
\delta H_{l+1}= L \cdot \delta H_l + N[\delta H_l] \;.
\end{equation}

\noindent Here $L$ is the linear approximation of the full
transformation in the neighborhood of the fixed point, and $N[\delta
H_l]$ contains all contributions to $\delta H_{l+1}$ of $\order(\delta
H_l^2)$ and higher.

The object of the renormalization group calculation is to compute
trajectories and this requires a representation for $\delta H_l$.  The
problem of computing trajectories is one of the most common in
physics, and a convenient basis for the representation of $\delta H_l$
is provided by the eigenoperators of $L$, since $L$ dominates the
transformation near the fixed point.  These eigenoperators and their
eigenvalues are found by solving

\begin{equation}
L \cdot O_m=\lambda_m O_m \;.
\end{equation}

\noindent If $H^*$ is Gaussian or near-Gaussian it is usually
straightforward to find $L$, and its eigenoperators and eigenvalues.
This is not typically true if $H^*$ contains strong interactions, but
in QCD we hope to use a perturbative renormalization group in the
regime of asymptotic freedom, and the QCD ultraviolet fixed point is
apparently a critical Gaussian fixed point.

Using the eigenoperators of $L$ as a basis we can represent $\delta
H_l$,

\begin{equation}
\delta H_l = \sum_{m\in R} \mu_{m_l}O_m +\sum_{m\in M} g_{m_l}O_m+
\sum_{m\in I} w_{m_l}O_m \;.
\end{equation}

\noindent Here the operators $O_m$ with $m\in R$ are {\it relevant}
(\ie, $\lambda_m>1$), the operators $O_m$ with $m\in M$ are {\it
marginal} (\ie, $\lambda_m=1$), and the operators with $m\in I$ are
either {\it irrelevant} (\ie, $\lambda_m<1$) or become irrelevant
after many applications of the transformation.  The motivation behind
this nomenclature is made clear by considering repeated application of
$L$, which causes the relevant operators to grow exponentially, the
marginal operators to remain unchanged in strength, and the irrelevant
operators to decrease in magnitude exponentially.  There are technical
difficulties associated with the symmetry of $L$ and the completeness
of the eigenoperators that I ignore\cite{wegner}.

$L$ depends both on the transformation and the fixed point, but there
are always an infinite number of irrelevant operators.  On the other
hand, transformations of interest for Euclidean lattice field theory
typically lead to a finite number of relevant and marginal operators.
One of the most serious problems for a perturbative light-front
renormalization group is that {\it an infinite number of relevant and
marginal operators are required.}  In the case of scalar field theory,
an infinite number of relevant and marginal operators arise because
the light-front cutoffs violate Lorentz covariance and cluster
decomposition. These are continuous symmetries, and their violation
leads to an infinite number of constraints on the Hamiltonian.  The
key to showing that the light-front renormalization group may not be
rendered useless by an infinite number of relevant and marginal
operators is the observation that both the strength and the evolution
of all but a finite number of relevant and marginal operators are
fixed by Lorentz covariance and cluster decomposition.  This has not
been proven to all orders, and this is not the last time I will state
something that has not been proven to all orders.

One alternative to employing symmetries such as covariance to fix the
strength of marginal and relevant counterterms is to insist that
non-canonical relevant and marginal variables are not independent
functions of the cutoff, but depend on the cutoff implicitly through
their dependence on canonical variables.  This requirement fixes the
manner in which the new variables evolve with the cutoff, and it also
fixes their value at all cutoffs once the values of the canonical
variables are chosen, if we insist that the new counterterms vanish
when canonical interactions are turned off.  The remarkable feature of
this procedure is that the value it gives to the new variables is
precisely the value required to restore Lorentz covariance and cluster
decomposition\cite{perryrg} in all cases studied to date.  This is
{\it coupling coherence}.

To simplify subsequent discussion, the statement that $\delta H_l$ is
small is assumed to mean that all masses and couplings in the
expansion of $\delta H_l$ are small.  The analysis itself should
signal when this assumption is naive. A rigorous discussion would
require consideration of the spectra of the eigenoperators. If the
coefficient of a single operator (\eg, a single mass) becomes large,
it may be straightforward to alter the analysis so that this
coefficient is included to all orders in an approximation of the
transformation, so that one perturbs only in the small coefficients.

For the purpose of illustration, let me assume that $\lambda_m=4$ for
all relevant operators, and $\lambda_m=1/4$ for all irrelevant
operators. The transformation can be represented by an infinite number
of coupled, nonlinear difference equations:

\begin{equation}
\mu_{m_{l+1}}=4 \mu_{m_l} + N_{\mu_m}[\mu_{m_l}, g_{m_l}, w_{m_l}] \;,
\end{equation}

\begin{equation}
g_{m_{l+1}}=g_{m_l} + N_{g_m}[\mu_{m_l}, g_{m_l}, w_{m_l}] \;,
\end{equation}

\begin{equation}
w_{m_{l+1}}={1 \over 4} w_{m_l} + N_{w_m}[\mu_{m_l}, g_{m_l},
w_{m_l}] \;.
\end{equation}

\noindent Sufficiently near a critical Gaussian fixed point, the
functions $N_{\mu_m}$, $N_{g_m}$, and $N_{w_m}$ should be adequately
approximated by an expansion in powers of $\mu_{m_l}$, $g_{m_l}$, and
$w_{m_l}$. The assumption that the Hamiltonian remains in the
neighborhood of the fixed point, so that all $\mu_{m_l}$, $g_{m_l}$,
and $w_{m_{l}}$ remain small, must be justified {\it a posteriori}.
Any precise definition of the neighborhood of the fixed point within
which all approximations are valid must also be provided {\it a
posteriori}.

Wilson has given a general discussion of how these equations are
solved\cite{wilsonrg}, and I repeat only the most important points. In
perturbation theory these equations are equivalent to an infinite
number of coupled, first-order, nonlinear differential equations. To
solve them we must specify `boundary' values for every variable,
possibly at different $l$, and then employ a {\it stable} numerical
algorithm to find the variables at all other values of $l$ for which
the trajectory remains near the fixed point. We want to apply the
transformation $\calN$ times, letting $\calN \rightarrow \infty$, and
adjusting the initial Hamiltonian so that this limit exists. Eq. (72)
must be solved by `integrating' in the exponentially stable direction
of decreasing $l$ (\ie, typically toward larger cutoffs), while Eq.
(74) must be solved in the direction of increasing $l$.  Eq. (73) is
linearly unstable in either direction.  The coupled equations must be
solved using an iterative algorithm.  Such systems of coupled
difference equations and the algorithms required for their solution
are familiar in numerical analysis.  In this context the need for
renormalization can be understood by considering the fact that the
renormalization group difference equations need to be solved over an
infinite number of scales in principle.

The final output of the renormalization group analysis is the cutoff
Hamiltonian $H_\calN$.  If this Hamiltonian is the final point in an
infinitely long trajectory of Hamiltonians, it will yield the same
observables below the final cutoff as $H_0$; but for an infinitely
long trajectory $H_0$ contains no cutoff, so $H_\calN$ {\it will yield
results that do not depend on the cutoff}.  It is for this reason that
$H_\calN$ and all other Hamiltonians on any infinitely long trajectory
are referred to as {\it renormalized Hamiltonians}.  How we solve
the final cutoff Hamiltonian problem using $H_\calN$ depends on the
theory.    For QCD, even if $H_\calN$ can be derived by purely
perturbative techniques, it will have to be solved non-perturbatively
because of confinement.  We must have an accurate approximation for
$H_\calN$; however, we do not necessarily need to explicitly construct
accurate approximations for all $H_l$.

The boundary values for the irrelevant variables should be set at
$l=0$, because we need to solve Eq. (74) in the direction of
increasing $l$. At large $l$ all variables are exponentially
insensitive to the irrelevant boundary values.  Therefore, they can be
chosen arbitrarily (universality); and the values of the irrelevant
variables at $l=\calN$ are output by Wilson's renormalization group.
This is one of the differences between Wilson's renormalization group
and the Gell-Mann--Low renormalization group in which irrelevant
variables are not treated.  Irrelevant operators are important in
$H_\calN$ unless the final cutoff is much larger than the scale of
physical interest.  The fact that they are irrelevant implies that
their final values are exponentially insensitive to their initial
values; and it implies that they are driven at an exponential rate
toward a function of the relevant and marginal variables, as discussed
below.  The fact that they are irrelevant does not necessarily imply
that they are unimportant.  This depends on how sensitive the physical
observables of interest are to physics near the scale of the cutoff.

The boundary values required by Eqs. (72) and (73) can be given at
$l=\calN$. Sufficiently far from $l=0$ the irrelevant variables are
exponentially driven to maintain polynomial dependence on relevant and
marginal variables, and sufficiently far from $l=\calN$ the relevant
variables are exponentially driven toward similar polynomial
dependence on the marginal variables.  While the calculation of
transient behavior near $l=0$ and $l=\calN$ usually requires a
numerical computation, the relevant and irrelevant variables are
readily approximated by polynomials that involve only marginal
variables in the intermediate region.  These polynomials are
determined by the expansions of $N_{\mu_m}$, $N_{g_m}$, and $N_{w_m}$;
and they can be fed back into Eq. (73) to find an approximate
equation for the marginal variables that requires direct knowledge
only of the marginal variables themselves.

Wilson has given an excellent discussion of the relationship between
divergences in standard perturbation theory (\eg, Feynman perturbation
theory) and the perturbative renormalization group\cite{wilsonrg}, and
I close this section by repeating the most salient points.  There are
usually no divergences encountered when a renormalization group
transformation is applied once; however, divergences can arise in the
form of powers of $l$ and exponents containing $l$, when $T$ is
applied a large number of times; and these divergences are directly
related to the divergences in Feynman perturbation theory. There are
no divergences apparent when the perturbative renormalization group
equations are solved using a stable numerical algorithm; however, if
we attempt to expand a coupling $g_l$ in powers of $g_0$, for example,
powers of $l$ appear.  As $l \rightarrow \infty$ these powers of $l$
lead to the divergences familiar in Feynman perturbation theory.

The perturbative renormalization group offers a significant
improvement over standard perturbation theory. In standard
perturbation theory we deal only with bare and renormalized
parameters.  Except in super-renormalizable theories, the ratio of
bare and renormalized parameters goes to infinity (or zero). If a
perturbative expansion of an observable in powers of the renormalized
parameters converges, the expansion for the same observable in terms
of the bare parameters cannot converge.  This leads to some
interesting departures from logic in standard perturbation
theory\cite{dirac3}. A small contribution of the renormalization group
is that logic may sometimes be restored to perturbation theory.

\section{Coupling Coherence}

This section develops the ideas behind coupling coherence, closely
following the original paper\cite{coupcoh} from which simple
examples are taken.  The basic idea behind coupling coherence was
first formulated by Oehme, Sibold, and Zimmerman\cite{coupcoh2}.  They
were interested in field theories where many couplings appear, such
as the standard model, and wanted to find some means of reducing the
number of couplings.  Wilson and I developed the ideas independently
in an attempt to deal with the functions that appear in marginal and
relevant light-front operators.

As argued above, the use of a Hamiltonian formulation, as opposed to a
Lagrangian formulation, severely limits one's choice of regulators.
The only practical regulators for non-perturbative Fock space
calculations employ cutoffs that violate explicit Lorentz covariance
and gauge invariance, opening a Pandora's box containing an infinite
number of relevant and marginal operators.  Counterterms must be added
to the Hamiltonian to restore these symmetries to observables, and to
obtain finite results. In perturbation theory, dimensional
power-counting separates these operators into `renormalizable' (i.e.,
relevant and marginal) and `non-renormalizable' (i.e., irrelevant).
Light-front divergences associated with transverse and longitudinal
momenta appear separately, and a straightforward power-counting
analysis requires us to scale these momenta separately.  Divergences
associated with large transverse momenta lead to counterterms that
contain functions of longitudinal momenta; and divergences associated
with small longitudinal momenta may lead to counterterms that contain
functions of transverse momenta.   The appearance of entire functions
of momenta in the counterterms, and the appearance of unfamiliar
`infrared' longitudinal divergences, significantly impacts
perturbative renormalization theory.

The puzzle is how to reconcile our knowledge from covariant
formulations of QCD that only one running coupling constant
characterizes the renormalized theory with the infinite set of
counterterms required by the light-front formulation. What happens in
perturbation theory when there are an infinite number of relevant and
marginal operators?  In particular, does the solution of the
perturbative renormalization group equations require an infinite
number of independent counterterms (\ie, independent functions of the
cutoff)? Coupling coherence provides the conditions under which a
finite number of running variables determines the renormalization
group trajectory of the renormalized Hamiltonian. To leading
nontrivial orders these conditions are satisfied by the counterterms
introduced to restore Lorentz covariance in scalar field theory and
gauge invariance in light-front gauge theories. In fact, the
conditions can be used to determine all counterterms in the
Hamiltonian, including relevant and marginal operators that contain
functions of longitudinal momentum fractions; and with no direct
reference to Lorentz covariance, this symmetry is restored to
observables by the resultant counterterms in scalar field
theory\cite{perryrg}.

In this section I use simple examples in which there are a finite
number of relevant and marginal operators \abinit, and use coupling
coherence to discover when only one or two of these may independently
run with the cutoff.  In general such conditions are met only when an
underlying symmetry exists.

Consider a theory in which two scalar fields interact,

\begin{equation}
V(\phi)={\lambda_1 \over 4!}\phi_1^4+{\lambda_2 \over 4!}\phi_2^4+
{\lambda_3 \over 4!}\phi_1^2 \phi_2^2 \;.
\end{equation}

\noindent Under what conditions will there be fewer than three running
coupling constants?  We can use a simple cutoff on Euclidean momenta,
$q^2<\Lambda^2$.  Letting $t=\ln(\Lambda/\Lambda_0)$, the
Gell-Mann--Low equations are

\begin{equation}
{\partial \lambda_1 \over \partial t} = 3 \zeta \lambda_1^2 + {1
\over 12} \zeta \lambda_3^2 + \order(2\;loop) \;,
\end{equation}

\begin{equation}
{\partial \lambda_2 \over \partial t} = 3 \zeta \lambda_2^2 + {1
\over 12} \zeta \lambda_3^2 + \order(2\;loop) \;,
\end{equation}

\begin{equation}
{\partial \lambda_3 \over \partial t} = {2 \over 3}
\zeta \lambda_3^2 +  \zeta \lambda_1 \lambda_3+\zeta \lambda_2 \lambda_3
+ \order(2\;loop) \;;
\end{equation}

\noindent where $\zeta=\hbar/(16\pi^2)$.  It is not important at this
point to understand how these equations are derived.

First suppose that $\lambda_1$ and $\lambda_2$ run separately, and ask
whether it is possible to find $\lambda_3(\lambda_1,\lambda_2)$ that
solves Eq. (78).  To one-loop order this leads to

\begin{equation}
\Bigl(3 \lambda_1^2 + {1 \over 12} \lambda_3^2 \Bigr) \;
{\partial \lambda_3 \over \partial \lambda_1} \;+\;
\Bigl(3 \lambda_2^2 + {1 \over 12} \lambda_3^2 \Bigr) \;
{\partial \lambda_3 \over \partial \lambda_2} =
{2 \over 3} \lambda_3^2 +  \lambda_1 \lambda_3+
\lambda_2 \lambda_3 \;.
\end{equation}

\noindent If $\lambda_1$ and $\lambda_2$ are independent, we can
equate powers of these variables on each side of Eq. (79).  If we allow
the expansion of $\lambda_3$ to begin with a constant, we find a
solution to Eq. (79) in which all powers of $\lambda_1$ and $\lambda_2$
appear. In this case a constant appears on the right-hand-sides of
Eqs. (76) and (77), and there will be no Gaussian fixed points for
$\lambda_1$ and $\lambda_2$. We are generally not interested in the
possibility that a counterterm does not vanish when the canonical
coupling vanishes, so we will simply discard this solution both here
and below.  We are interested in the conditions under which one
variable ceases to be independent, and the appearance of such an
arbitrary constant indicates that the variable remains independent
even though its dependence on the cutoff is being reparameterized in
terms of other variables.

If we do not allow a constant in the solution, we find that
$\lambda_3=\alpha \lambda_1+\beta \lambda_2+\order(\lambda^2)$. When
we insert this in Eq. (79) and equate powers on each side, we obtain
three coupled equations for $\alpha$ and $\beta$.  These equations
have no solution other than $\alpha=0$ and $\beta=0$, so we conclude
that if $\lambda_1$ and $\lambda_2$ are independent functions of $t$,
$\lambda_3$ will also be an independent function of $t$ unless the two
fields decouple.

Assume that there is only one independent variable, $\lt=\lambda_1$,
so that $\lambda_2$ and $\lambda_3$ are functions of $\lt$.  In this
case we obtain two coupled equations,

\begin{equation}
\Bigl(3 \lt^2+{1 \over 12} \lambda_3^2 \Bigr) \; {\partial \lambda_2
\over \partial \lt} = 3 \lambda_2^2+{1 \over 12} \lambda_3^2 \;,
\end{equation}

\begin{equation}
\Bigl(3 \lt^2+{1 \over 12} \lambda_3^2 \Bigr) \; {\partial \lambda_3
\over \partial \lt} = {2 \over 3} \lambda_3^2 + \lt \lambda_3+
\lambda_2 \lambda_3 \;.
\end{equation}

\noindent If we again exclude a constant term in the expansions of
$\lambda_2$ and $\lambda_3$ we find that the only non-trivial
solutions to leading order are $\lambda_2=\lt$, and either $\lambda_3
= 2 \lt$ or $\lambda_3=6 \lt$.  If $\lambda_3=2\lt$,

\begin{equation}
V(\phi)={\lt \over 4!}\;\bigl(\phi_1^2+\phi_2^2 \bigr)^2 \;,
\end{equation}

\noindent and we find the $O(2)$ symmetric theory.  If
$\lambda_3=6\lt$,

\begin{equation}
V(\phi)={\lt \over 2 \cdot 4!}\;\Bigl[\bigl(\phi_1+\phi_2\bigr)^4+
\bigl(\phi_1-\phi_2\bigr)^4\Bigr] \;,
\end{equation}

\noindent and we find two decoupled scalar fields.  Therefore,
$\lambda_2$ and $\lambda_3$ do not run independently with the cutoff
if there is a symmetry that relates their strength to $\lambda_1$.

The condition that a limited number of variables run with the cutoff
does not only reveal symmetries broken by the regulator, it may also
be used to uncover symmetries that are broken by the vacuum. Consider
a scalar field theory with

\begin{equation}
V(\phi)={r \Lambda^2 \over 2}\phi^2+{g \Lambda \over 3!}\phi^3+{\lambda
\over 4!} \phi^4 \;.
\end{equation}

\noindent  The renormalization group equations for this model depend
on whether we normal-order the Hamiltonian, on whether we keep
zero-modes in light-front field theory, etc.  No physical result
should depend on such choices, but the Hamiltonian itself is not
physical. To be explicit, use a cutoff on Euclidean momenta,
$q^2<\Lambda^2$; and do not normal-order the Hamiltonian, since
normal-ordering is not typically maintained by Euclidean
renormalization group transformations.  In this case, we must keep
tadpole insertions coming from the $\phi^3$ interaction and the
renormalization group equations are

\begin{equation}
{\partial r \over \partial t}=-2 r-\zeta{\lambda \over 1+r}+\zeta {g^2
\over (1+r)^2}+\zeta {g^2 \over r(1+r)} +\order(2\;loop)\;,
\end{equation}

\begin{equation}
{\partial g \over \partial t} = -g + 3\zeta {\lambda g \over (1+r)^2}
+\zeta {\lambda g \over r(1+r)} +\order(2\;loop)\;,
\end{equation}

\begin{equation}
{\partial \lambda \over \partial t} = 3 \zeta {\lambda^2 \over
(1+r)^2}+\order(2\;loop) \;.
\end{equation}

Under what conditions is $g$ a function of $\lambda$ and $r$?
Substituting $g(\lambda,r)$ into Eq. (86) we have

\begin{eqnarray}
\Biggl[-2 r-\zeta{\lambda \over 1+r}+\zeta {g^2 \over (1+r)^2}+\zeta {g^2
\over r(1+r)}\Biggr]\;{\partial g \over \partial r} +
\Biggl[3 \zeta {\lambda^2 \over (1+r)^2}\Biggr]\;{\partial g \over
\partial \lambda} \nonumber \\ = -g + 3\zeta
{\lambda g \over (1+r)^2} +\zeta {\lambda g \over r(1+r)} \;.
\end{eqnarray}

\noindent If we allow $g$ to depend explicitly on $\zeta$ (\ie,
explicitly on $\hbar$ in a loop expansion), we will find that
$g(\lambda,r)=f(\lambda)\sqrt{r}+\order(\hbar)$ and $f(\lambda)$ will
not be completely determined.  However, if we insist that $g$ contain
no explicit dependence on $\zeta$ we find a unique solution,
$g=\sqrt{3\lambda r}$.  Using Eq. (84) we find that with this value
of $g$,

\begin{equation}
V(\psi)={\lambda \over 4!} \Biggl(\psi^2-{3 r \Lambda^2 \over
\lambda}\Biggr)^2 \;,
\end{equation}

\noindent where $\psi=\phi+\sqrt{3r\Lambda^2/\lambda}$.  In short we
discover spontaneous symmetry breaking.  In light-front Hamiltonian field
theory it is natural to normal-order the Hamiltonian and drop all
zero-modes, in which case the second and fourth terms on the
right-hand side of Eq. (85), and the final term on the right-hand-side
of Eq. (86) drop out.  One can easily verify that the final result for
$g$ remains unchanged.

This last example is of some interest in light-front field theory,
because it is difficult to reconcile vacuum symmetry breaking with the
requirements that we work with a trivial vacuum and drop zero-modes
in practical non-perturbative Hamiltonian calculations.  Of course,
the only way that we can build vacuum symmetry breaking into the
theory without including a nontrivial vacuum as part of the state
vectors is to include symmetry breaking interactions in the
Hamiltonian and work in the hidden symmetry phase.  The problem then
becomes one of finding all necessary operators without sacrificing
predictive power.  The renormalization group specifies what operators
are relevant, marginal, and irrelevant; and coupling coherence
provides one way to fix the strength of the symmetry-breaking
interactions in terms of the symmetry-preserving interactions.  This
does not solve the problem of how to treat the vacuum in light-front
QCD by any means, because we have only studied perturbation theory;
but this result is encouraging and should motivate further
investigation.

Hopefully these examples have clarified the use of coupling
coherence.  In calculations below coupling coherence is used to find
the QED and QCD Hamiltonians to second order in the couplings, but
before this can be done we need a renormalization group appropriate
for light-front Hamiltonians.  The first step is to define the
renormalization group transformation, which is done in the next
section.

\section{The Similarity Transformation}

Stan G{\l}azek and Ken Wilson studied the problem of small energy
denominators, which are already apparent in Wilson's first complete
non-perturbative renormalization group calculations\cite{wilson70}.
These problems plague most formalisms designed to produce effective
Hamiltonians that can be used in model spaces, such as Bloch's
method\cite{bloch}, where the problem is associated with `intruder
states.'  G{\l}azek and Wilson came up with a 
solution\cite{glazek1,glazek2}. Instead of truncating a theory by
introducing a simple
energy cutoff, which leads to pathological interactions in the
effective Hamiltonian for states near the cutoff, they introduced
cutoffs in the interactions that remove direct couplings between
states that differ drastically in energy.  This cutoff removes no
states from the theory, so we still encounter the entire Fock space
and must find approximations that employ a subspace.

In this section I want to give an oversimplified discussion of the
similarity transformation, using sharp cutoffs that must eventually be
replaced with smooth cutoffs that require a more complicated
formalism.  G{\l}azek and Wilson have given a more complete
discussion, and I refer the interested reader to their
papers\cite{glazek1,glazek2}.

Suppose we have a Hamiltonian,

\begin{equation}
H^\lzero=H_0^\lzero+V^\lzero \;,
\end{equation}

\noindent where $H^\lzero_0$ is diagonal.  The cutoff $\lzero$
indicates that $\langle \phi_i|V^\lzero|\phi_j\rangle=0$ if $|E_{0
i}-E_{0 j}|>\lzero$. I should note that $\lzero$ is defined differently in
this section from other sections.  We want to use a similarity
transformation, which automatically leaves all eigenvalues and other
physical matrix elements invariant, that lowers this cutoff to
$\lone$.  This similarity transformation will constitute the first
step in a renormalization group transformation, with the second step
being a rescaling of energies that returns the cutoff to its original
numerical value.

The transformed Hamiltonian is

\begin{equation}
H^\lone=e^{i R}\bigl(H_0^\lzero+V^\lzero\bigr)e^{-i R} \;,
\end{equation}

\noindent where $R$ is a hermitian operator. If $H^\lzero$ is already
diagonal, $R=0$.  Thus, if $R$ has an expansion in powers of $V$, it
starts at first order and we can expand the exponents in powers of $R$
to find the perturbative approximation of the transformation.

We must adjust $R$ so that the matrix elements of $H^\lone$ vanish for
all states that satisfy $\lone<|E_{0 i}-E_{0 j}|<\lzero$.  We insist
that this happens to each order in perturbation theory.  Consider such
a matrix element,

\begin{eqnarray}
\langle \phi_i|H^\lone|\phi_j\rangle &=& \langle \phi_i|
 e^{i R}\bigl(H_0^\lzero+V^\lzero\bigr)e^{-i R} |\phi_j\rangle
\nonumber \\ &=& \langle \phi_i| (1+i R+\cdot\cdot\cdot)
(H_0^\lzero+V^\lzero\bigr) (1 -i R-\cdot\cdot\cdot) |\phi_j\rangle
\nonumber \\ &=& \langle \phi_i|H_0^\lzero |\phi_j\rangle  +
\langle \phi_i|V^\lzero+i\Bigl[R,H_0^\lzero\Bigr] |\phi_j\rangle +
\cdot\cdot\cdot \;.
\end{eqnarray}
\noindent The last line contains all terms that appear in first-order
perturbation theory.  Since $\langle \phi_i|H_0^\lzero |\phi_j\rangle
=0$ for these off-diagonal matrix elements, we can satisfy our new
constraint using

\begin{equation}
\langle \phi_i|R |\phi_j\rangle = {i \langle \phi_i|V^\lzero
|\phi_j\rangle \over E_{0 j}-E_{0 i}} \;+\;\order(V^2) \;.
\end{equation}

\noindent This fixes the matrix elements of $R$ when $\lone<|E_{0
i}-E_{0 j}|<\lzero$ to first order in $V^\lzero$.  I will assume that
the matrix elements of $R$ for $|E_{0i}-E_{0j}|<\lone$ are zero to
first order in $V$, and fix these matrix elements to second order
below.

Given $R$ we can compute the nonzero matrix elements of $H^\lone$.  To
second order in $V^\lzero$ these are

\begin{eqnarray}
H^\lone_{ab} &=& \langle \phi_a| H_0+V + i\Bigl[R,H_0\Bigr]
+i\Bigl[R,V \Bigr] -{1 \over 2} \Bigl\{R^2,H_0\Bigr\}
+ R H_0 R + \order(V^3) |\phi_b\rangle \nonumber \\
&=& \langle \phi_a|H_0+V + i\Bigl[R_2,H_0\Bigr]|\phi_b\rangle
\nonumber \\ &&
-{1 \over 2} \sum_k \Theta_{a k} \Theta_{k b} V_{ak} V_{kb} \Biggl[ {1
\over E_{0k}-E_{0a}}+{1 \over E_{0k}-E_{0b}} \Biggr] \nonumber \\
&&-\sum_k\Biggl[\Theta_{ak}\bigl(1-\Theta_{kb}\bigr) {V_{ak}
V_{kb} \over E_{0k}-E_{0a}} + \Theta_{kb}\bigl(1-\Theta_{ak}\bigr)
{V_{ak} V_{kb} \over E_{0k}-E_{0b}} \Biggr]+\order(V^3) \nonumber
\\
&=& \langle \phi_a|H_0+V+ i\Bigl[R_2,H_0\Bigr] |\phi_b\rangle
\nonumber \\&&+{1 \over 2} \sum_k \Theta_{a k} \Theta_{k b}
V_{ak} V_{kb} \Biggl[ {1 \over E_{0k}-E_{0a}}+{1 \over E_{0k}-E_{0b}}
\Biggr] \times \nonumber \\
&&~~~~~\Biggl[\theta(|E_{0a}-E_{0k}|-|E_{0b}-E_{0k}|) -
\theta(|E_{0b}-E_{0k}|- |E_{0a}-E_{0k}|)\Biggr] \nonumber \\
&&-\sum_k V_{ak} V_{kb} \biggl[ \Theta_{ak} {
\theta(|E_{0a}-E_{0k}|-|E_{0b}-E_{0k}|) \over E_{0k} - E_{0a} } +
\nonumber \\
&&~~~~~~~~~~~~~~~~~~~~~~~~ \Theta_{kb} { \theta(|E_{0b}-E_{0k}|-
|E_{0a}-E_{0k}|) \over E_{0k} - E_{0b} } \Biggr] \;.
\end{eqnarray}

\noindent I have dropped the $\lzero$ superscript
on the right-hand side of this equation and used subscripts to
indicate matrix elements. The operator $\Theta_{ij}$ is one if
$\lone<|E_{0i}-E_{0j}|<\lzero$ and zero otherwise.  It should also be
noted that $V_{ij}$ is zero if $|E_{0i}-E_{0j}|>\lzero$.  All energy
denominators involve energy differences that are at least as large as
$\lone$, and this feature persists to higher orders in perturbation
theory; which is the main motivation for choosing this transformation.
$R_2$ is second-order in $V$, and we are still free to choose its matrix
elements; however, we must be careful not to introduce small energy
denominators when choosing $R_2$.\footnote{In earlier versions of these
lectures I made the mistake of choosing an $R_2$ that contains small
energy denominators.  This leads to errors throughout the original text,
although all substantive results survive the correction.}

The matrix element $\langle \phi_a| i\Bigl[R_2, H_0 \Bigr] |\phi_b
\rangle = i (E_{0b}-E_{0a}) \langle \phi_a| R_2 | \phi_b \rangle $
must be specified.  I will choose this matrix element to cancel the
first sum in the final right-hand side of Eq. (94).  The motivation for
this choice will become clear later, after I introduce coupling
coherence. At this order, the possibility of coupling coherence requires
that the hamiltonian depend only on a single scale and not on the entire
sequence of scales.  For this to happen the sums that appear in the
transformed hamiltonian must exactly complement sums that appeared in the
initial interaction.  Examples will clarify this point.  To cancel the
first sum in the final right-hand side of Eq. (94) requires

\begin{eqnarray}
\langle \phi_a| R_2 | \phi_b \rangle &&= {i \over 2} \sum_k \Theta_{ak}
\Theta_{kb} V_{ak} V_{kb} \Biggl[ {1 \over E_{0k}-E_{0a}}+{1 \over
E_{0k}-E_{0b}} \Biggr] \times \nonumber \\
&&\Biggl[\theta(|E_{0a}-E_{0k}|-|E_{0b}-E_{0k}|) -
\theta(|E_{0b}-E_{0k}|- |E_{0a}-E_{0k}|)\Biggr] \;.
\end{eqnarray}

\noindent This choice differs from the  transformations given by
Wilson and Glazek\cite{glazek1,glazek2,nonpert}. No small energy
denominator appears in $R_2$ because it is being used to cancel a term
that involves a large energy difference.  If we tried to use $R_2$ to
cancel the remaining sum also, we would find that it includes matrix
elements that diverge as $E_{0a}-E_{0b}$ goes to zero, and this is not
allowed.

The non-vanishing matrix elements of $H^{\Lambda_1}$ are now
completely determined to $\order(V^2)$,

$~$

\begin{eqnarray}
H^\lone_{ab} &=&
\langle \phi_a|H_0+V^\lzero|\phi_b\rangle
\nonumber \\&& - \sum_k V_{ak} V_{kb} \biggl[ \Theta_{ak} {
\theta(|E_{0a}-E_{0k}|-|E_{0b}-E_{0k}|) \over E_{0k} - E_{0a} } +
\nonumber \\
&&~~~~~~~~~~~~~~~~~~~~~~~~ \Theta_{kb} { \theta(|E_{0b}-E_{0k}|-
|E_{0a}-E_{0k}|) \over E_{0k} - E_{0b} } \Biggr]
 \;.
\end{eqnarray}

\noindent Let me again mention that $V^\lzero_{ij}$ is zero if
$|E_{0i}- E_{0j}|>\lzero$, so there are implicit cutoffs that result
from previous transformations.

As a final word of caution, I should mention that the use of step
functions produces long-range pathologies in the interactions that
lead to infrared divergences in gauge theories.  One must replace the
step functions with smooth functions to avoid this problem.  This problem
will not show up in any calculations in this paper.

Such expressions are common in many-body theory
and this second-order result is quite simple.  It is a straightforward
exercise to invent diagrams that correspond to each term, but no one
has reduced the calculation of the Hamiltonian to all orders to a set
of diagrammatic rules.

Coupling coherence requires that the new Hamiltonian be identical to
the original Hamiltonian at second order, except $\lzero \rightarrow
\lone$; because running couplings do not affect this order.  It is
fairly easy to find at least two coupling coherent solutions to this
order, and a higher order calculation is required to choose between
them in principle.  This issue will arise in the discussion of QED and
QCD, where it is easily resolved without further calculations.

\section{The Light-Front Renormalization Group}

In this section I want to use the similarity transformation to form a
perturbative light-front renormalization group for scalar field
theory.  Ideally we would begin with a set of rules for constructing
Hamiltonians from derivatives and field operators, but the similarity
transformation I have derived requires us to deal with matrix elements
of the Hamiltonian.  It is possible to write the Hamiltonian in terms
of `free' particle creation and annihilation operators.   If we want
to stay as close as possible to the canonical construction of field
theories, we can:

\begin{itemize}
\item Write a set of `allowed' operators using powers of
derivatives and field operators.
\item Introduce `free' particle creation and annihilation
operators, and expand all field operators in this basis.
\item Introduce cutoffs on the Fock space transition operators.
\end{itemize}

Instead of following this program I will skip to the final step, and
simply write a Hamiltonian to initiate the analysis.

\begin{eqnarray}
H & = &\qquad  \int \dqt_1 \; \dqt_2 \; (16 \pi^3)
\delta^3(q_1-q_2) \; u_2(-q_1,q_2)
\;a^\dagger(q_1) a(q_2) \nonumber \\
&&+{1 \over 6} \int \dqt_1\; \dqt_2\; \dqt_3\; \dqt_4 \; (16 \pi^3)
\delta^3(q_1+q_2+q_3-q_4) \nonumber \Theta(q_4^--q_3^--q_2^--q_1^-) \\
&&\qquad\qquad\qquad\qquad\qquad
u_4(-q_1,-q_2,-q_3,q_4)\; a^\dagger(q_1) a^\dagger(q_2)
a^\dagger(q_3)
a(q_4)  \nonumber \\
&&+{1 \over 4} \int \dqt_1\; \dqt_2\; \dqt_3\; \dqt_4 \; (16 \pi^3)
\delta^3(q_1+q_2-q_3-q_4) \Theta(q_4^-+q_3^--q_2^--q_1^-) \nonumber \\
&&\qquad\qquad\qquad\qquad\qquad u_4(-q_1,-q_2,q_3,q_4)\; a^\dagger(q_1)
a^\dagger(q_2) a(q_3) a(q_4)  \nonumber \\
&&+{1 \over 6} \int \dqt_1\; \dqt_2\; \dqt_3\; \dqt_4 \; (16 \pi^3)
\delta^3(q_1-q_2-q_3-q_4) \Theta(q_4^-+q_3^-+q_2^--q_1^-) \nonumber \\
&&\qquad\qquad\qquad\qquad\qquad u_4(-q_1,q_2,q_3,q_4)\; a^\dagger(q_1)
a(q_2) a(q_3) a(q_4)  \nonumber \\
&&+ \qquad {\cal O}(\phi^6) \;,
\end{eqnarray}

\noindent where,

\begin{equation}
\dqt={dq^+ d^2q_\perp \over 16\pi^3 q^+} \;,
\end{equation}

\begin{equation}
q_i^-={q_{i\perp}^2 \over q_i^+} \;,
\end{equation}

\noindent and,

\begin{equation}
\Theta(Q^-)=\theta\Biggl({\Lambda^2 \over P^+}-|Q^-| \Biggr) \;.
\end{equation}

\noindent Here and below I simply assume that no operators appear that
break the discrete $\phi \rightarrow -\phi$ symmetry.  The functions
$u_2$ and $u_4$ are not yet determined.  If we assume locality in the
transverse direction, these functions can be expanded in powers of
their transverse momentum arguments.  Note that to specify the cutoff
both transverse and longitudinal momentum scales are required, and in
this case the longitudinal momentum scale is independent of the
particular state being studied.  Note also that $P^+$ breaks
longitudinal boost invariance and that a change in $P^+$ can be
compensated by a change in $\Lambda^2$.  This may have important
consequences, because the Hamiltonian should be a fixed point with
respect to changes in the cutoff's longitudinal momentum scale, since
this scale invariance is protected by Lorentz covariance\cite{perryrg}.

I have specified the similarity transformation in terms of matrix
elements, and will work directly with matrix elements, which are easily
computed in the free particle Fock space basis.  In order to study the
renormalization group transformation I will assume that the
Hamiltonian includes only the interactions shown above.  A single
transformation will produce a Hamiltonian containing products of
arbitrarily many creation and annihilation operators, but it is not
necessary to understand the transformation in full detail.

I will {\it define} the full renormalization group transformation as:
(i) a similarity transformation that lowers the cutoff in $\Theta$,
(ii) a rescaling of all transverse momenta that returns the cutoff to
its original numerical value, (iii) a rescaling of the creation and
annihilation operators by a constant factor $\zeta$, and (iv) an
overall constant rescaling of the Hamiltonian to absorb a
multiplicative factor that results from the fact that it has the
dimension of transverse momentum squared.  These rescaling operations
are introduced so that it may be possible to find a fixed point
Hamiltonian.  This rescaling operation will also allow us to directly
compare the matrix elements of the transformed Hamiltonian with those
of the original Hamiltonian and see why power counting reveals what
operators are renormalizable in perturbation theory.

To find the critical Gaussian fixed point we need to study the
linearized approximation of the full transformation, as discussed in
the section on Wilson's renormalization group.  In general the
linearized approximation can be extremely complicated, but near a
critical Gaussian fixed point it is particularly simple in light-front
field theory with zero modes removed, because tadpoles are excluded.
We have already seen that the similarity transformation does not
produce any first order change in the Hamiltonian, so the first order
change is determined entirely by the rescaling operation.  If we let

\begin{equation}
\Lambda_1 = \eta \Lambda_0 \;,
\end{equation}

\begin{equation}
{\bf p}_{i\perp} = \eta {\bf p}_{i\perp}' \;,
\end{equation}

\noindent and,

\begin{equation}
a_p = \zeta a_{p'} \;,\;\;\;a_p^\dagger = \zeta a_{p'}^\dagger \;,
\end{equation}

\noindent to first order the transformed Hamiltonian is

\begin{eqnarray}
H & = &\qquad  \zeta^2 \int \dqt_1 \; \dqt_2 \; (16 \pi^3)
\delta^3(q_1-q_2) \; u_2(q_i^+, \eta q_i^\perp)
\;a^\dagger(q_1) a(q_2) \nonumber \\
&&+{1 \over 6} \eta^4 \zeta^4 \int \dqt_1\; \dqt_2\; \dqt_3\; \dqt_4 \; (16
\pi^3)
\delta^3(q_1+q_2+q_3-q_4) \nonumber \Theta(q_4^--q_3^--q_2^--q_1^-) \\
&&\qquad\qquad\qquad\qquad\qquad
u_4(q_i^+, \eta q_i^\perp)\; a^\dagger(q_1) a^\dagger(q_2)
a^\dagger(q_3)
a(q_4)  \nonumber \\
&&+{1 \over 4} \eta^4 \zeta^4 \int \dqt_1\; \dqt_2\; \dqt_3\; \dqt_4 \; (16
\pi^3)
\delta^3(q_1+q_2-q_3-q_4) \Theta(q_4^-+q_3^--q_2^--q_1^-) \nonumber \\
&&\qquad\qquad\qquad\qquad\qquad u_4(q_i^+, \eta q_i^\perp)\; a^\dagger(q_1)
a^\dagger(q_2) a(q_3) a(q_4)  \nonumber \\
&&+{1 \over 6} \eta^4 \zeta^4 \int \dqt_1\; \dqt_2\; \dqt_3\; \dqt_4 \; (16
\pi^3)
\delta^3(q_1-q_2-q_3-q_4) \Theta(q_4^-+q_3^-+q_2^--q_1^-) \nonumber \\
&&\qquad\qquad\qquad\qquad\qquad u_4(q_i^+, \eta q_i^\perp)\; a^\dagger(q_1)
a(q_2) a(q_3) a(q_4)  \nonumber \\
&&+ \qquad \order(\phi^6) \;,
\end{eqnarray}

\noindent where I have simplified my notation for the arguments
appearing in the functions $u_i$. An overall factor of $\eta^2$ that
results from the engineering dimension of the Hamiltonian has been
removed.  The Gaussian fixed point is found by insisting that the
first term remains constant, which requires

\begin{equation}
\zeta^2 u_2^*(q^+, \eta {\bf q}_\perp)=u_2^*(q^+,{\bf q}_\perp) \;.
\end{equation}

\noindent I have used the fact that $u_2$ actually depends on only one
momentum other than the cutoff.

The solution to this equation is a monomial in ${\bf q}_\perp$ which
depends on $\zeta$,

\begin{equation}
u_2^*(q^+,{\bf q}_\perp)=f(q^+)\bigl({\bf q}_\perp \bigr)^n \;,\;\;\;
\zeta=\Biggl({1 \over \eta}\Biggr)^{(n/2)} \;.
\end{equation}

\noindent The solution depends on our choice of $n$ and to obtain the
appropriate free particle dispersion relation we need to choose $n=2$,
so that

\begin{equation}
u_2^*(q^+,{\bf q}_\perp)=f(q^+) {\bf q}_\perp^2 \;\;,\;\;\;
\zeta={1 \over \eta} \;.
\end{equation}

\noindent $f(q^+)$ is allowed because the cutoff scale $P^+$ allows us
to form the dimensionless variable $q^+/P^+$, which can enter the
one-body operator.  We will see that this happens in QED and QCD. Note
that the constant in front of each four-point interaction becomes one,
so that their scaling behavior is determined entirely by $u_4$. If we
insist on transverse locality (which may be violated because we remove
zero modes), we can expand $u_4$ in powers of its transverse momentum
arguments, and discover powers of $\eta q_i^\perp$ in the transformed
Hamiltonian. Since we are lowering the cutoff, $\eta<1$, and each
power of transverse momentum will be suppressed by this factor. This
means increasing powers of transverse momentum are increasingly {\it
irrelevant}, as defined in the section on Wilson's renormalization
group.

I will not go through a complete derivation of the eigenoperators of
the linearized approximation to the renormalization group
transformation about the critical Gaussian fixed point, but the
derivation is simple\cite{perryrg}.  Increasing powers of transverse
derivatives and increasing powers of creation and annihilation
operators lead to increasingly irrelevant operators.  The irrelevant
operators are called `non-renormalizable' in old-fashioned Feynman
perturbation theory.  Their magnitude decreases at an exponential rate
as the cutoff is lowered, which means that they {\it increase} at an
exponential rate as the cutoff is raised and produce increasingly
large divergences if we try to follow their evolution perturbatively
in this exponentially unstable direction.

The only relevant operator is the mass operator,

\begin{equation}
\mu^2 \int \dqt_1 \; \dqt_2 \; (16 \pi^3) \delta^3(q_1-q_2) \;
\;a^\dagger(q_1) a(q_2) \;,
\end{equation}

\noindent while the fixed point Hamiltonian is marginal (of course),
and the operator in which $u_4=\lambda$ (a constant) is marginal.  A
$\phi^3$ operator would also be relevant.

The next logical step in a renormalization group analysis is to study
the transformation to second order in the interaction, keeping the
second-order corrections from the similarity transformation.  I will
compute the correction to $u_2$ to this order and refer the interested
reader to Ref. \cite{perryrg} for more complicated examples.  However,
I warn the reader that in this reference a Bloch transformation is
used, which leads to pathologies apparently not produced by the
similarity transformation.

The matrix element of the one-body operator between single
particle states is

\begin{equation}
\langle p'|h|p\rangle = \langle 0|\;a(p')\; h\;
a^\dagger(p)\;|0\rangle = (16 \pi^3) \delta^3(p'-p)\;u_2(-p,p) \;.
\end{equation}

\noindent Thus, we easily determine $u_2$ from the matrix element.  It
is easy to compute matrix elements between other states.  We computed
the matrix elements of the effective Hamiltonian generated by the
similarity transformation when the cutoff is lowered in Eq. (96), and
now we want to compute the second-order term generated by the
four-point interactions above.  There are additional corrections to
$u_2$ at second-order in the interaction if $u_6$, {\it etc.} are
nonzero.

Before rescaling we find that the transformed Hamiltonian contains

\begin{eqnarray}
&&(16 \pi^3)\delta^3(p'-p)\;\delta v_2(-p,p)=
\nonumber \\ &&~~~~~
{1 \over 3!} \int \dkt_1 \dkt_2 \dkt_3 \; \theta\bigl
(\Lambda_0 -| p^- - k_1^- -k_2^- -k_3^-|\bigr) \nonumber \\
&&~~~~~
\theta\bigl( | p^- - k_1^- -k_2^- -k_3^-| - \eta \Lambda_0\bigr) ~
{\langle p'| V|k_1,k_2,k_3\rangle \langle k_1,k_2,k_3|V|p\rangle \over
p^--k_1^--k_2^--k_3^-} \;,
\end{eqnarray}

\noindent where $p^-={\bf p}_\perp^2/p^+$, {\it etc.} One can readily
verify that

\begin{equation}
\langle p'|V|k_1,k_2,k_3\rangle = (16 \pi^3)
\delta^3(p'-k_1-k_2-k_3) \; \Theta\bigl(p^--k_1^--k_2^--k_3^-\bigr)\;
u_4(-p',k_1,k_2,k_3) \;,
\end{equation}

\begin{equation}
\langle k_1,k_2,k_3|V|p\rangle= (16 \pi^3) \delta^3(p-k_1-k_2-k_3) \;
\Theta\bigl(p^--k_1^--k_2^--k_3^-\bigr)\; u_4(-k_1,-k_2,-k_3,p) \;.
\end{equation}

\noindent This leads to the result,

\begin{eqnarray}
\delta v_2(-p,p)&=&{1 \over 3!} \int \dkt_1 \dkt_2 \dkt_3 \;
(16 \pi^3)\delta^3(p-k_1-k_2-k_3) \nonumber \\
&&\theta\bigl(\Lambda_0 -| p^- - k_1^- -k_2^- -k_3^-|\bigr)~
\theta\bigl( | p^- - k_1^- -k_2^- -k_3^-| - \eta \Lambda_0\bigr)
\nonumber \\ &&~~~~~~~
{u_4(-p,k_1,k_2,k_3)\; u_4(-k_1,-k_2,-k_3,p) \over
p^--k_1^--k_2^--k_3^-} \;.
\end{eqnarray}

\noindent
To obtain $\delta u_2$ from $\delta v_2$ we must rescale the momenta,
the fields, and the Hamiltonian.  The final result is

\begin{eqnarray}
\delta u_2(-p,p)&=&{1 \over 3!} \int \dkt_1 \dkt_2 \dkt_3 \;
(16 \pi^3)\delta^3(p-k_1-k_2-k_3) \nonumber \\
&&\theta\Biggl({\Lambda_0 \over \eta} -| p^- - k_1^- -k_2^-
-k_3^-|\Biggr)~
\theta\bigl( | p^- - k_1^- -k_2^- -k_3^-| - \Lambda_0\bigr)
\nonumber \\ &&~~~~~~~
{u_4(p',k_1',k_2',k_3')\; u_4(-k_1',-k_2',-k_3',p') \over
p^--k_1^--k_2^--k_3^-} \;,
\end{eqnarray}

\noindent where $p^{+'}=p^+$, $p^{\perp'}=\eta p^\perp$,
$k_i^{+'}=k_i^+$, and $k_i^{\perp'}=\eta k_i^\perp$.

\section{QED and QCD Hamiltonians}

This is a dull section filled with details.  I follow the pioneering
work of Brodsky and Lepage (see the references in \cite{qcd2}) to
develop a canonical QCD Hamiltonian (from which it is easy to
determine the QED Hamiltonian), and provide many details required to
use it.  My derivation is not intended to be rigorous, and there is no
discussion of constrained quantization or gauge fixing of zero modes.

Before proceeding, let me mention that if we start with a
Hamiltonian that includes the free quark and gluon kinetic energies,
and the quark-gluon vertex alone, it is possible to derive the rest of
the canonical Hamiltonian using coupling coherence.  This is not an
easy task, but one finds that the three- and four-gluon interactions,
as well as the instantaneous gluon exchange interactions and suitably
defined self-induced inertias, must be added or new couplings and
masses in addition to the canonical coupling appear and run
independently with the cutoff.  Of course, it is easiest to start with
all terms in the canonical Hamiltonian through ${\cal O}(g)$, because
non-canonical counterterms appear at ${\cal O}(g^2)$ and must be
determined by non-canonical means in any case.

I repeat my recommendation of the papers by Zhang and
Harindranath\cite{Zh 93a}, where elegant techniques for constrained
quantization developed by Faddeev and Jackiw\cite{jackiw} are used to
derive a formal expression for the QCD Hamiltonian, taking special
care to discuss infrared divergences.  They use a chiral
representation of the Dirac gamma matrices, which is particularly
well-suited to the isolation of the two dynamical components of Dirac
spinors; and they provide many example one-loop calculations that are
useful to anyone who wants to start doing light-front calculations.
Perhaps the first useful discussion of non-covariant divergences and
counterterms is provided by Charles Thorn (see Thorn's paper in
\cite{qcd2}) in his paper on asymptotic freedom in light-cone gauge.
His work has not received the attention that it deserves.  For a
recent discussion of the counterterms required to restore covariance
and gauge invariance in QED, see the paper by Mustaki, Pinsky,
Shigemitsu, and Wilson\cite{lfqed}.  Of course, the work of Brodsky
and Lepage is also a useful starting point.

I will use $A^+=0$ gauge, and since I drop zero modes I do not
directly confront the fact that this condition cannot be implemented
for gauge field zero modes. I use the Bjorken and Drell conventions
for gamma matrices\cite{bj},  largely because these are the
conventions I have used until recently in my calculations.  The gamma
matrices are

\begin{equation}
\gamma^0=\left(\begin{array}{cc}
                1 & 0 \\ 0 & -1
                \end{array}\right)~,~~~~
                                 \gamma^k=\left(\begin{array}{cc}
                                         0 & \sigma_k \\ -\sigma_k & 0
                                           \end{array}\right)\;,
\end{equation}

\noindent where $\sigma_k$ are the Pauli matrices.  This leads to

\begin{equation}
\gamma^+=\left(\begin{array}{cc}
                1 & \sigma_3 \\ -\sigma_3 & -1
                \end{array}\right)~,~~~~
                                 \gamma^-=\left(\begin{array}{cc}
                                       1 & -\sigma_3 \\ \sigma_3 & -1
                                           \end{array}\right)\;.
\end{equation}

\noindent Useful identities for many calculations are $\gamma^+
\gamma^- \gamma^+=4 \gamma^+$, and $\gamma^- \gamma^+ \gamma^-=4
\gamma^-$.

The operator that projects onto the dynamical fermion degree of
freedom is

\begin{equation}
\Lambda_+= {1 \over 2} \gamma^0 \gamma^+={1 \over 4} \gamma^- \gamma^+=
                                        {1 \over 2} \left(\begin{array}{cc}
                                         1 & \sigma_3 \\ \sigma_3 & 1
                                         \end{array}\right) \;,
\end{equation}

\noindent and the complement projection operator is

\begin{equation}
\Lambda_-= {1 \over 2} \gamma^0 \gamma^-={1 \over 4} \gamma^+ \gamma^-=
                                         {1 \over 2} \left(\begin{array}{cc}
                                         1 & -\sigma_3 \\ -\sigma_3 & 1
                                         \end{array}\right) \;.
\end{equation}

The Dirac spinors $u(p,\sigma)$ and $v(p,\sigma)$ satisfy

\begin{equation}
(\pslash-m) u(p,\sigma)=0 \;,\;\;\; (\pslash+m) v(p,\sigma)=0 \;,
\end{equation}

\noindent and,

\begin{equation}
\overline{u}(p,\sigma)u(p,\sigma')=-\overline{v}(p,\sigma)
v(p,\sigma')=2 m \delta_{\sigma \sigma'}
\;,
\end{equation}

\begin{equation}
\overline{u}(p,\sigma) \gamma^\mu u(p,\sigma')=
\overline{v}(p,\sigma) \gamma^\mu v(p,\sigma')=
2 p^\mu \delta_{\sigma \sigma'}
\;,
\end{equation}

\begin{equation}
\sum_{\sigma=\pm {1 \over 2}} u(p,\sigma) \overline{u}(p,\sigma)
= \pslash+m \;,\;\;\; \sum_{\sigma=\pm {1 \over 2}}
v(p,\sigma) \overline{v}(p,\sigma)= \pslash-m
\;.
\end{equation}

There are only two physical gluon (photon) polarization vectors,
$\epsilon_{1\perp}$ and $\epsilon_{2\perp}$; but it is sometimes
convenient (and dangerous once covariance and gauge invariance are
violated) to use $\epsilon^\mu$, where

\begin{equation}
\epsilon^+=0 \;,\;\;\;\epsilon^-={2 {\bf q}_\perp \cdot \epsilon_\perp
\over q^+} \;.
\end{equation}

\noindent It is often possible to avoid using an explicit
representation for $\epsilon_\perp$, but completeness relations are
required,

\begin{equation}
\sum_{\lambda} \epsilon^\mu_\perp(\lambda) \epsilon^{*\nu}_\perp(\lambda)
= -g_\perp^{\mu \nu} \;,
\end{equation}

\noindent so that,

\begin{equation}
\sum_{\lambda} \epsilon^\mu(\lambda) \epsilon^{*\nu}(\lambda)
= -g_\perp^{\mu \nu} + {1 \over q^+} \bigl(\eta^\mu q_\perp^\nu + \eta^\nu
q_\perp^\mu \bigr) + { {\bf q}_\perp^2 \over (q^+)^2} \eta^\mu \eta^\nu \;,
\end{equation}

\noindent where $\eta^+=\eta^1=\eta^2=0$ and $\eta^-=2$.  One often
encounters diagrammatic rules in which the gauge propagator is written
so that it looks covariant; but this is dangerous in loop calculations
because such expressions require one to add and subtract terms that
contain severe infrared divergences.

The QCD Lagrangian density is

\begin{equation}
{\cal L}=-{1 \over 2} Tr F^{\mu \nu} F_{\mu \nu} + \overline{\psi}
\Bigl(i \Dslash -m\Bigr)\psi \;,
\end{equation}

\noindent where $F^{\mu \nu}=\partial^\mu A^\nu-\partial^\nu A^\mu+i g
\bigl[A^\mu,A^\nu\bigr]$ and $i D^\mu=i \partial^\mu-g A^\mu$.  The
SU(3) gauge fields are $A^\mu=\sum_a A^\mu_a T^a$, where $T^a$ are
one-half the Gell-Mann matrices, $\lambda^a$,  and satisfy $Tr~ T^a
T^b= 1/2~ \delta^{ab}$ and $\bigl[T^a,T^b\bigr]=i f^{abc} T^c$.

The dynamical fermion degree of freedom is $\psi_+=\Lambda_+ \psi$,
and this can be expanded in terms of plane wave creation and
annihilation operators at $x^+=0$,

\begin{equation}
\psi_+^r(x)=\sum_{\sigma=\pm 1/2} \int_{k^+>0}
{dk^+ d^2k_\perp \over 16\pi^3 k^+}
\Bigl[b^r(k,\sigma) u_+(k,\sigma) e^{-ik\cdot x} +
d^{r\dagger}(k,\sigma) v_+(k,\sigma)
e^{ik\cdot x}\Bigr] \;,
\end{equation}

\noindent where these field operators satisfy

\begin{equation}
\Bigl\{\psi_+^r(x),\psi_+^{s\dagger}(y)\Bigr\}_{x^+=y^+=0}=\Lambda_+
\delta_{rs} \delta^3(x-y) \;,
\end{equation}

\noindent and the creation and annihilation operators satisfy

\begin{equation}
\Bigl\{b^r(k,\sigma),b^{s\dagger}(p,\sigma')\Bigr\}=
\Bigl\{d^r(k,\sigma),d^{s\dagger}(p,\sigma')\Bigr\}
= 16\pi^3 k^+ \delta_{rs} \delta_{\sigma \sigma'} \delta^3(k-p) \;.
\end{equation}

\noindent The indices $r$ and $s$ refer to SU(3) color.  In general,
when momenta are listed without specification of components, as in
$\delta^3(p)$, I am referring to $p^+$ and ${\bf p}_\perp$.

The transverse dynamical gluon field components can also be expanded
in terms of plane wave creation and annihilation operators,

\begin{equation}
A_\perp^{ic}(x)=\sum_\lambda\int_{k^+>0}
{dk^+ d^2k_\perp \over 16\pi^3 k^+}
\Bigl[a^c(k,\lambda) \epsilon_\perp^i(\lambda) e^{-ik\cdot x}
+a^{c\dagger}(k,\lambda) \epsilon_\perp^{i*}(\lambda) e^{ik\cdot x}
\Bigr] \;.
\end{equation}

\noindent The superscript $i$ refers to the transverse dimensions $x$
and $y$, and the superscript $c$ is for SU(3) color.  If required the
physical polarization vector can be represented

\begin{equation}
{\bf \epsilon}_\perp(\uparrow)=-{1 \over \sqrt{2}} (1,i) \;,\;\;\;
{\bf \epsilon}_\perp(\downarrow)={1 \over \sqrt{2}} (1,-i) \;.
\end{equation}

\noindent The quantization conditions are

\begin{equation}
\Bigl[A^{ic}_{\perp}(x),\partial^+ A^{jd}_{\perp}(y)
\Bigr]_{x^+=y^+=0} =i \delta^{ij} \delta_{cd} \delta^3(x-y) \;,
\end{equation}

\begin{equation}
\Bigl[a^c(k,\lambda),a^{d\dagger}(p,\lambda')\Bigr] = 16\pi^3 k^+
\delta_{\lambda \lambda'} \delta_{cd} \delta^3(k-p) \;.
\end{equation}

The classical equations for $\psi_-=\Lambda_- \psi$ and $A^-$ do
not involve time-derivatives, so these variables can be eliminated in
favor of dynamical degrees of freedom.  This formally yields

\begin{eqnarray}
\psi_-&=&{1 \over i\partial^+} \Bigl[i {\bf \alpha}_\perp \cdot {\bf
D}_\perp+\beta m\Bigr] \psi_+ \nonumber \\
&=&\psit_- - {g \over i\partial^+} {\bf \alpha}_\perp \cdot {\bf
A}_\perp \psi_+ \;,
\end{eqnarray}

\noindent where the variable $\psit_-$ is defined on the second
line to separate the interaction-dependent part of $\psi_-$; and

\begin{eqnarray}
A^{a-}&=&{2 \over i\partial^+} i {\bf \partial}_\perp
\cdot {\bf A}^a_\perp +
{2 i g f^{abc} \over (i\partial^+)^2}\Biggl\{\bigl(i\partial^+
A^{bi}_\perp\bigr)A^{ci}_\perp +2 \psi_+^\dagger T^a \psi_+\Biggr\}
\nonumber \\
&=&\At^{a-}+{2 i g f^{abc} \over (i\partial^+)^2}
\Biggl\{\bigl(i\partial^+ A^{bi}_\perp\bigr) A^{ci}_\perp
+2 \psi_+^\dagger T^a \psi_+ \Biggr\}
\;,
\end{eqnarray}

\noindent where the variable $\At^-$ is defined on the second line to
separate the interaction-dependent part of $A^-$.

Given these replacements, we can follow a canonical procedure to
determine the Hamiltonian.  This path is full of difficulties that
I ignore, because ultimately I will use coupling coherence to refine
the definition of the Hamiltonian and determine the non-canonical
interactions that are inevitably produced by the violation of explicit
covariance and gauge invariance.  For my purposes it is sufficient to
write down a Hamiltonian that can serve as a starting point:

\begin{equation}
H=H_0+V \;,
\end{equation}

\begin{eqnarray}
H_0&=&\int d^3x \Bigl\{ Tr\bigl(\partial^i_\perp A^j_\perp
\partial^i_\perp A^j_\perp\bigr)+\psi_+^\dagger \bigl(i \alpha^i_\perp
\partial^i_\perp+\beta m\bigr) \psi_+\Bigr\} \nonumber \\
&=&\sum_{colors} \int {dk^+ d^2k_\perp \over 16\pi^3 k^+} \Biggl\{
\sum_\lambda {{\bf k}_\perp^2 \over k^+} a^\dagger(k,\lambda)
a(k,\lambda) \nonumber \\
&&~~~~~~~~~~~+ \sum_\sigma {{\bf k}_\perp^2+m^2 \over k^+}
\Bigl(b^\dagger(k,\sigma) b(k,\sigma)+d^\dagger(k,\sigma) d(k,\sigma)
\Bigr) \Biggr\} \;.
\end{eqnarray}

\noindent In the last line the `self-induced inertias' ({\it i.e.},
one-body operators produced by normal-ordering $V$) are not included.
It is difficult to regulate the field contraction encountered in
normal-ordering in a manner exactly consistent with the cutoff
regulation of contractions encountered later.  Coupling coherence
avoids this issue and produces the correct one-body counterterms with
no discussion of normal-ordering required.

The interactions are complicated and are most easily
written using the variables, $\psit=\psit_- + \psi_+$,
and $\At$, where $\At^+=0$, $\At^-$ is defined above, and
$\At^i_\perp=A^i_\perp$.  Using these variables we have

\begin{eqnarray}
V&=\int d^3x\Biggl\{& g \overline{\psit} \gamma_\mu \At^\mu \psit +
2 g~ Tr\Bigl(i\partial^\mu \At^\nu \Bigl[
\At_\mu, \At_\nu \Bigr] \Bigr) \nonumber \\
&&-{g^2 \over 2} Tr\Bigl( \Bigl[\At^\mu,\At^\nu\Bigr] \Bigl[
\At_\mu, \At_\nu \Bigr] \Bigr) + g^2 \overline{\psit} \gamma_\mu
\At^\mu {\gamma^+ \over 2i\partial^+} \gamma_\nu \At^\nu \psit
\nonumber \\
&&+{g^2 \over 2} \overline{\psi} \gamma^+ T^a \psi {1 \over
(i\partial^+)^2} \overline{\psi} \gamma^+ T^a \psi \nonumber \\
&&-g^2 \overline{\psi} \gamma^+ \Biggl({1 \over (i\partial^+)^2}
\Bigl[i\partial^+ \At^\mu, \At_\mu \Bigr] \Biggr) \psi \nonumber \\
&&+g^2 ~Tr \Biggl(\Bigl[i\partial^+ \At^\mu, \At_\mu \Bigr] {1 \over
(i\partial^+)^2} \Bigl[i\partial^+ \At^\nu, \At_\nu \Bigr] \Biggr)
~~~\Biggr\}\;.
\end{eqnarray}

\noindent The commutators in this expression are SU(3) commutators
only.  The potential algebraic complexity of calculations becomes
apparent when one systematically expands every term in $V$ and
replaces:

\begin{equation}
\psit^- \rightarrow {1 \over i\partial^+}\Bigl(i {\bf \alpha}_\perp \cdot
{\bf \partial}_\perp+\beta m\Bigr) \psi_+ \;,
\end{equation}

\begin{equation}
\At^- \rightarrow {2 \over i\partial^+} i{\bf \partial}_\perp \cdot
{\bf A}_\perp \;;
\end{equation}

\noindent and then expands $\psi_+$ and ${\bf A}_\perp$ in terms of
creation and annihilation operators.  It rapidly becomes evident that
one should avoid such explicit expansions if possible.

\section{Positronium: QED in 3+1 Dimensions}

In this section I finally come to a `new' calculation, although my
goals are quite modest.  I will first outline a procedure that can be
used to compute the masses and eigenstates of QED bound states to
arbitrarily high accuracy (in principle).  I will then show that the
leading result for the binding energy of positronium is exactly the
Bohr energy, and that the necessary Coulomb interaction is
automatically produced as a two-body interaction by the similarity
transformation and coupling coherence.

The leading results for positronium in light-front QED have already
been obtained by others.  Brodsky, Pauli, and Tang\cite{tang} showed
how to set up the positronium bound state problem in DLCQ with a
Tamm-Dancoff truncation\cite{tamm} to two sectors of Fock space: (i)
the electron-positron sector, and (ii) the electron-positron-photon
sector.  Krautgartner, Pauli, and Wolz\cite{kraut} derived the
nonrelativistic continuum limit for the calculation analytically, and
with a prescription for handling an infrared divergence they showed
that to leading order the binding energy for positronium results. Kaluza
and Pauli\cite{kaluza1} showed that reasonably accurate numerical results
can be obtained with proper renormalization of the Hamiltonian, although
logarithmic `divergences' remain in their calculations.  The divergences
that remain are four-fermion interactions which are
sector-dependent\cite{tamm}. These divergences are not a serious problem
if one uses a sufficiently small cutoff, weak coupling, and does not seek
precise answers.  Kaluza and Pirner\cite{kaluza2} extended the work in
Ref. \cite{kraut} through the fine structure.

There are no doubt references that I have missed, and Brodsky and
Lepage certainly understood the important ingredients for a
light-front positronium calculation long ago.

Several important ingredients for an `all-orders' calculation are
missing or not clear in the above work.  First, no general method
for computing non-canonical counterterms is given, and as one
consequence sector-dependent coupling counterterms due to the
Tamm-Dancoff truncation are sometimes missed (see \cite{ghpsw} for a
good discussion of the problems). Second, when the DLCQ calculations
were used to derive an integral equation for the electron-positron
bound state in the continuum limit\cite{kraut}, the full energy
appeared in the interaction kernel and had to be replaced by hand with
a free energy to avoid new infrared divergences.  It was not clear how
to carry this prescription to higher orders.  Nonetheless, the DLCQ
calculations have laid the groundwork upon which my calculations are
based; and to the order I work in this paper the differences are not
important.

The small new step taken here is the use of a renormalization group
computational framework that produces the `leading' result ({\it
i.e.}, the Bohr energy) and indicates how to systematically improve
this result.

The first step is to compute a renormalized cutoff Hamiltonian as a
power series in the coupling $e$. Starting with the canonical
Hamiltonian as a `seed,' this is done with the similarity
renormalization group\cite{glazek1,glazek2} and coupling
coherence\cite{coupcoh,perryrg}.  The result is an apparently unique
perturbative series,

\begin{equation}
H^\Lambda_N=h_0 + e_\Lambda h_1 +e_\Lambda^2 h_2 +
\cdot\cdot\cdot + e_\Lambda^N h_N \;.
\end{equation}

\noindent If we use the resultant Hamiltonian to compute scattering
observables to ${\cal O}(e^N)$, they should respect all symmetries of
Lagrangian perturbation theory such as Lorentz covariance and gauge
invariance (for gauge invariant matrix elements).  This has not been
proven to all orders, but I have found no exceptions.

Here $e_\Lambda$ is the running coupling constant, and all remaining
dependence on $\Lambda$ in the operators $h_i$ must be explicit. In
principle I must also treat $m_\Lambda$, the electron running mass, as
an independent function of $\Lambda$; but this will not affect the
results to the order I compute here. We must calculate the
Hamiltonian to a fixed order, and systematically improve the
calculation later by including higher order terms.  This starting
point should be completely adequate for QED, but must be refined for
QCD to include non-perturbative condensate effects; because it will
never produce a chiral-symmetry breaking operator, for example.

Having obtained the Hamiltonian to some order in $e$, the next step is
to split it into two parts,

\begin{equation}
H^\Lambda=\H_0+\V \;.
\end{equation}

\noindent As discussed before, $\h0$ must be accurately solved
non-perturbatively, producing a zeroth order approximation for the
eigenvalues and eigenstates.  The greatest ambiguities in the
calculation appear in the choice of $\h0$, which requires one of
science's most powerful computational tools, trial and error.

In QED and QCD {\it I assume that all interactions in $\H_0$ preserve
particle number}, with all interactions that involve particle creation
and annihilation in $\V$.  This assumption is consistent with the
original hypothesis that a constituent picture will emerge, but it
should emerge as a valid approximation.

The final step before the loop is repeated, starting with a more
accurate approximation for $H^\Lambda$, is to compute corrections from
$\V$ in bound state perturbation theory.  There is no reason
to compute these corrections to arbitrarily high order, because the
initial Hamiltonian contains errors that limit the accuracy we can
obtain in bound state perturbation theory.

One test of the choice of $\H_0$ is provided by the cutoff.
If an exact calculation were performed the results would be
independent of $\Lambda$; but the truncation of the series for
$H^\Lambda$ and the non-perturbative behavior of $\H_0$ will
introduce $\Lambda$ dependence. For example, if the cutoff is
extremely large and $\H_0$ does not include any operators
involving photon emission and absorption, the zeroth order bound state
calculations will contain arbitrarily large errors.  If the cutoff is
made too small ({\it e.g.}, much smaller than $\alpha m$ in QED),
arbitrarily large errors will also result.  The study of these errors
is both interesting and important, but I will have little more to say
here.

In this section I: (i) compute $H^\Lambda$ to ${\cal O}(e^2)$, (ii)
assume the cutoff to be ${\cal O}(\alpha^{3/4} m)$ for non-perturbative
analyses, (iii) include the most infrared singular two-body
interactions in $\H_0$, and (iv) estimate the binding energy for
positronium to ${\cal O}(\alpha^2 m)$.

In QED the natural low-energy transverse momentum scale emerges easily
in a renormalization group analysis, because it is set by the electron
mass when the coupling is small.  In QCD it is not obvious how the
scale $\Lambda_{QCD}$ emerges, and it is easiest to assume that this
happens and use the resultant non-perturbative scaling properties of
the theory near $\Lambda_{QCD}$ to justify the assumption {\it a
posteriori}.

Since $\H_0$ is assumed to include interactions that preserve particle
number, the zeroth order positronium ground state will be a pure
electron-positron state.  We only need  one- and two-body
interactions; {\it i.e.}, the electron self-energy and the
electron-positron interaction.  The canonical interactions can be
found in Eq. (138), and the second-order change in the Hamiltonian is
given in Eq. (96). The shift due to the bare electron mixing with
electron-photon states to lowest order is

\begin{equation}
\delta \Sigma_p=e^2 \int {dk_1^+ d^2k_{1\perp} \over 16 \pi^3}
{\theta(k_1^+) \theta(k_2^+) \over k_1^+ k_2^+} {\overline{u} (p)
D_{\mu\nu}(k_1) \gamma^\mu \bigl(\kslash_2+m\bigr) \gamma^\nu u(p)
\over p^--k_1^--k_2^-} \;,
\end{equation}

\noindent where,

\begin{equation} k_2^+=p^+-k_1^+ \;,\;\;\;{\bf k}_{2\perp}={\bf
p}_\perp-{\bf k}_{1\perp} \;,
\end{equation}

\begin{equation}
k_i^-={k_{i\perp}^2 + m_i^2 \over k_i^+} \;,
\end{equation}

\begin{equation}
D_{\mu\nu}(k)=-g_{\perp \mu \nu} + {k_\perp^2 \over (k^+)^2} \eta_\mu
\eta_\nu + {1 \over k^+} \Bigl(k_{\perp \mu} \eta_\nu + k_{\perp \nu}
\eta_\mu\Bigr) \;,
\end{equation}

\begin{equation}
\eta_\mu a^\mu = a^+ \;.
\end{equation}

\noindent I have not yet displayed the cutoffs.  To evaluate the
integrals it is easiest to use Jacobi variables $x$ and ${\bf s}$ for
the relative electron-photon motion,

\begin{equation}k_1^+=x p^+\;,\;\;\;{\bf k}_{1\perp}=x {\bf
p}_\perp+{\bf s} \;,
\end{equation}

\noindent which implies

\begin{equation}k_2^+=(1-x) p^+\;,\;\;\;{\bf k}_{2\perp}=(1-x) {\bf
p}_\perp-{\bf s} \;.
\end{equation}

The second-order change in the electron self-energy becomes

\begin{eqnarray}
\delta \Sigma_p &=& -{e_\Lambda^2 \over p^+} \int {dx d^2s \over 16\pi^3}
\theta\Biggl(y \Lambda_0^2-{s^2 +x^2 m^2 \over x (1-x)}\Biggr)
\theta\Biggl({s^2 +x^2 m^2 \over x (1-x)}-y \Lambda_1^2\Biggr)
\nonumber \\
&&~~~\Biggl({1 \over s^2+x^2 m^2}\Biggr)~
\overline{u}(p,\sigma) \Biggl\{ 2 (1-x) \pslash -2 m + \nonumber \\
&&~~~~~~~~~~~~~~~~~~~~~~{\gamma^+ \over p^+}
\Biggl[ {2 s^2 \over x^2}+{2s^2 \over 1-x}+{x (2-x) m^2 \over 1-x} \Biggr]
\Biggr\} u(p,\sigma) \;,
\end{eqnarray}

\noindent where $y=p^+/P^+$.

Our choice of a cutoff that depends on light-front free energies was
dictated by the desire to avoid introducing spectator dependence and
to produce a well-behaved perturbative expansion for the transformed
Hamiltonian.  This forces us to introduce a scale, $\Lambda^2/P^+$,
and $P^+$ violates explicit longitudinal boost covariance.

It is straightforward in this case to determine the self-energy
required by coupling coherence.  Since the electron-photon coupling
does not run until third order, to second order the self-energy must
exactly reproduce itself with $\Lambda_0 \rightarrow \Lambda_1$.
For the self-energy to be finite we must assume that $\delta \Sigma$
reduces a positive self-energy, so that

\begin{eqnarray}
\Sigma^{\Lambda}_{coh}(p)&=& {e_\Lambda^2 \over p^+}
\int {dx d^2s \over 16\pi^3} \theta\bigl(yx-\epsilon\bigr)
\theta\Biggl(y \Lambda^2-{s^2 +x^2 m^2 \over x (1-x)}\Biggr)
\Biggl({1 \over s^2+x^2 m^2}\Biggr) \nonumber
\\&&
\overline{u}(p,\sigma)
\Biggl\{ 2 (1-x) \pslash -2 m +{\gamma^+ \over p^+}
\Biggl[ {2 s^2 \over x^2}+{2s^2 \over 1-x}+
{x (2-x) m^2 \over 1-x} \Biggr]
\Biggr\} u(p,\sigma) \nonumber \\
&=&{e_\Lambda^2 \over 8\pi^2 p^+}
\Biggl\{2 y \Lambda^2 \ln\Biggl(
{ y^2 \Lambda^2 \over (y\Lambda^2 + m^2)\epsilon}\Biggr)
-{3 \over 2} y \Lambda^2+{1 \over 2}
{ym^2 \Lambda^2 \over y \Lambda^2+m^2} \nonumber
\\ &&~~~~~~~~~~~~~~~~~~~~~~~~+ 3 m^2
\ln\Biggl( {m^2 \over y \Lambda^2 + m^2} \Biggr) \Biggl\} + {\cal
O}(\epsilon/y) \;.
\end{eqnarray}

I have been forced to introduce {\it a second cutoff},

\begin{equation}
x p^+ > \epsilon P^+ \;,
\end{equation}

\noindent because after the ${\bf s}$ integration is completed we are
left with a logarithmically divergent $x$ integration.  Other choices
for this second infrared cutoff are possible and lead to similar
results.  I believe that these additional $\ln(\epsilon)$ divergences
occur only when a photon is emitted and absorbed in immediately
successive vertices. In other words, they resemble
super-renormalizable divergences, which can be renormalized
non-perturbatively.  However, I do not subtract these divergences.

The electron and photon (quark and gluon) `mass' operators, are a
function of a longitudinal momentum scale introduced by the cutoff,
and there is an exact scale invariance required by longitudinal boost
invariance.  Here I mean by `mass operator' the one-body operator
when the transverse momentum is zero, even though this does not agree
with the free mass operator because it includes longitudinal momentum
dependence.  The cutoff violates boost invariance and the mass
operator is required to restore the symmetry.  The only local scale is
the parton longitudinal momentum, $p^+$; and the bare mass term scales
as $1/p^+$.  There is not always a single particle that naturally sets
the scale for a global cutoff.  A mass term that scales as $1/p^+$
causes wee partons to decouple from valence partons, and removes the
infrared singularities from QCD\cite{nonpert}.  My simple confinement
mechanism requires these infrared singularities to cancel rather than
disappear, and it is important that the gluon `mass' scales as $1/P^+$
where $P^+$ is a scale fixed by the cutoff and might be chosen to be
the total longitudinal momentum of the state.  Physical results cannot
depend on the cutoff scale, so the natural scale for single hadron
problems is the momentum of the hadron.

We must interpret this new infrared divergence, because we have no
choice about whether it is in the Hamiltonian if we use coupling
coherence.  We can only choose between putting the divergent operator
in $\H_0$ or in $\V$.  I make different choices in QED and QCD, and
the arguments are based on physics.

{\it This divergent electron `mass' is a complete lie.}  We encounter
a term proportional to $e_\Lambda^2 \Lambda^2 \ln(1/\epsilon)/P^+$
when the scale is $\Lambda$; however, we can reduce this scale as far
as we please in perturbation theory.  Photons are massless, so the
electron will continue to dress itself with small-x photons to
arbitrarily small $\Lambda$. Since I believe that this divergent
self-energy is exactly canceled by mixing with small-x photons, and
that this mixing can be treated perturbatively in QED, I simply put
the divergent electron self-energy in $\V$, which is treated
perturbatively.

These divergences place a severe constraint on our calculation.  In
perturbation theory they should cancel in gauge-invariant matrix
elements, and {\it any non-perturbative approximation we choose must
maintain this cancellation.} {\it In QED I will put all infrared
singular operators in $\V$}, and the cancellations should proceed in
bound state perturbation theory exactly as before if they are
`super-renormalizable.'

Before deriving the electron-positron interaction, I should mention
that the photon self-energy is not divergent, although it does contain
a `mass' term that is proportional to $\alpha \Lambda^2/P^+$.  This
bare mass is exactly canceled by mixing with electron-positron pairs
and more complicated states at higher order, leaving a massless
physical photon.  If the cutoff is extremely large, the photon bare
mass must be tuned very precisely to get accurate low energy results,
which is one reason that we need to use a cutoff near the scale of
physical interest in practical calculations.  The photon mass should
be placed in $\V$ and treated perturbatively for a ground state
positronium calculation.

There are two time-ordered `diagrams' involving photon exchange
between an electron with initial momentum $p_1$ and final momentum
$p_2$, and a positron with initial momentum $k_1$ and final momentum
$k_2$.  Using Eq. (96), we find the required matrix element of $\delta
H$,

\begin{eqnarray}
\delta H &=& -{e^2 \over q^+} D_{\mu
\nu}(q) \overline{u}(p_2,\sigma_2) \gamma^\mu u(p_1,\sigma_1)
\overline{v}(k_1,\lambda_1) \gamma^\nu v(k_2,\lambda_2) \nonumber \\
&&~~\theta\bigl(|q^+|-\epsilon P^+\bigr)
\theta\Biggl({\Lambda_0^2 \over P^+} - \mid p_1^- -p_2^-
-q^- \mid\Biggr) \theta\Biggl({\Lambda_0^2 \over P^+} -
\mid k_2^- -k_1^- -q^- \mid\Biggr) \nonumber \\
&&~~\Biggl[ {\theta\bigl(|p_1^- -p_2^- -q^-| -\Lambda_1^2 / {\cal P}^+
\bigr)
\;\; \theta\bigl(|p_1^- -p_2^- -q^-|- |k_2^- -k_1^- -q^-| \bigr)
\over p_1^- -p_2^- -q^-} \nonumber \\ &&~~~~
+{\theta\bigl(|k_2^- -k_1^- -q^-| - \Lambda_1^2 / {\cal P}^+ \bigr)
\;\; \theta\bigl( |k_2^- -k_1^- -q^-| - |p_1^- -p_2^- -q^-| \bigr)
\over k_2^- -k_1^- -q^-} \Biggr]
\nonumber \\
&&~~~~~~~~~~~\theta\Biggl(\Lone-\mid p_1^-+k_1^--p_2^--k_2^- \mid
\Biggr) \;, \nonumber \\
~
\end{eqnarray}

\noindent where $q^+=p_1^+-p_2^+\;,\;\;{\bf q}_\perp={\bf p}_{1\perp}
- {\bf p}_{2\perp}$, and $q^-=q_\perp^2/q^+$.

I have used the second cutoff on longitudinal momentum that I was
forced to introduce when computing the change in the self-energy.  We
will see in the section on confinement that it is essential to include
this cutoff everywhere consistently.  In QED this point is not
immediately important, because all infrared singular interactions,
including the infrared divergent self-energy, are put in $\V$ and
treated perturbatively.  Since I do not compute perturbative
corrections for positronium, I do not encounter the problem in QED.

To determine the interaction that must be added to the Hamiltonian to
maintain coupling coherence, we must again find an interaction that
added to $\delta V$ reproduces itself with $\Lambda_0 \rightarrow
\Lambda_1$ everywhere.  The coupling coherent interaction generated by
the first terms in $\delta H$ are not uniquely determined at this
order. There is some ambiguity because we can obtain coupling
coherence either by having $\delta H$ increase the strength of an
operator by adding additional phase space strength, or we can have
$\delta H$ reduce the strength of an operator by subtracting phase
space strength.  The ambiguity is resolved in higher orders, so I will
simply state the result.  If an instantaneous photon-exchange
interaction is present in $H$, we want $\delta H$ to cancel part of
this marginal operator and to increase the strength of a new
photon-exchange interaction. This new interaction reproduces the
effects of high energy photon exchange removed by the cutoff.  The
result is

\begin{eqnarray}
V_{coh}^{\Lambda} &=& - {e_{\Lambda}^2 \over q^+}  D_{\mu
\nu}(q) \overline{u}(p_2,\sigma_2) \gamma^\mu u(p_1,\sigma_1)
\overline{v}(k_1,\lambda_1) \gamma^\nu v(k_2,\lambda_2) \nonumber \\
&&~~~~~~~\theta\bigl(|q^+|-\epsilon P^+\bigr)~
\theta\Biggl(\Lam-\mid p_1^-+k_1^--p_2^--k_2^- \mid
\Biggr) \nonumber \\
&&~~\Biggl[ {\theta\bigl(|p_1^- -p_2^- -q^-| -\Lambda^2 / {\cal P}^+ \bigr)
\;\; \theta\bigl(|p_1^- -p_2^- -q^-|- |k_2^- -k_1^- -q^-| \bigr)
\over p_1^- -p_2^- -q^-} \nonumber \\ &&~~~~
+{\theta\bigl(|k_2^- -k_1^- -q^-| - \Lambda^2 / {\cal P}^+ \bigr)
\;\; \theta\bigl( |k_2^- -k_1^- -q^-| - |p_1^- -p_2^- -q^-| \bigr)
\over k_2^- -k_1^- -q^-} \Biggr]  \;. \nonumber \\
\end{eqnarray}

This matrix element exactly reproduces photon exchange above the
cutoff.  The cutoff removes the direct coupling of electron-positron
states to electron-positron-photon states whose energy differs by more
than the cutoff, and coupling coherence dictates that the result of
this mixing should be replaced by a direct interaction between the
electron and positron.  We could obtain this result by much simpler
means at this order by simply demanding that the Hamiltonian produce
the `correct' scattering amplitude at $\order(e^2)$ with the cutoffs
in place.  Of course, this procedure requires us to provide the
`correct' amplitude, but this is easily done in perturbation theory.

This interaction is rather unusual, because it arises from a tree
diagram.  In a Euclidean or equal-time renormalization group
calculation, tree diagrams are usually ignored.  To understand this
note that the renormalization group is generating interactions between
low energy states produced by their mixing with high energy states,
which means one needs to study transitions from low energy to high
energy and back to low energy at second order.  In equal-time
coordinates the intermediate state will have high energy only if the
photon has a large momentum; but in this case either the incoming or
outgoing electron-positron pair will have a large relative momentum
and cannot be a low energy state.  In light-front coordinates the
intermediate state can have high energy because the photon has a small
longitudinal momentum.  The exchange of a small longitudinal momentum
photon has negligible effect on the relative longitudinal momentum of
the pair, so a counterterm is produced.  This counterterm is
effectively local because of the cutoffs\cite{perryrg}.

$V^\Lambda_{coh}$ is non-canonical, and we will see that it is
responsible for producing the Coulomb interaction.  This same operator
is produced by directly eliminating the electron-positron-photon
sector\cite{kraut}.  However, such a direct elimination leads to the
bound state energy appearing in the energy denominator, producing
infrared divergences that depend on the interactions.  Using a
similarity transformation and coupling coherence we can produce the
required energy-independent interaction without eliminating any Fock
space sectors.

Almost all of this interaction is irrelevant (in the renormalization
group sense), and the marginal part serves to exactly cancel the
canonical instantaneous photon exchange interaction above the cutoff.
Irrelevant operators typically have a small effect on the
renormalization group trajectory of Hamiltonians for large cutoffs;
however, as the cutoff approaches the bound state scale their effects
become increasingly important.  We need some guidance to decide which
irrelevant operators are most important, and in light-front field
theory we can use longitudinal momentum dependence, since it is only
the transverse momentum dependence that affects whether an operator is
relevant, marginal or irrelevant.

{\it I assume that it is the most infrared singular part of $V_{coh}$
that is important.}  As explained above, this operator receives
substantial strength only from the exchange of photons with small
longitudinal momentum; so we expect inverse $q^+$ dependence to
indicate `strong' interactions between low-energy pairs.  So the part
of $V_{coh}$ that is included in $\H_0$ is

\begin{eqnarray}
\tilde{V}_{coh}^{\Lambda} &=& - e_{\Lambda}^2
{q_\perp^2 \over (q^+)^3}~
\overline{u}(p_2,\sigma_2) \gamma^+ u(p_1,\sigma_1)
\overline{v}(k_1,\lambda_1) \gamma^+ v(k_2,\lambda_2) \nonumber \\
&&~~~~~~~\theta\bigl(|q^+|-\epsilon P^+\bigr)~
\theta\Biggl(\Lam-\mid p_1^-+k_1^--p_2^--k_2^- \mid
\Biggr) \nonumber \\
&&~~\Biggl[ {\theta\bigl(|p_1^- -p_2^- -q^-| -\Lambda^2 / {\cal P}^+ \bigr)
\;\; \theta\bigl(|p_1^- -p_2^- -q^-|- |k_2^- -k_1^- -q^-| \bigr)
\over p_1^- -p_2^- -q^-} \nonumber \\ &&~~~~
+{\theta\bigl(|k_2^- -k_1^- -q^-| - \Lambda^2 / {\cal P}^+ \bigr)
\;\; \theta\bigl( |k_2^- -k_1^- -q^-| - |p_1^- -p_2^- -q^-| \bigr)
\over k_2^- -k_1^- -q^-} \Biggr]   \nonumber \\
&=& - 4 e_{\Lambda}^2 \sqrt{p_1^+ p_2^+ k_1^+ k_2^+}
{q_\perp^2 \over (q^+)^3} \delta_{\sigma_1 \sigma_2}
\delta_{\lambda_1 \lambda_2} \nonumber \\
&&~~~~~~~\theta\bigl(|q^+|-\epsilon P^+\bigr)~
\theta\Biggl(\Lam-\mid p_1^-+k_1^--p_2^--k_2^- \mid
\Biggr) \nonumber \\
&&~~\Biggl[ {\theta\bigl(|p_1^- -p_2^- -q^-| -\Lambda^2 / {\cal P}^+ \bigr)
\;\; \theta\bigl(|p_1^- -p_2^- -q^-|- |k_2^- -k_1^- -q^-| \bigr)
\over p_1^- -p_2^- -q^-} \nonumber \\ &&~~~~
+{\theta\bigl(|k_2^- -k_1^- -q^-| - \Lambda^2 / {\cal P}^+ \bigr)
\;\; \theta\bigl( |k_2^- -k_1^- -q^-| - |p_1^- -p_2^- -q^-| \bigr)
\over k_2^- -k_1^- -q^-} \Biggr]  \;. \nonumber \\
\end{eqnarray}

The Hamiltonian is almost complete to second order in the
electron-positron sector, and only the instantaneous photon exchange
interaction must be added.  The matrix element of this interaction is

\begin{eqnarray}
V_{instant} &=& - e_{\Lambda}^2 \Biggl({1 \over q^+}\Biggr)^2
\overline{u}(p_2,\sigma_2)
\gamma^+ u(p_1,\sigma_1) \overline{v}(k_1,\lambda_1) \gamma^+
v(k_2,\lambda_2) \nonumber \\
&&~~~~~~~~~\times
\theta\Biggl(\Lam-\mid p_1^-+k_1^--p_2^--k_2^- \mid
\Biggr) \nonumber \\
&=& - 4 e_{\Lambda}^2 \sqrt{p_1^+ p_2^+ k_1^+ k_2^+}  \Biggl({1 \over
q^+}\Biggr)^2 \delta_{\sigma_1 \sigma_2}
\delta_{\lambda_1 \lambda_2} \nonumber \\
&&~~~~~~~~~~~~\times
\theta\Biggl(\Lam-\mid p_1^-+k_1^--p_2^--k_2^- \mid
\Biggr)
\;.
\end{eqnarray}

\noindent The only cutoff that appears is the cutoff directly run by
the similarity transformation that prevents the initial and final
states from differing in energy by more than $\Lambda^2/P^+$.

This brings us to a final subtle point.  Since there are no cutoffs in
$V_{instant}$ that directly limit the momentum exchange, the matrix
element diverges as $q^+ \rightarrow 0$.  Consider $\tilde{V}_{coh}$
in this same limit,

\begin{eqnarray}
\tilde{V}_{coh}^{\Lambda} &\rightarrow&
4 e_{\Lambda}^2 \sqrt{p_1^+ p_2^+ k_1^+ k_2^+} \Biggl({1 \over
q^+}\Biggr)^2  \delta_{\sigma_1 \sigma_2}
\delta_{\lambda_1 \lambda_2} \nonumber \\
&&~~~~ \times \theta\Biggl(\mid q^- \mid - {\Lambda^2 \over P^+}
\Biggr)  ~ \theta\Biggl(\Lam-\mid p_1^-+k_1^--p_2^--k_2^- \mid \Biggr)
\;.
\end{eqnarray}

\noindent This means that as $q^+ \rightarrow 0$, $V_{coh}$ partially
screens $V_{instant}$, leaving the original operator multiplied by
$\theta(\Lambda^2/P^+-|q^-|)$.  However, even after this partial
screening, the matrix elements of the remaining part of $V_{instant}$
between bound states diverge and we must introduce the same infrared
cutoff used for the self-energy to regulate these divergences. This is
explicitly shown in the section on confinement. However, all
divergences from $V_{instant}$ are exactly canceled by the exchange
of massless photons, which persists to arbitrarily small cutoff.  This
cancellation is exactly analogous to the cancellation of the infrared
divergence of the self-energy, and will be treated in the same way.
The portion of $V_{instant}$ that is not canceled by
$\tilde{V}_{coh}$ will be included in $\V$, the perturbative part of
the Hamiltonian.  We will not encounter this interaction until we also
include photon exchange below the cutoff perturbatively, so all
infrared divergences should cancel in this bound state perturbation
theory.  I repeat that this is not guaranteed for arbitrary choices of
$\H_0$, and we are not free to simply cancel these divergent
interactions with counterterms because coupling coherence completely
determines the Hamiltonian.

I should also mention that if we do not include the canonical
instantaneous photon exchange interaction in the seed Hamiltonian,
coupling coherence will generate it.

We now have the complete interaction that I include in $\H_0$. 
Letting $\H_0=h_0+\v_0$, where $h_0$ is the free hamiltonian, I add
parts of $V_{instant}$ and $V_{coh}$ to obtain

\begin{eqnarray} 
\v_0 &=& - 4 e_{\Lambda}^2 \sqrt{p_1^+ p_2^+ k_1^+ k_2^+}
\delta_{\sigma_1 \sigma_2} \delta_{\lambda_1 \lambda_2}
\theta\Biggl(\Lam-\mid p_1^-+k_1^--p_2^--k_2^- \mid
\Biggr) \times \nonumber \\ 
&& \Biggl\{ 
\Biggl[ {q_\perp^2 \over (q^+)^3} { 1 \over p_1^- -p_2^- -q^-}
+ {1 \over (q^+)^2} \Biggr]  \theta\bigl(|p_1^- -p_2^- -q^-|
-\Lambda^2 / {\cal P}^+ \bigr) \nonumber \\
&&~~~~~~~~~~~~~~~~~~~~~~~~~~~~ \times \theta\bigl(|p_1^- -p_2^-
-q^-|-  |k_2^- -k_1^- -q^-| \bigr) \nonumber \\ 
&&+
\Biggl[ {q_\perp^2 \over (q^+)^3} {1 \over k_2^- -k_1^- -q^-} 
+ {1 \over (q^+)^2} \Biggr]   \theta\bigl(|k_2^- -k_1^- -q^-| 
- \Lambda^2 / {\cal P}^+ \bigr) \nonumber \\
&&~~~~~~~~~~~~~~~~~~~~~~~~~~~~ \times \theta\bigl( |k_2^- -k_1^-
-q^-| - |p_1^- -p_2^- -q^-| \bigr) \Biggr\} \;.
\end{eqnarray}

\noindent There is remaining freedom in the cutoff dependence of
$\H_0$ that should be used to obtain an efficient analysis of QED
bound states, but I will not exploit this freedom here.  Since we can
always add and subtract any operator we please, we can use this
freedom to decrease cutoff dependence.  For example, there are several
sharp cutoffs in $\v_0$ that result from my simple choice for the
similarity transformation.  We can add and subtract the same operator
with a smoother cutoff to improve the analysis.

In order to present an analytic analysis I will make assumptions that
are not necessary in a numerical analysis and that can be justified $a~
posteriori$.  First I will assume that the electron and positron
momenta can be arbitrarily large, but that in low-lying states their
relative momenta satisfy

\begin{equation}
|{\bf p}_\perp-{\bf k}_\perp| \ll m \;,
\end{equation}

\begin{equation}
|p^+-k^+| \ll p^++k^+ \;.
\end{equation}

\noindent It is essential that the condition for longitudinal momenta
not involve the electron mass, because masses have the scaling
dimensions of transverse momenta and not longitudinal momenta.  As
above, I use $p$ for the electron momenta and $k$ for the positron
momenta.  To be even more specific, I will assume that

\begin{equation}
|{\bf p}_\perp-{\bf k}_\perp| \sim \alpha m \;,
\end{equation}

\begin{equation}
|p^+-k^+| \sim \alpha (p^++k^+) \;.
\end{equation}

\noindent  This allows us to use power counting to evaluate the
perturbative strength of operators for small coupling, which may prove
essential in the analysis of QCD\cite{nonpert}. Note that these
conditions allow us to infer

\begin{equation}
|{\bf p}_{1\perp}-{\bf p}_{2\perp}| \sim \alpha m \;,
\end{equation}

\begin{equation}
|p_1^+-p_2^+| \sim \alpha (p^++k^+) \;.
\end{equation}

Given these order of magnitude estimates for momenta, we can
drastically simplify the free energies in the kinetic energy operator
and the energy denominators in $\v_0$.  We can use transverse boost
invariance to choose a frame in which

\begin{equation}
p_i^+=y_i P^+ \;,\;\;p_{i\perp}=\kappa_i \;,\;\;\;\;k_i^+=(1-y_i)P^+
\;,\;\;k_{i\perp}=-\kappa_i \;,
\end{equation}

\noindent so that

\begin{eqnarray}
P^+(p_1^--p_2^--q^-) &=& {\kappa_1^2+m^2 \over y_1} - {\kappa_2^2+m^2
\over y_2} - {(\kappa_1-\kappa_2)^2 \over y_1-y_2} \nonumber
\\&=&-4 m^2 (y_1-y_2) - {(\kappa_1-\kappa_2)^2 \over y_1-y_2} +
\order(\alpha^2 m^2) \;.
\end{eqnarray}

\noindent To leading order all energy denominators are the same.
Each  energy denominator is $\order(\alpha m^2)$, which is large in
comparison to the binding energy we will find.  This is important,
because the bulk of the photon exchange that is important for the low
energy bound state involves intermediate states that have larger
energy than the differences in constituent energies in the region of
phase space where the wave function receives most of its strength.
This allows us to use a perturbative renormalization group to compute
the dominant effective interactions.

There are similar simplifications for all energy denominators.  After
making these approximations we find that the matrix element of
$\v_0$ is

\begin{eqnarray}
\v_0&=&4 e_\Lambda^2
\sqrt{y_1 y_2 (1-y_1) (1-y_2)} ~\theta\Bigl( 4 m^2
|y_1-y_2| + {(\kappa_1-\kappa_2)^2 \over |y_1-y_2|}
-\Lambda^2\Bigr) \nonumber \\
&&~~~~~\theta\Bigl(\Lambda^2-4\mid \kappa_1^2+4 m^2(1-2y_1)^2-
\kappa_2^2-4m^2(1-2y_2)^2 \mid\Bigr) \nonumber \\
&&~~~~~~~~~~~~~~~~~~~ {(\kappa_1-\kappa_2)^2
\over (y_1-y_2)^2} \Biggl[{1 \over 4 m^2 (y_1-y_2)^2+(\kappa_1 -
\kappa_2)^2}-{1 \over (\kappa_1-\kappa_2)^2}\Biggr] \nonumber \\
&=&-16 e_\Lambda^2 m^2 \sqrt{y_1 y_2 (1-y_1) (1-y_2)}~ \theta\Bigl( 4
m^2 |y_1-y_2| + {(\kappa_1-\kappa_2)^2
\over |y_1-y_2|}-\Lambda^2\Bigr) \nonumber
\\ &&~~~~\theta\Bigl(\Lambda^2-4\mid \kappa_1^2+4 m^2(1-2y_1)^2-
\kappa_2^2-4m^2(1-2y_2)^2 \mid\Bigr) \nonumber \\
&&~~~~~~~~~~~~~~~~~~~~~~\Biggl[
{1 \over 4 m^2 (y_1-y_2)^2+(\kappa_1-\kappa_2)^2} \Biggr] \;.
\end{eqnarray}

\noindent In principle the electron-positron annihilation graphs
should also be included at this order, but the resultant effective
interactions do not diverge as $q^+ \rightarrow 0$, so I include such
effects perturbatively in $\V$.

At this point we can complete the zeroth order analysis of positronium
using the state,

\begin{eqnarray}
|\Psi(P)\rangle &=& \sum_{\sigma \lambda} \int {dp^+ d^2p_\perp \over
16\pi^3 p^+} {dk^+ d^2k_\perp \over 16\pi^3 k^+} \sqrt{p^+ k^+}
16\pi^3 \delta^3(P-p-k) \nonumber \\
&&~~~~~~~~~~~~~~\phi(p,\sigma;k,\lambda) b^\dagger(p,\sigma)
d^\dagger(k,\lambda) |0\rangle \;,
\end{eqnarray}

\noindent where $\phi(p,\sigma;k,\lambda)$ is the wave function for
the relative motion of the electron and positron, with the
center-of-mass momentum being $P$.  We need to choose the longitudinal
momentum appearing in the cutoff, and I will use the natural scale
$P^+$.  The matrix element of $\H_0$ is

\begin{eqnarray}
&&\langle \Psi(P)|\H_0|\Psi(P')\rangle = 16\pi^3 \delta^3(P-P') \times
\nonumber \\
&&~~~\Biggl\{ \int {dy d^2\kappa \over 16\pi^3}
\Bigl[4m^2+4\kappa^2+4m^2(1-2y)^2\Bigr] |\phi(\kappa,y)|^2 \nonumber \\
&&~~~~~-16 e^2 m^2 \int {dy_1 d^2\kappa_1 \over 16\pi^3} {dy_2
d^2\kappa_2 \over 16\pi^3} \theta\Bigl(4 m^2|y_1-y_2| + {(\kappa_1 -
\kappa_2)^2 \over |y_1-y_2|}-\Lambda^2\Bigr) \nonumber \\
&&~~~~~~~~~~~\theta\Bigl(\Lambda^2-4\mid \kappa_1^2+4 m^2(1-2y_1)^2-
\kappa_2^2-4m^2(1-2y_2)^2 \mid\Bigr) \nonumber \\
&&~~~~~~~~~~~~~~~\Biggl[{1 \over 4m^2|y_1-y_2|^2+(\kappa_1-\kappa_2)^2}
\Biggr] \phi^*(\kappa_2,y_2) \phi(\kappa_1,y_1) \Biggr\} \;.
\end{eqnarray}

\noindent I have chosen a frame in which $P_\perp=0$ and used the
Jacobi coordinates defined above, and indicated only the electron
momentum in the wave function since momentum conservation fixes the
positron momentum.  I have also dropped the spin indices because the
interaction in $\H_0$ is independent of spin.

If we vary this expectation value subject to the constraint that the
wave function is normalized we obtain the equation of motion,

\begin{eqnarray}
M^2 \phi(\kappa_1,y_1) &=& (4 m^2 - 4 m E + E^2) \phi(\kappa_1,y_1)
\nonumber \\&=& \Bigl[4m^2+4\kappa_1^2+4m^2(1-2y_1)^2\Bigr]
\phi(\kappa_1,y_1) \nonumber \\
&&-16 e^2 m^2 \int {dy_2 d^2\kappa_2 \over 16\pi^3}
\theta\Bigl(4 m^2|y_1-y_2| + {(\kappa_1 -
\kappa_2)^2 \over |y_1-y_2|} -\Lambda^2\Bigr) \nonumber \\
&&~~~~~~\theta\Bigl(\Lambda^2-4\mid \kappa_1^2+4 m^2(1-2y_1)^2-
\kappa_2^2-4m^2(1-2y_2)^2 \mid\Bigr) \nonumber \\
&&~~~~~~~~~~\Biggl[{1 \over 4m^2|y_1-y_2|^2+(\kappa_1-\kappa_2)^2}
\Biggr] \phi(\kappa_2,y_2) \;.
\end{eqnarray}

\noindent $E$ is the binding energy, and we can drop the $E^2$ term
since it will be $\order(\alpha^4)$.

I do not think that it is possible to solve this equation analytically
with the cutoffs in place, and with the light-front kinematic
constraints $-1 \le 1-2y_i \le 1$.  In order to determine the binding
energy to leading order, we need to evaluate the regions of phase
space removed by the cutoffs.

If we want to find a cutoff for which the ground state is dominated by
the electron-positron component of the wave function, we need the first
cutoff to remove the important part of the electron-positron-photon
phase space.  Using the `guess' that $|\kappa|=\order(\alpha m)$ and
$1-2y=\order(\alpha)$, this requires

\begin{equation}
\Lambda^2<\alpha m^2 \;.
\end{equation}

\noindent On the other hand, we cannot allow the cutoff to remove the
region of the electron-positron phase space from which the wave
function receives most of its strength.  This requires

\begin{equation}
\Lambda^2>\alpha^2 m^2 \;.
\end{equation}

\noindent A compromise that satisfies both requirements is

\begin{equation}
\Lambda^2=\alpha^{3/2} m^2 \;.
\end{equation}

We are able to satisfy both requirements in QED because $\alpha$ is
small.  In QCD these competing constraints may require us to relax our
assumption that ground state hadrons are well-approximated by the
minimal valence configuration alone.  This is an extremely important
point that cannot be reasonably analyzed in QCD with simple power counting
arguments.  This problem will be illuminated by better approximations
in QED that allow further Fock space sectors to enter the state
non-perturbatively, and that thereby yield accurate results over a
wider range of cutoffs.  Another possibility is to use our freedom to
add and subtract operators to the Hamiltonian.  We can avoid using
interactions in $\H_0$ that are pinched between two cutoffs by adding
and subtracting operators that do not have the lower cutoff, for
example.

If we choose $\Lambda=\alpha^{3/4} m$, we can remove both cutoffs and
allow $-\infty<1-2y_i<\infty$ as a first approximation for the
integral equation, without altering the leading power law behavior of
the binding energy.  These approximations will produce $\order(e^5)$
errors in the leading analysis, but these errors become important when
we go to the next order and must be fixed.  There are a number of
ways to do this in QED.

With these final approximations the analytic solution is

\begin{equation}
\phi(\kappa,y)= \sqrt{8 \pi \alpha^5 m^6}
\Bigl(\kappa^2+(1-2y)^2m^2+mE\Bigr)^{-2} \;,
\end{equation}

\begin{equation}
E={1 \over 4} \alpha^2 m \;.
\end{equation}

\noindent This is exactly the Bohr energy, and the wave function is
related to the Fourier transform of the nonrelativistic positronium
ground state wave function.  Note that this wave function is
normalized only to leading order.

I hope to have accomplished two objectives in this section.  First, I
wanted to use a familiar bound state problem to illustrate the
formalism.  At first, the arguments seem extremely complicated; but
this is largely because I started with the full field theory and
systematically derived a leading approximation for the ground state.
The reader should try doing this in equal-time field theory,
especially in a gauge other than Coulomb gauge.  My second objective
was to lay the groundwork for the discussion of confinement in QCD.
To $\order(\alpha)$ the coupling coherent QCD Hamiltonian is almost
identical to the QED Hamiltonian, and the only difference in my
analysis of QCD is my choice of $\H_0$, which is based on physics.


\section{Asymptotic Freedom}

Before I proceed to the discussion of confinement, I must discuss
asymptotic freedom in Hamiltonian light-front QCD briefly.  In order
to derive a constituent picture, we must replace almost all Fock space
components that appear in hadronic wave functions when the cutoff is
arbitrarily large with effective operators in the Hamiltonian and
other observables.  It has long been understood that degrees of
freedom that couple perturbatively can be removed from the wave
functions and replaced by effective interactions, and we will
certainly want to remove as many degrees of freedom as possible in
this manner.  Since QCD is asymptotically free\cite{free}, we might
hope to remove an infinite number of scales perturbatively, as is done
in lattice gauge theory.

The first problem we must face before exploiting asymptotic freedom is
that asymptotic freedom is known from Lagrangian calculations in which
covariance and gauge invariance are maintained manifestly.  Moreover,
the coupling runs with the four-momentum squared, and this is not a
variable that we encounter directly in light-front Hamiltonian
calculations.  To exploit asymptotic freedom we need the coupling to
run with the cutoff that is run by the perturbative light-front
renormalization group, and we need to find that the coupling becomes
arbitrarily small as this cutoff is taken to infinity.  If this
happens, we can hope to exploit coupling coherence in the regime of
asymptotic freedom to approximate the light-front Hamiltonian.  Of
course, we need to reduce the cutoff as far as possible so that
hadronic states can be dominated by low-energy few-parton components,
and we will be limited by the fact that the coupling grows as the
cutoff is reduced.

The results I show in this section do not come from a third-order
similarity transformation, because the appropriate calculation has not
been completed yet.  I will show the results of a third-order,
time-ordered calculation of the quark-gluon vertex\cite{Pe 93a}. If
the same cutoffs are used, the similarity results should be identical.

As discussed in the last section, renormalization is complicated in
light-cone gauge by infrared singularities that are commonly thought
to be spurious.  In Lagrangian QCD these result from inverse powers of
longitudinal momenta in the gluon propagator. Light-front infrared
singularities are typically avoided in Lagrangian QCD by using a pole
prescription developed by Mandelstam\cite{mand} and
Leibbrandt\cite{lei}. Bassetto, Nardelli and Soldati\cite{bassetto}
have provided a broad discussion of renormalization in light-cone
gauge, and I refer the interested reader to their excellent work. What
is most important for my discussion is that the Mandelstam-Leibbrandt
pole prescription prevents us from directly relating light-front
time-ordered diagrams to Feynman diagrams, because the poles are
inappropriately placed to complete $k^-$ integrals first.

Simple arguments from second-order perturbation theory\cite{lee}
indicate that the so-called spurious infrared divergences play an
important role in understanding asymptotic freedom in Hamiltonian
light-front QCD, so we should not expect any reasonable regulator to
remove them.

The first one-loop calculations relevant to asymptotic freedom in
light-cone gauge were completed by Thorn (see Thorn in \cite{qcd2}),
who showed that the four-gluon vertex displays asymptotic freedom to
leading order.  The first calculations for the quark-gluon vertex in
Lagrangian QCD were completed by Curci, Furmanski, and
Petronzio\cite{curci}.

In order to study contributions from various regions of phase space, I
will cut off transverse and longitudinal momenta separately, so that

\begin{equation}
k^+>\epsilon,~
|{\bf k_\perp}|<\Lambda\;.
\end{equation}

\noindent In addition to these cutoffs, I regularize the familiar
infrared divergences that occur in gauge theories for small $|{\bf
k_\perp}|$ by requiring all loop momenta to satisfy $|{\bf
k_\perp}|>\mu~.$  In a renormalization group transformation this type
of infrared problem is usually avoided, because there are always lower
and upper cutoffs. I assume that $\mu$ is much larger than all other
masses in the problem, and consider only the leading contribution to the
running coupling constant from degrees of freedom with large transverse
momenta.

We must compute a complete set of one-loop corrections to the
off-shell quark-gluon vertex function, $\Gamma(E;p_1,p_2;q)$, with $E$
being the off-shell energy. These corrections are divided into five
parts, so that $\Gamma=\Gamma_1+\cdot\cdot\cdot+\Gamma_5$.  $\Gamma_1$
includes all diagrams in which there is a gluon loop attached to the
quark line, and the emitted gluon attaches to the quark outside the
loop.  `Mass' subtractions are made to isolate the vertex
renormalization.  $\Gamma_2$ consists of diagrams in which the
outgoing gluon attaches to the quark inside the loop, and $\Gamma_3$
includes the diagrams in which the outgoing gluon attaches to the
gluon in the loop.  $\Gamma_4$ includes diagrams in which the outgoing
gluon is dressed by a fermion loop, plus the appropriate `mass'
counterterms.  Finally, $\Gamma_5$ includes diagrams in which the
outgoing gluon is dressed by a gluon loop, plus the appropriate `mass'
counterterms.  The number of diagrams required is much larger than
needed in a manifestly covariant, gauge invariant Lagrangian
calculation.

For massless quarks, the ultraviolet divergent parts of the vertex
corrections are

\begin{equation}
\tilde\Gamma_1(p_1,p_2;q)={g^3 \over 8 \pi^2} \overline{u}(p_2)
\epslash(q) u(p_1)  T^a
(-C_F) \Bigl\{2\; \logpp -3\Bigr\} \;\loglmu \;,
\end{equation}

\begin{equation}
\tilde\Gamma_2(p_1,p_2;q)={g^3 \over 8 \pi^2} \overline{u}(p_2)
\epslash(q) u(p_1)  T^a
(C_F-{C_A \over 2}) \Bigl\{2 \;\logpp -3\Bigr\}~\loglmu \;,
\end{equation}

\begin{equation}
\tilde\Gamma_3(p_1,p_2;q)={g^3 \over 8 \pi^2} \overline{u}(p_2)
\epslash(q) u(p_1)  T^a
({C_A \over 2}) \Bigl\{2 \;\logpp -3+4\; \logq
 \Bigr\}\; \loglmu  \;,
\end{equation}

\begin{equation}
\tilde\Gamma_4(p_1,p_2;q) = {g^3 \over 8 \pi^2} \overline{u}(p_2)
\epslash(q) u(p_1)  T^a  \bigl({-N_f \over 3} \bigr) \loglmu,
\end{equation}

\begin{equation}
\tilde\Gamma_5(p_1,p_2;q)={g^3 \over 8 \pi^2} \overline{u}(p_2)
\epslash(q) u(p_1)  T^a
(C_A/2) \Bigl\{{11 \over 3}-4 \; \logq  \Bigr\}\;\loglmu \;.
\end{equation}

\noindent For SU(3), $C_F=4/3$ and $C_A=3$; while $N_f$ is the number
of light quark flavors.  The sum of these individual contributions to
the ultraviolet divergent part of the vertex is

\begin{equation}
\tilde\Gamma(p_1,p_2;q)={g^3 \over 16 \pi^2} \overline{u}(p_2)
\epslash(q) u(p_1)  T^a
\Bigl\{{11 C_A \over 3}-{2 N_f \over 3}\Bigr\}\;\loglmu \;.
\end{equation}

All light-front infrared divergences cancel.  Note however, there are
infrared divergent terms of ${\cal O}(\mu\;ln(\epsilon)/\Lambda)$ that
do not diverge as $\Lambda \rightarrow \infty$ with $\epsilon$ fixed
($i.e.$, ultraviolet irrelevant). These are not shown, they do not
cancel, and they are energy-dependent. If a vertex cutoff that limits
light-front energy transfer is used, $\Lambda$ and $\epsilon$ are
related.  As a result of this relationship,  $ln(\epsilon)/\Lambda$ is
a small number when $\Lambda$ is large, and there is no ambiguity
regarding the dominant contributions to vertex renormalization at the
one-loop level.  This is a clear illustration of why we must choose
light-front energy cutoffs to obtain a well-behaved perturbative
expansion for the QCD Hamiltonian. The serious light-front infrared
divergences that forced me to introduce a second cutoff for the
electron self-energy {\it do not} appear in the vertex correction. Eq.
(182) is exactly what is obtained in Lagrangian QCD\cite{free}.

A Ward identity we might naively expect to find,
$\tilde\Gamma_1+\tilde\Gamma_2+\tilde\Gamma_3=0$, does not hold.
There are infrared divergences that violate this relationship, and it
is not just renormalization of the gluon propagator that leads to
asymptotic freedom. In fact, $\tilde\Gamma_5$ contains the logarithmic
divergence normally associated with asymptotic freedom, but it also
contains stronger divergences that have the wrong sign for asymptotic
freedom, as first noted by Thorn.  These extra divergences are not an
artifact of the cutoffs that can be removed.  They are an inevitable
consequence of using a Hamiltonian formulation and a gauge in which no
ghosts occur. This is easily understood by considering second-order
perturbation theory\cite{lee}.  It is not possible for gluon
wave-function renormalization to lead to asymptotic freedom in a
Hamiltonian formulation with a positive metric Fock space. Vacuum
polarization can only screen the bare charge.  Any regulator that
alters this result will also force us to introduce ghosts, with
disastrous effects for non-perturbative Fock space calculations.

While infrared divergences cancel in the ultraviolet divergent
corrections associated with asymptotic freedom, they are a sign of
singular long-range interactions that should significantly modify
perturbation theory at some point.  There is no reason to expect that
gluons will continue to screen the singular interactions in the QCD
Hamiltonian as the cutoff is lowered. While it is reasonable to assume
that cancellations are maintained in QED by massless photons, such
massless excitations must be removed from the spectrum in QCD
non-perturbatively. At this time it is not clear whether the infrared
divergences are disasters or welcome precursors of non-perturbative
physics that may appear in relatively simple operators.  I take the
latter view in the next section.

\section{A New Confinement Mechanism}

This section relies heavily on the discussion of positronium, because
I only require the QCD Hamiltonian determined to $\order(\alpha)$ to
discuss a simple confinement mechanism which appears naturally in
light-front QCD.  To this order the QCD Hamiltonian in the
quark-antiquark sector is almost identical to the QED Hamiltonian in
the electron-positron sector.  Of course the QCD Hamiltonian differs
significantly from the QED Hamiltonian in other sectors, and this is
essential for justifying my choice of $\H_0$ for non-perturbative
calculations.

The basic strategy for doing a sequence of (hopefully) increasingly
accurate QCD bound state calculations is identical to the strategy for
doing QED calculations up to the point I will reach in this paper.  I
use coupling coherence to find an expansion for $H^\Lambda$ in powers
of the QCD coupling constant to a finite order.  I then divide the
Hamiltonian into a non-perturbative part, $\h0$, and a perturbative
part, $\V$.  The division is based on the physical argument
that adding a parton in an intermediate state should require more
energy than indicated by the free Hamiltonian, and that as a result
these states will `freeze out' as the cutoff approaches
$\Lambda_{QCD}$. When this happens the evolution of the Hamiltonian as
the cutoff is lowered further changes qualitatively, and operators
that were consistently canceled over an infinite number of scales also
freeze, so that their effects in the few parton sectors can be studied
directly.  A one-body operator and a two-body operator arise in this
fashion, and serve to confine both quarks and gluons.

The simple confinement mechanism I outline is certainly not the final
story, but it may be the seed for the full confinement mechanism.  One
of the most serious problems we face when looking for non-perturbative
effects such as confinement is that the search itself depends on the
effect.  A candidate mechanism must be found and then shown to
self-consistently produce itself as the cutoff is lowered towards
$\Lambda_{QCD}$.

Once we find a candidate confinement mechanism, it is possible
to study heavy quark bound states with little modification of the QED
strategy.  Of course the results in QCD will differ from those in
QED because of the new choice of $\H_0$, and in higher orders
because of the gluon interactions.  When we move on to light quark
bound states, it becomes essential to introduce a mechanism for chiral
symmetry breaking\cite{mustaki,nonpert}.  Chiral symmetry breaking
cannot show up in a perturbative coupling coherence analysis unless
the cutoffs violate chiral symmetry.  The best we can hope for is
that the pion mass-squared will turn out to be negative, because this
is a clear signal that symmetry breaking will occur\cite{harivary}.
In this case, the chiral symmetry breaking operators, whose origin is
vacuum structure in equal-time calculations, will serve to raise the
pion mass to zero.  I will discuss this briefly in the last section.

When we compute the QCD Hamiltonian to $\order(\alpha)$, several
significant new features appear.  First are the familiar gluon
interactions.  In addition to the many gluon interactions found in the
canonical Hamiltonian, there are modifications to the instantaneous
gluon exchange interactions, just as there were modifications to the
electron-positron interaction.  For example, a Coulomb interaction
will automatically arise.  In addition the gluon self-energy will
differ drastically from the photon self-energy.

Recall that the photon develops a self-energy because it mixes with
electron-positron pairs, and this self energy is $\order(\alpha
\Lambda^2/P^+)$.  For small cutoffs this small bare self-energy
is exactly canceled when the photon is allowed to mix with pairs
below the cutoff.  I will not go through the calculation, but because
the gluon also mixes with gluon pairs in QCD, the gluon self-energy
acquires an infrared divergence, just as the electron did in QED.  In
QCD both the quark and gluon self-energies are proportional to $\alpha
\Lambda^2 \ln(1/\epsilon)/P^+$, where $\epsilon$ is the secondary
cutoff on parton longitudinal momenta introduced in section 11.  This
means that even when the primary cutoff $\Lambda^2$ is finite, the
energy of a single quark or a single gluon is infinite, because we are
supposed to let $\epsilon \rightarrow 0$.  One can easily argue that
this result is meaningless, because the relevant matrix elements of
the Hamiltonian are not even gauge invariant; however, since we must
live with a variational principle when doing Hamiltonian calculations,
this result may be useful.

In QED I argued that the bare electron self-energy was a complete lie,
because the bare electron mixes with photons carrying arbitrarily small
longitudinal momenta to cancel this bare self-energy and produce a
finite mass physical electron.  However, in QCD there is no reason to
believe that this perturbative mixing continues to arbitrarily small
cutoffs.  {\it There are no massless gluons in the world.}  In this
case, {\it the free QCD Hamiltonian is a complete lie and cannot be
trusted at small energies.}

On the other hand, coupling coherence gives us no choice about the
quark and gluon self-energies as computed in perturbation theory.
These self-energies appear because of the behavior of the theory at
extremely high energies.  The question is not whether large
self-energies appear in the Hamiltonian.  The question is whether
these self-energies are canceled by mixing with low-energy
multi-gluon states.  I argue that this cancellation does not occur,
and that {\it the infrared divergent quark and gluon self-energies
should be included in $\H_0$.}  The transverse scale for these energies
is the running scale $\Lambda$, and over many orders of magnitude we
should see the self-energies canceled by mixing.  However, as the
cutoff approaches $\Lambda_{QCD}$, I speculate that these
cancellations cease to occur because bound states involving gluons
arise and produce a mass gap.

But if the quark and gluon self-energies diverge, and the divergences
cannot be canceled by mixing between sectors with an increasingly
large number of partons, how is it possible to obtain finite mass
hadrons?  The parton-parton interaction also diverges, and one of my
main objectives in this section is to show that the infrared
divergence in the two-body interaction exactly cancels the infrared
divergence in the one-body operator for color singlet states.

Of course, the cancellation of infrared divergences is not enough to
obtain confinement.  The cancellation is exact regardless of the
relative motion of the partons in a color singlet state, and
confinement requires a residual interaction.  I will show that the
$\order(\alpha)$ QCD Hamiltonian produces a logarithmic potential in
both longitudinal and transverse directions.  I will not discuss
whether a logarithmic confining potential is reasonable, but to the
best of my knowledge there is no rigorous demonstration that the
confining interaction should be linear, and a logarithmic potential
may certainly be of interest phenomenologically for heavy quark bound
states\cite{quigg,quarkonia}.

The calculation of how the quark self-energy changes when a similarity
transformation lowers the cutoff on energy transfer is almost
identical to the electron self-energy calculation. Following the steps
in the section on positronium, we find the one-body operator required
by coupling coherence,

\begin{eqnarray}
\Sigma^{\Lambda}_{coh}(p)&=&
{g^2 C_F \over 8\pi^2 p^+} \Biggl\{2 y \Lambda^2 \ln\Biggl({ y^2
\Lambda^2 \over (y \Lambda^2+m^2) \epsilon }
\Biggr) -{3 \over 2} y
\Lambda^2+{1 \over 2} {y m^2 \Lambda^2 \over y \Lambda^2+m^2} \nonumber \\
&&~~~~~~~~~~~~~~~+ 3 m^2
\ln\Biggl( {m^2 \over y \Lambda^2 + m^2} \Biggr) \Biggl\} + {\cal
O}(\epsilon/y) \;.
\end{eqnarray}

The calculation of the quark-antiquark interaction required
by coupling coherence is also nearly identical to the QED
calculation.  Keeping only the infrared singular parts of the
interaction, as was done in a first calculation for QED,

\begin{eqnarray}
\tilde{V}_{coh}^{\Lambda} &=&
 - 4 g_{\Lambda}^2 C_F \sqrt{p_1^+ p_2^+ k_1^+ k_2^+}
{q_\perp^2 \over (q^+)^3}  \delta_{\sigma_1 \sigma_2}
\delta_{\lambda_1 \lambda_2} \nonumber \\
&&~~~~~~~\theta\bigl(|q^+|-\epsilon P^+\bigr)~
\theta\Biggl(\Lam-\mid p_1^-+k_1^--p_2^--k_2^- \mid
\Biggr) \nonumber \\
&&~~\Biggl[ {\theta\bigl(|p_1^- -p_2^- -q^-| -\Lambda^2 / {\cal P}^+ \bigr)
\;\; \theta\bigl(|p_1^- -p_2^- -q^-|- |k_2^- -k_1^- -q^-| \bigr)
\over p_1^- -p_2^- -q^-} \nonumber \\ &&~~~~
+{\theta\bigl(|k_2^- -k_1^- -q^-| - \Lambda^2 / {\cal P}^+ \bigr)
\;\; \theta\bigl( |k_2^- -k_1^- -q^-| - |p_1^- -p_2^- -q^-| \bigr)
\over k_2^- -k_1^- -q^-} \Biggr]  \;. \nonumber \\
\end{eqnarray}

The instantaneous gluon exchange interaction is

\begin{eqnarray}
V^\Lambda_{instant} &=& 
- 4 g_{\Lambda}^2 C_F \sqrt{p_1^+ p_2^+ k_1^+ k_2^+}
\Biggl({1 \over q^+}\Biggr)^2 \delta_{\sigma_1 \sigma_2}
\delta_{\lambda_1 \lambda_2} \nonumber \\
&&~~\times \theta\bigl(|q^+|-\epsilon P^+\bigr)~
\theta\Biggl(\Lam-\mid p_1^-+k_1^--p_2^--k_2^- \mid
\Biggr)
\;.
\end{eqnarray}

\noindent Just as in QED the coupling coherent interaction induced by
gluon exchange above the cutoff partially cancels instantaneous gluon
exchange.  For the discussion of confinement the part of $V_{coh}$
that remains is not important, because it produces a Coulomb
interaction.  However, the part of the instantaneous interaction that
is not canceled is

\begin{eqnarray}
\tilde{V}^\Lambda_{instant} &=& - 4 g_{\Lambda}^2 C_F
\sqrt{p_1^+ p_2^+ k_1^+ k_2^+}
\Biggl({1 \over q^+}\Biggr)^2 \delta_{\sigma_1 \sigma_2}
\delta_{\lambda_1 \lambda_2} \nonumber \\
&&\times
\theta\Biggl(\Lam-\mid p_1^-+k_1^--p_2^--k_2^- \mid \Biggr)
\theta\bigl(|p_1^+-p_2^+|-\epsilon P^+\bigr) \nonumber \\
&&\times \Biggl[ \theta\bigl(\Lam -|p_1^- -p_2^- -q^-|  \bigr)
\;\; \theta\bigl(|p_1^- -p_2^- -q^-|- |k_2^- -k_1^- -q^-| \bigr) +
\nonumber \\
&&\theta\bigl(\Lam-|k_2^- -k_1^- -q^-| \bigr)
\;\; \theta\bigl( |k_2^- -k_1^- -q^-| - |p_1^- -p_2^- -q^-| \bigr) \Biggr]
\;.
\end{eqnarray}

Note that this interaction contains a cutoff that projects onto
exchange energies below the cutoff, because the interaction has been
screened by gluon exchange above the cutoffs. This interaction can
become important at long distances, if parton exchange below the
cutoff is dynamically suppressed.  In QED I argued that this singular
long range interaction is exactly canceled by photon exchange below
the cutoff, because such exchange is not suppressed no matter how low
the cutoff becomes.  Photons are massless and experience no
significant interactions, so they are exchanged to arbitrarily low
energies as effectively free photons.  This cannot be the case for
gluons.

For the discussion of confinement, I will place only the most singular
parts of the quark self-energy and the quark-antiquark interaction in
$\h0$.  To see that all infrared divergences cancel and that the
residual long range interaction is logarithmic, we can study the
matrix element of these operators for a quark-antiquark state,

\begin{eqnarray}
|\Psi(P)\rangle &=& \sum_{\sigma \lambda} \sum_{rs}
\int {dp^+ d^2p_\perp \over
16\pi^3 p^+} {dk^+ d^2k_\perp \over 16\pi^3 k^+} \sqrt{p^+ k^+}
16\pi^3 \delta^3(P-p-k) \nonumber \\
&&~~~~~~~~~~~~~~\phi(p,\sigma,r;k,\lambda,s) b^{r\dagger}(p,\sigma)
d^{s\dagger}(k,\lambda) |0\rangle \;,
\end{eqnarray}

\noindent where $r$ and $s$ are color indices and I will choose $\phi$to be
a color singlet and drop color indices.  The cancellations we
find do not occur for the color octet configuration.  The matrix
element is,

\begin{eqnarray}
\langle \Psi(P)|\H_0|\Psi(P')\rangle &=& 16\pi^3 P^+ \delta^3(P-P') \times
\nonumber \\
&\Biggl\{& \int {dy d^2\kappa \over 16\pi^3}
\Biggl[{g_\Lambda^2 C_F \Lambda^2 \over 2\pi^2 P^+} \; \ln\bigl({1 \over
\epsilon}\bigr) \Biggr] |\phi(\kappa,y)|^2 \nonumber \\
&-&{4 g_\Lambda^2 C_F \over P^+} \int {dy_1 d^2\kappa_1 \over 16\pi^3}
{dy_2 d^2\kappa_2 \over 16\pi^3} \theta\Biggl(\Lambda^2-\mid
{\kappa_1^2+m^2 \over y_1(1-y_1)}-{\kappa_2^2+m^2 \over y_2(1-y_2)}
\mid\Biggr) \nonumber \\
&&~~\times \Biggl[
\theta\Biggl(\Lambda^2- \mid {\kappa_1^2+m^2 \over y_1} - {\kappa_2^2
+m^2 \over y_2}-{(\kappa_1-\kappa_2)^2 \over |y_1-y_2|}\mid \Biggr)
\nonumber \\ &&~~~~~~
\times \theta\Biggl(\mid {\kappa_1^2+m^2 \over y_1} - {\kappa_2^2
+m^2 \over y_2}-{(\kappa_1-\kappa_2)^2 \over |y_1-y_2|}\mid -
\nonumber \\ &&~~~~~~~~~~~~~~~~~~~~~~~~~~
\mid {\kappa_2^2+m^2 \over 1-y_2} - {\kappa_1^2 +m^2
\over 1-y_1}-{(\kappa_1-\kappa_2)^2 \over |y_1-y_2|}
\mid\Biggr)  \nonumber \\ && ~~~~~~ +
\theta\Biggl(\Lambda^2-\mid {\kappa_2^2+m^2 \over 1-y_2} - {\kappa_1^2 +m^2
\over 1-y_1}-{(\kappa_1-\kappa_2)^2 \over |y_1-y_2|}
\mid\Biggr)  \nonumber \\ && ~~~~~~
\times \theta\Biggl(\mid {\kappa_2^2+m^2 \over 1-y_2} - {\kappa_1^2 +m^2
\over 1-y_1}-{(\kappa_1-\kappa_2)^2 \over |y_1-y_2|}\mid -
\nonumber \\ &&~~~~~~~~~~~~~~~~~~~~~~~~~~
\mid {\kappa_1^2+m^2 \over y_1} - {\kappa_2^2
+m^2 \over y_2}-{(\kappa_1-\kappa_2)^2 \over |y_1-y_2|}\mid \Biggr)
\Biggr]
\nonumber \\ &&~~
\times \theta\Bigl(\mid y_1-y_2 \mid -\epsilon\Bigr)
\Biggl({1 \over y_1-y_2 }\Biggr)^2 \phi^*(\kappa_2,y_2)
\phi(\kappa_1,y_1)
\Biggr\}.
\end{eqnarray}

\noindent Here I have chosen a frame in which the center-of-mass
transverse momentum is zero, assumed that the longitudinal momentum
scale introduced by the cutoffs is that of the bound state, and used
Jacobi coordinates,

\begin{equation}
p_i^+=y_i P^+\;,\;\;p_{i\perp}=\kappa_i\;\;;\;\;\; k_i^+=(1-y_i)P^+\;,
\;\; k_{i\perp}=-\kappa_i \;.
\end{equation}

The first thing I want to do is show that the last term is divergent
and the divergence exactly cancels the first term.  My demonstration
is not elegant, but it is straightforward.  The divergence results
from the region $y_1 \sim y_2$.  In this region the second and third
cutoffs restrict $(\kappa_1-\kappa_2)^2$ to be small compared to
$\Lambda^2$, so we should change variables,

\begin{equation}
Q={\kappa_1+\kappa_2 \over 2} \;,\;\;Y={y_1+y_2 \over 2} \;\;;\;\;\;
q=\kappa_1-\kappa_2 \;,\;\;y=y_1-y_2 \;.
\end{equation}

\noindent Using these variables we can approximate the above
interaction near $q=0$ and $y=0$.  The double integral becomes

\begin{eqnarray}
{-4 g_\Lambda^2 C_F \over P^+} \int {dY d^2Q \over 16\pi^3} {dy d^2q \over
16\pi^3} \theta(1-Y) \theta(Y) |\phi(Q,Y)|^2 \nonumber \\
~~~~~\theta\Biggl(\Lambda^2-{q^2 \over |y|}\Biggr) \theta\bigl(
|y|-\epsilon\bigr) \theta\bigl(\eta-|y|\bigr)
~ \Biggl({1 \over y}\Biggr)^2 \;,
\end{eqnarray}

\noindent where $\eta$ is an arbitrary constant that restricts $|y|$
from becoming large. Completing the $q$ and $y$ integration we get

\begin{equation}
-{g_\Lambda^2 C_F \Lambda^2 \over 2\pi^2 P^+} \ln\Biggl({1 \over
\epsilon}\Biggr) \int {dY d^2Q \over 16\pi^3}
\theta(1-Y) \theta(Y) |\phi(Q,Y)|^2 \;.
\end{equation}

The divergent part of this exactly cancels the first term on the
right-hand side of Eq. (188).  This cancellation occurs for any state,
and this cancellation is unusual because it is between the expectation
value of a one-body operator and the expectation value of a two-body
operator.  The cancellation resembles what happens in the Schwinger
model and is easily understood.  It results from the fact that a color
singlet has no color monopole moment.  If the state is a color octet
the divergences are both positive and cannot cancel.  Since the
cancellation occurs in the matrix element, {we can let $\epsilon
\rightarrow 0$ before diagonalizing} $\H_0$.

The fact that the divergences cancel exactly does not indicate that
confinement occurs.  This requires the residual interactions to
diverge at large distances, which means small momentum transfer.
Equivalently, we need the color dipole self-energy to diverge if the
color dipole moment diverges because the partons separate to large
distance.

My analysis of the residual interaction is neither elegant nor
complete.  I will only establish that the interaction is logarithmic
in the longitudinal direction at zero transverse separation and
logarithmic in the transverse direction at zero longitudinal
separation.  I will not compute the range of the interaction or
consider separations in both longitudinal and transverse directions.
The analysis is complicated by the step function cutoffs, by the
presence of the infrared divergence analyzed above, and by the lack of
simple variables that display a kinematic rotational invariance.

In order to avoid the infrared divergence, I will compute spatial
derivatives of the potential.  This removes the constant infrared
divergence, but it also hides dependence on the range of the
interaction; both of which are easily seen by considering derivatives
of $c_1+c_2 \ln(r/a)$.

Let us first consider the potential in the longitudinal direction.
Given a momentum space expression, we can set $x_\perp=0$ and the
Fourier transform of the longitudinal interaction requires the
transverse momentum integral of the potential,

\begin{equation}
{\partial \over \partial z} V(z) = \int {d^3 q \over (2\pi)^3} i q_z
V(q_\perp,q_z) e^{i q_z z} \;.
\end{equation}

\noindent We are interested only in the long range potential, so we
can assume that $q_z$ is arbitrarily small during the analysis and
approximate the step functions accordingly.

For our interaction this leads to

\begin{eqnarray}
{\partial \over \partial x^-} V(x^-) &=& -4 g_\Lambda^2 C_F P^+ \int {dq^+
d^2q_\perp \over 16\pi^3} \theta\bigl(P^+-|q^+|\bigr) \nonumber \\
&&~~~~~\theta\bigl(\Lam-{q_\perp^2 \over |q^+|}\bigr) \Bigl( {1
\over q^+} \Bigr)^2 (i q^+) e^{i q^+ x^-} \;.
\end{eqnarray}

\noindent Completing the $q_\perp$ integration we have

\begin{eqnarray}
{\partial \over \partial x^-} V(x^-) &=&  - {i g_\Lambda^2 C_F \Lambda^2
\over 4\pi^2} \int dq^+ \theta\bigl(P^+-|q^+|\bigr) {q^+ \over |q^+|}
e^{i q^+ x^-} \nonumber \\
&=& {g_\Lambda^2 C_F \Lambda^2 \over 2\pi^2} \int_0^{P^+} dq^+
\sin\bigl({q^+ x^-}\bigr) \nonumber \\
&=& {g_\Lambda^2 C_F \Lambda^2 \over 2\pi^2} \Biggl( {1 \over x^-} -
{\cos\bigl( P^+ x^- \bigr) \over x^-}\Biggr) \nonumber \\
&=& {g_\Lambda^2 C_F \Lambda^2 \over 2\pi^2}  {\partial \over \partial x^-}
\;
\ln\bigl(|x^-|) + short~range \;.
\end{eqnarray}

\noindent To see that the term involving a cosine in the next-to-last
line produces a short range potential, simply integrate it.  At large
$|x^-|$, which is the only place we can trust our approximations for
the original integrand, this yields a logarithmic potential, as
promised.

Next consider the potential in the transverse direction.  Here we can
set $x^-=0$ and get

\begin{eqnarray}
{\partial \over \partial x_\perp^i} V(x_\perp) &=& -4 g_\Lambda^2 C_F P^+
\int {dq^+ d^2 q_\perp \over 16\pi^3} \theta\bigl(P^+-|q^+|\bigr) \nonumber
\\ &&~~~~~\theta\Biggl(\Lam-{q_\perp^2 \over |q^+|}\Biggr) \Biggl( {1
\over q^+}\Biggr)^2 (i q_\perp^i) e^{i {\bf q}_\perp \cdot {\bf
x}_\perp} \;.
\end{eqnarray}

\noindent Here I have used the fact that the integration is dominated
by small $q^+$ to simplify the integrand again.  Completing the $q^+$
integration this becomes

\begin{eqnarray}
{\partial \over \partial x_\perp^i} V(x_\perp) &=& -{i g_\Lambda^2 C_F
\Lambda^2 \over 2\pi^3} \int d^2q_\perp {q_\perp^i \over q_\perp^2} e^{
i {\bf q}_\perp \cdot {\bf x}_\perp}~+~short~range \nonumber \\
&=& {g_\Lambda^2 C_F \Lambda^2 \over \pi^2} {x_\perp^i \over x_\perp^2}
\nonumber \\
&=& {g_\Lambda^2 C_F \Lambda^2 \over \pi^2} {\partial \over \partial
x_\perp^i} \;\ln(|{\bf x_\perp}|) \;.
\end{eqnarray}

\noindent Once again, this is the derivative of a logarithmic
potential, as promised.  The strength of the long-range logarithmic
potential is not spherically symmetrical in these coordinates, with
the potential being larger in the transverse than in the longitudinal
direction.  A simple numerical analysis indicates that the departure
goes like $\cos^2(\theta)$.

This violation of simple nonrelativistic rotational invariance has
interesting implications for the constituent wave function. It
survives when the quark mass is taken into account.  For massive
quarks in the weak coupling limit, low-lying states are
nonrelativistic Coulombic bound states, and the long-range potential
is not important until the bound state radius approaches the distance
at which the logarithmic potential is as strong as the Coulomb
potential.  Even for low-lying states, however, the presence of a {\it
weak} long-range potential that violates rotational invariance affects
the spectrum and the states.  We can study how violations cancel
perturbatively as a clue to how the symmetry might be dynamically
restored for the low-lying states non-perturbatively.  Perturbatively
this depends on a cancellation between the instantaneous interaction
and gluon exchange, and it becomes highly non-trivial to show that
mixing with low-lying states containing an extra gluon can restore
rotational invariance non-perturbatively.  I believe it requires a
calculation that allows string-like gluon configurations between heavy
quarks when they are separated by a distance comparable to the size of
the lowest-lying bound state containing an extra gluon.

Had we computed the quark-gluon or gluon-gluon interaction, we would
find essentially the same residual long range two-body interaction in
every Fock space sector.  In QCD gluons have a divergent self-energy
and experience divergent long range interactions with other partons if
we use coupling coherence.  In this sense, the assumption that gluon
exchange below some cutoff is suppressed is consistent with the
Hamiltonian that results from this assumption.  To show that gluon
exchange is suppressed when $\Lambda \rightarrow \Lambda_{QCD}$,
rather than some other scale ({\it i.e.}, zero as in QED), a
non-perturbative calculation of gluon exchange is required.  This is
exactly the calculation bound state perturbation theory produces, and
bound state perturbation theory suggests how the perturbative
renormalization group calculation might be modified to generate these
confining interactions self-consistently.

If perturbation theory, which produced this potential, continues to be
reasonable, this long range potential will be exactly canceled in QCD
just as it is in QED. We need this exact cancellation of new forces to
occur at short distances and turn off at long distances, if we want
asymptotic freedom to give way to a simple constituent confinement
mechanism. At short distances the divergent self-energies and two-body
interactions cannot be ignored, but they should exactly cancel
pairwise if these divergences appear only when gluons are emitted and
absorbed in immediately successive vertices, as I have speculated.
The residual interaction must be analyzed more carefully at short
distances, but in any case a logarithmic potential is less singular
than a Coulomb interaction, which does not disturb asymptotic
freedom.  This is the easy part of the problem.  The hard part is
discovering how the potential can survive at any scale.

A perturbative renormalization group will not solve this problem.  The
key I have suggested is that interactions between quarks and gluons in
the intermediate states required for cancellation of the potential
will eventually produce a non-negligible energy gap.  I am unable to
detail this mechanism without an explicit calculation, but let me
sketch a naive picture of how this might happen.

Focus on the quark-antiquark-gluon intermediate state, which mixes
with the quark-antiquark state to screen the long range potential.
The free energy of this intermediate state is always higher than that
of the quark-antiquark free energy, as is shown using a simple
kinematic argument\cite{nonpert}.  However, if the gluon is massless,
the energy gap between such states can be made arbitrarily small.  As
we use a similarity transformation to run a vertex cutoff on energy
transfer, mixing persists to arbitrarily small cutoffs since the gap
can be made arbitrarily small.  Wilson has suggested using a gluon
mass to produce a gap that will prevent this mixing from persisting as
the cutoff approaches $\Lambda_{QCD}$\cite{nonpert}, and I am
suggesting a slightly different mechanism.

If we allow two-body interactions to act in both sectors to all
orders, even the Coulomb interaction can produce quark-antiquark and
quark-antiquark-gluon bound states.  In this respect QCD again differs
qualitatively from QED because the photon cannot be bound and the
energy gap is always arbitrarily small even when the electron-positron
interaction produces bound states.  If we assume that a fixed energy
gap arises between the quark-antiquark bound states and the
quark-antiquark-gluon bound states, and that this gap establishes the
important scale for non-perturbative QCD, these states must cease to
mix as the cutoff goes below the gap.

Of course, this qualitative change in the evolution of the Hamiltonian
will not occur all at once.  Long before a bound state energy gap is
encountered directly, the continuum density of states will be shifted
from the free density and alter the mixing.  Note that small shifts
will be unable to alter the cancellation of infrared divergences,
because these effects are more singular than all others.

If we want to include the effect of confining interactions at any
scale, it is not possible to use momentum eigenstates, because their
energy diverges.  Since asymptotic freedom is most easily seen in
momentum space (which diagonalizes the fixed point Hamiltonian), any
new basis used to include confinement should closely resemble momentum
eigenstates at short distances and high energies.  This suggests a
wave packet (or wavelet) analysis, as long advocated by
Wilson\cite{nonpert}.

An important ingredient for any calculation that tries to establish
confinement self-consistently is a {\it seed mechanism}, because it is
possible that it is the confining interaction itself which alters the
evolution of the Hamiltonian so that confinement can arise.  I have
proposed a simple seed mechanism whose perturbative origin is
appealing because this allows the non-perturbative evolution induced
by confinement to be matched on to the perturbative evolution required
by asymptotic freedom.

\section{Future Prospects}

Instead of summarizing what I have said, I want to use this final
section to discuss the large number of problems left unsolved.

The strategy I have outlined will no doubt evolve as new results
reveal its flaws. While it is possible that a constituent picture will
eventually emerge from QCD, it seems unlikely that it will take the
simple form envisioned in this paper.  The quarks and gluons I have
employed are still very close to the bare quarks and gluons one sees
when studying QCD at very large cutoffs, near the free field fixed
point.  I have painted a picture in which the Hamiltonian evolves
perturbatively over an infinite number of scales, with each scale
providing a rather simple layer of structure for the partons and
slightly altering the vertices and direct interactions, slowly
building up new interactions required by the hidden symmetries of the
theory.

I have assumed that this smooth evolution changes rather abruptly at
some scale near $\Lambda_{QCD}$, giving way to a new evolution in
which relatively simple non-canonical parton structure and few-body
interactions begin to dominate the dynamical structure of QCD.  I have
guessed that a few simple operators that display singular infrared
behavior will emerge after being canceled perturbatively over many
scales to prevent many-parton intermediate states from continuing to
cancel their effects.  A constituent picture emerges because these
operators require only a few constituents to produce non-perturbative
effects, and they self-consistently turn off the many-parton effects
that would block a simple constituent picture from emerging.

Such an abrupt change is too naive, and must give way to a more
reasonable evolution in which singular interactions gradually cause
the QCD Hamiltonian to depart from its perturbative trajectory.
Whether such a gradual change can be self-consistently approximated by
considering a sequence of structural layers, each of which immediately
requires only a few partons, is not clear.  Perturbative evolution
never requires us to explicitly consider the effects of more than a
few partons on subsequent parton or interaction structure, because at
any order of perturbation theory a limited number of partons can be
generated.  The strength of the renormalization group lies in its
ability to unravel an arbitrarily large number of scales when there is
some sort of self-similarity that can be uncovered.  This
self-similarity is most easily revealed when the Hamiltonian
trajectory lies near a fixed point so that the Hamiltonian quite
clearly continues to reproduce itself in form.  It is likely that new
ideas will be required to solve QCD.  For example, instead of staying
near a fixed point, we might imagine a physical trajectory that lies
near a reference trajectory, allowing us to use a new set of
perturbative renormalization group equations that describe the
evolution of the deviations of the physical trajectory from the
reference trajectory.

My outline of the three renormalization problems encountered in
Hamiltonian light-front field theory is motivated by the study of
perturbation theory, in which momentum eigenstates are reasonable
approximations.  As the Hamiltonian leaves the vicinity of the
critical Gaussian fixed point, this ceases to be true, and the power
counting rules that govern my classification of light-front
renormalization problems may cease to be relevant.

There is little work on light-front infrared renormalization, and
field theorists typically avoid formalisms that cause the infrared
region of phase space to appear radically different from other
regions. Wilson speculates that infrared renormalization hides complex
structure that may prove central to the solution of
QCD\cite{nonpert}.  This speculation seems strange if one only studies
perturbation theory, where singular infrared effects are typically
canceled; and it is difficult to understand how field theory can
exploit vastly different infrared longitudinal structure and
ultraviolet transverse structure, and still maintain both boost and
rotational symmetries.  Progress in this direction is slowed by a
common unwillingness to explore totally new territory where old
maps are of limited use and where we must expect to get lost often
while slowly developing new maps.

Perhaps the weakest point in my description of light-front
renormalization problems is the problem of exactly zero longitudinal
momentum.  There are no `vacuum modes' in the formulation I use, so
there can be no direct confrontation of dynamical effects produced by
degrees of freedom with zero longitudinal momentum.  Asymptotic
freedom severely constrains the operators the vacuum can produce,
unless they are relevant, marginal, or nonlocal.  If they are relevant
or marginal, we can hope to discover them when doing hadronic
structure calculations, and control them using a renormalization group.

Perhaps the most interesting example is provided by chiral symmetry
breaking.  As far as I know, the perturbative renormalization group
will not produce a chiral symmetry breaking operator.  While it is
true that a quark mass is produced, this operator does not break
chiral symmetry in light-front coordinates\cite{nonpert,mustaki}.
The only operator that violates chiral symmetry in the canonical
Hamiltonian is a relevant quark-gluon vertex that is proportional to a
quark mass.  One would normally assume that this operator is only
responsible for explicit chiral symmetry breaking, and is not produced
by spontaneous chiral symmetry breaking.  One radical possibility is
that the pion will emerge as a massless particle without requiring any
chiral symmetry breaking operator in the Hamiltonian, because chiral
symmetry takes a new form in light-front coordinates.  I do not think
that this is likely.

If the chiral symmetry breaking operator is local in the transverse
direction, it should be relevant.  The reason is simple.  If the
operator produces observable hadronic mass splittings at the scale
$\Lambda_{QCD}$, it cannot be infinitesimal at this scale.  Any finite
irrelevant operator tends to blow up exponentially in magnitude as one
runs the cutoff to larger values, and this makes it difficult to
imagine how asymptotic freedom could survive.  While marginal
operators do not blow up exponentially, their strength tends to
decrease only logarithmically and typically they increase
logarithmically.  Again, since asymptotic freedom is itself only a
logarithmic effect, it is hard to see how it can survive the
introduction of a new marginal operator.  On the other hand, a
relevant operator of reasonable strength at small cutoffs will
decrease in magnitude at an exponential rate as the cutoff is raised,
causing no trouble for asymptotic freedom.

If the chiral symmetry breaking operator is nonlocal, it will be much
more difficult to find it, and our best hope is to study longitudinal
scaling properties of the theory.  The renormalization group machinery
I developed is useful for running a transverse cutoff, but we must run
a longitudinal momentum cutoff to understand the longitudinal scaling
properties of QCD.  It is fairly easy to show that a longitudinal
renormalization group will not have a simple perturbative realization
with a free Hamiltonian fixed point\cite{perryrg}.  When a
longitudinal cutoff is changed, states with nearly the same energies
that coupled with the initial cutoff cease to couple with the new
cutoff, and small energy denominators appear in the perturbative
expansion of the transformation.  This does not mean no longitudinal
renormalization group exists, but it means that we must develop a
non-perturbative renormalization group to study longitudinal scaling.

There are certainly many calculations in 1+1 dimensions that have not
been done, and many of these can serve as publishable pedagogical
exercises.  However, 1+1 dimensions differs drastically from 3+1
dimensions and can even be more difficult in some cases because there are
an infinite number of renormalizable operators.  In the Schwinger
model, perhaps the most interesting remaining problem is to show that
the vacuum structure of the theory that I ignore has some effect on
finite momentum observables ({\it i.e.}, the stuff a 1+1 dimensional
experimentalist could see, and not just the stuff a 1+1 dimensional
theorist might worry about).  If this can be shown, the Schwinger
model might provide guidance for how we should fix vacuum counterterms
in QCD.

There are a large number of simple 3+1 dimensional problems that have
not been done, even in perturbation theory.  Many theorists have been
interested in how to map light-front perturbation theory onto
manifestly covariant Feynman perturbation theory.  In the process of
drawing such maps, however, little or no attention is paid to the need
to perform non-perturbative calculations, and tricks are used that can
only be applied in perturbation theory.  Moreover, these maps do not
exploit the unique strengths of light-front coordinates and may even
seem to indicate that these strengths are flaws.  Pick any theory, use
a cutoff that violates symmetries, and you will have a difficult time
finding calculations in the literature.

The non-perturbative renormalization group pioneered by Wilson is in
its infancy and is not even completely defined yet.  Very few
theorists have applied this type of formalism to light-front problems,
but I believe this situation will change, because there are no other
realistic alternatives.  What we need are tractable non-perturbative
renormalization group calculations.  I have mentioned a possibility I
would like to explore.  If we do not insist that $\H_0$ is a fixed
point, but allows it to contain interactions that do not change
particle number, a non-perturbative renormalization group might be
developed that preserves many of the features of the perturbative
renormalization group.  The problem with this type of idea is that
once $\H_0$ contains an interaction, even the linear approximation of
the transformation requires the non-perturbative solution of many-body
problems.  Such problems are typically intractable, so we must
approximate the linear transformation.  Wilson's phase space cell
techniques may prove useful, especially if we tailor the interactions
to take a simple form in a wavelet basis, as is done in shell model
calculations for example.

Coupling coherence is a recent development and very few calculations
have been completed.  The most important issue is whether it can be
applied usefully to non-perturbative renormalization group
calculations.  The only perturbative calculations I have completed
that are non-perturbative in the couplings use an $\hbar$ expansion.
Since the basic idea requires all new counterterms to go to zero as a
canonical coupling goes to zero, the statement of the principle is
perturbative in nature.  In order to use the idea non-perturbatively,
it may be necessary to construct a trajectory that extends to a
Gaussian fixed point.  If this is not possible, other ideas may be
required to fix all counterterms, such as the restoration of
symmetries violated by the cutoffs.

There are many applications of coupling coherence in QED and QCD that
should be studied.  For example, in QED one can perform many
perturbative calculations to test whether coupling coherence restores
gauge invariance.  This happens to lowest order in electron-electron
scattering, and to lowest order coupling coherence supplies the
counterterms required to insure that the physical photon is massless,
but higher order calculations must be completed and additional
observables checked.

Coupling coherence can be applied to operators other than the
Hamiltonian.  An important example is the electromagnetic current
itself.  After a similarity transformation has been defined to bring
the Hamiltonian towards the diagonal, it must be applied to all other
observables.  After applying a similarity transformation to the
canonical current operator, one will find many new operators.  The
coherent current operator will not change in form, but will depend on
the cutoff only because the physical couplings and masses depend on
the cutoff.  This means, for example, that the current operator must
contain two-body terms, and these new terms can be tested by determining
whether the new current is conserved.  As an additional interesting
example, one can check that the axial anomaly is correctly produced.

There should be many applications of coupling coherence outside of
light-front field theory.  For example, one might perform
non-perturbative tests of the idea on the lattice.  Chiral symmetries
might supply important examples.

The similarity transformation is also a recent development that has
been applied to very few problems.  It is remarkable to me that this
formalism was not developed long ago by many-body theorists, who have
long used effective Hamiltonian techniques, and who have long faced
the problems posed by small energy denominators.  The similarity
transformation may have important applications in the nuclear shell
model, for example; and in condensed matter systems.

My examples were all perturbative, and it is important to find
non-perturbative approximations for the transformation.  It is likely
that the formalism developed by G{\l}azek and Wilson will evolve as
new applications appear.  Even in my perturbative studies I have found
it convenient to modify their formulation so that coupling coherence
is possible, and there are likely to be many similarity
transformations from which we can choose depending on the problem.
Any problem for which non-perturbative renormalization group
calculations exist is a candidate for application of the similarity
transformation, with Wilson's own work on fixed source
theory\cite{wilson70} and the Kondo problem\cite{wilsonrg} providing two
obvious examples.

The number of problems that we can study in QED is extremely large.
Most of these problems are of interest for the refinement of the
formalism, both analytically and numerically; but it is possible that
light-front field theory will offer non-perturbative approximations
not available in other formalisms, as Brodsky has advocated for
years.  In the immediate future, it is important to thoroughly
understand the cutoff dependence of the non-perturbative
approximations I have suggested, and to demonstrate that we can
readily obtain fine structure for positronium.  The integral equations
I have set up in this paper need to be solved numerically, and
extended to higher orders.  Of particular interest are excited states
with non-zero angular momentum, because rotational invariance is
violated when the second-order Hamiltonian is treated
non-perturbatively.  These violations should be small for low-lying
states, and the leading perturbative corrections should start to
restore rotational invariance.  In both QED and QCD, rotational
invariance will provide an important test for the approximations and
can be used to fix the strength of interactions non-perturbatively.

Hydrogen offers a kinematic scenario that resembles the heavy plus
light quark bound states, and the non-perturbative approximations
required to solve it accurately will differ from those required for
positronium.  An ambitious goal is to reproduce the Lamb shift.  The
basic strategy has been known since Bethe's
calculation\cite{schwinger}, but an accurate reproduction of the Lamb
shift by a calculation that has not been specifically tailored for
this purpose alone is quite demanding.

There are few perturbative light-front QCD calculations in existence,
even at the level of one loop; and the ones that do exist need to be
reproduced using the similarity transformation and coupling coherence.
These calculations offer serious warm-up exercises for anyone who wants
to study QCD non-perturbatively, at the minimum.  If it is possible to
carry such calculations to two loops, many lessons will no doubt be
learned.  Thorn discusses the interesting new mechanisms required for
asymptotic freedom to arise in gluon self-interactions (see Thorn in
Ref. \cite{qcd2}), and I have further emphasized the role of infrared
cancellations in the quark-gluon vertex.  I do not know whether it
will be possible to complete all-orders proofs for light-front QCD
with regulators that violate covariance and gauge invariance, but
these would certainly be useful in establishing the connection of the
light-front formalism with the well-established Lagrangian formalism.

Finally, the study of non-perturbative light-front QCD\cite{nonpert}
has barely been initiated.  The simple integral equations that result
from minimizing the approximate energy I have derived need to be
solved numerically for heavy quark systems, for glueballs, and for
lighter quarks; if only to see whether these simple ideas provide a
reasonable starting point. As in QED, it is important to see whether
the results display an approximate rotational invariance, since this
symmetry is violated in the non-perturbative calculations.

While this is being done, it is also necessary to carry the
calculation of the QCD Hamiltonian to third and fourth order, to see
whether the program is practical, whether the results improve, and
whether the important non-perturbative aspects of the second-order
calculation survive.

One of the primary advantages of deriving a constituent picture from
QCD is that eigenstates need not be completely determined by the
minimal valence configuration.  While the ground state for any set of
quantum numbers should be dominated by the minimal valence
configuration, the program may still be practical if states containing
a few additional partons are important for precision.  Extra partons,
especially gluons, should be especially important in excited states.
My guess is that even the first excitation will begin to contain extra
gluons, and it is even possible that hadrons in a fixed angular
momentum multiplet will have different constituent structure.  Since
the rotation operators contain interactions that change particle
number, it is possible for one polarization to be dominated by the
minimal valence configuration while another includes an additional
gluon, for example.

Ideally the simple second order Hamiltonian will produce reasonable
results, and hopefully these results can be significantly improved by
altering the strength of the second order operators by hand, using
covariance as a guide.

At this point we simply do not know what will happen.  Even if a
constituent picture does arise in light-front QCD, it may differ from
the constituent quark model in many significant ways.

I wish I could offer more answers.  It would be nice to have at least
a few solid calculations that fit experimental data, even if the
results are no better than those obtained by other methods.  The fact
of the matter is that the constituent quark model, the bag model, the
Skyrme model, Schwinger-Dyson calculations, lattice QCD, and other
approaches to the problem of bound states in QCD, are far ahead of
light-front QCD when it comes to producing numbers.  Hopefully I have
shown that this program is rapidly approaching the point where numbers
will be produced.

Finally I come to my main conclusion.  This is an exciting {\it new}
field, especially for young theorists.  There is opportunity where
there are many new problems that require developments at the level of
{\it fundamental} field theory.  While many may see this need as a
weakness in this field, I hope that many young theorists will see it
as the field's most appealing feature.  {\it The added bonus is that
these new problems are most interesting in QCD!}

\hspace{3in}
\section{Acknowledgments}

I want to thank the organizers of Hadrons '94, especially Victoria
Herscovitz and Cesar Vasconcellos, for their exceptional hospitality
and for the opportunity to visit their beautiful country.  I also
extend warm thanks to Gastao Krein for his hospitality and for
providing me the chance to calculate with few interruptions for an
entire week! Many of the results in this paper were finalized in
S{\~a}o Paulo.  I also want to thank Ken Wilson, who has patiently
taught me much of what makes sense in this paper.  I have also learned
a great deal from Stan G{\l}azek and Avaroth Harindranath, and I have
had fruitful discussions with many other theorists.  Finally, it is a
pleasure to thank my students Brent Allen, Martina Brisudov{\'a}, and
Billy Jones, who have corrected and extended much of this work. This work
was supported by the National Science Foundation under Grant Nos.
PHY-9102922 and PHY-9409042, and the Presidential Young Investigator
Program under Grant PHY-8858250.


\end{document}